\documentclass[fleqn,usenatbib]{mnras}

\usepackage{newtxtext,newtxmath}

\usepackage[T1]{fontenc}

\DeclareRobustCommand{\VAN}[3]{#2}
\let\VANthebibliography\thebibliography
\def\thebibliography{\DeclareRobustCommand{\VAN}[3]{##3}\VANthebibliography}

\usepackage{graphicx}
\usepackage{amsmath}
\usepackage{threeparttable}
\usepackage{multirow}
\usepackage{inputenc}
\usepackage{xcolor}
\usepackage{adjustbox}
\usepackage{chemformula}

\title[Degeneracies associated with \textit{HST} WFC3 transmission spectra of exoplanetary atmospheres]{Parameter degeneracies associated with interpreting \textit{HST} WFC3 transmission spectra of exoplanetary atmospheres}

\author[A. Novais et al.]{
Aline Novais$^{1}$\thanks{E-mail: aline.novais@fysik.lu.se}
Chloe Fisher$^{2}$,
Luan Ghezzi$^{3}$,
Daniel Kitzmann$^{4,5}$,
Brian Thorsbro$^{1,6}$, and
Kevin Heng$^{7,8,9,10}$
\\
$^{1}$Division of Astrophysics, Department of Physics, Lund University, Box 118, 221 09, Lund, Sweden\\
$^{2}$University of Oxford, Denys Wilkinson Building, Keble Road, Oxford, OX1 3RH, United Kingdom\\
$^{3}$Valongo Observatory, Federal University of Rio de Janeiro, Ladeira do Pedro Antonio, 43, 20080-090, Rio de Janeiro, Brazil\\
$^{4}$Space Research and Planetary Sciences, Physics Institute, University of Bern, Gesellschaftsstrasse 6, CH-3012 Bern, Switzerland\\
$^{5}$Center for Space and Habitability, University of Bern, Gesellschaftsstrasse 6, CH-3012 Bern, Switzerland\\
$^{6}$Observatoire de la Côte d’Azur, CNRS UMR 7293, BP4229, Laboratoire Lagrange, F-06304 Nice Cedex 4, France\\
$^{7}$University Observatory Munich, Ludwig Maximilian University, Scheinerstrasse 1, D-81679 Munich, Germany\\
$^{8}$ARTORG Center for Biomedical Engineering Research, University of Bern, Murtenstrasse 50, CH-3008, Bern, Switzerland\\
$^{9}$University College London, Department of Physics \& Astronomy, Gower St, London, WC1E 6BT, United Kingdom\\
$^{10}$Astronomy \& Astrophysics Group, Department of Physics, University of Warwick, Coventry CV4 7AL, United Kingdom
}

\date{Accepted 2025 March 5. Received 2025 March 4; in original form 2023 October 25}

\pubyear{2025}

\begin{document}
\label{firstpage}
\pagerange{\pageref{firstpage}--\pageref{lastpage}}
\maketitle

\begin{abstract}
The Wide Field Camera 3 (WFC3) instrument on the \textit{Hubble Space Telescope} has provided an abundance of exoplanet spectra over the years. These spectra have enabled analysis studies using atmospheric retrievals to constrain the properties of these objects. However, follow-up observations from the \textit{James Webb Space Telescope} have called into question some of the results from these older datasets, and highlighted the need to properly understand the degeneracies associated with retrievals of WFC3 spectra. In this study, we perform atmospheric retrievals of 38 transmission spectra from WFC3 and use model comparison to determine the complexity required to fit the data. We explore the effect of retrieving system parameters such as the stellar radius and planet's surface gravity, and thoroughly investigate the degeneracies between individual model parameters -- specifically the temperature, abundance of water, and cloud-top level. We focus on three case studies (HD 209458b, WASP-12b, and WASP-39b) in an attempt to diagnose some of the issues with these retrievals, in particular the low retrieved temperatures when compared to the equilibrium values. Our study advocates for the careful consideration of parameter degeneracies when interpreting retrieval results, as well as the importance of wider wavelength coverage to break these degeneracies, in agreement with previous studies. The combination of data from multiple instruments, as well as analysis from multiple data reductions and retrieval codes, will allow us to robustly characterise the atmosphere of these exoplanets. 
\end{abstract}


\section{Introduction}

Transmission spectra measure wavelength-dependent transit radii of exoplanets, which in principle encode information containing the cross sections and relative abundances of chemical species \citep{SeagerSasselov00, Brown01}. Interpreting these spectra to extract chemical abundances is a process known as atmospheric retrieval (first introduced to the study of exoplanetary atmospheres by \citealt{MadhusudhanSeager09}), which involves using modern Bayesian techniques to implement Occam's Razor \citep{BennekeSeager12}. For a recent review of atmospheric retrieval, see \citet{BarstowHeng20}.

Several retrieval works in the past 30 years have attempted to characterise space-observed atmospheres using transmission data from \textit{Hubble Space Telescope} (\textit{HST}) and \textit{Spitzer Space Telescope} \citep[e.g.][amongst many others]{Kreidberg+14, Sing+16, Barstow+17, Tsiaras+18, Wakeford+18, Pinhas+19, WelbanksMadhusudhan19}. Physical approximations such as isothermal and isobaric transit chords, and grey clouds (or even no clouds) were often used to estimate atmospheric properties. However, the low resolution and signal-to-noise ratio, together with the short wavelength range covered by \textit{HST} and \textit{Spitzer} data, have proven to be strong limitations against properly modelling the structure and composition of exoplanet atmospheres. As a result, retrieved atmospheric parameters have been shown to be strongly dependent on the wavelength of data and the choice of model. For instance, the Wide Camera Field 3 (WFC3) spectral range mainly covers \ch{H2O} features, failing to provide confident detections of other expected molecules, such as carbon-bearing species in particular. Therefore, retrievals can result in very high \ch{H2O} abundances, due to the lack of continuum opacities \citep{BarstowHeng20}. Another example is the cloud treatment, as a model with clouds will ``hide'' part of the atmosphere, allowing the chemical abundances to be higher than in the cloud-free case \citep{SeagerSasselov00}. For these reasons, retrievals from low-resolution transmission data often show degeneracies (i.e. ambiguity) between parameters \citep{BennekeSeager12, Griffith14, HengKitzmann17, FisherHeng18, WelbanksMadhusudhan19}.

Additionally, a constant issue in the exoplanet community is the strong disagreement between outcomes for the same planet, not only as a result of the degeneracies between parameters, but also due to different model implementations used by individual studies, as highlighted by \citet{BarstowHeng20}. Furthermore, it can be difficult to identify degeneracies, in particular when they occur between multiple parameters, which requires careful consideration of their posterior distributions.

Since the first studies on the equations governing the transmission spectra of exoplanets, degeneracies between atmospheric parameters have been considered. \citet{Brown01} noted the similar effects from lower clouds, higher temperatures, increased abundances and turbulent velocities, which all increase the strength of the spectral features. \citet{LecavelierDesEtangs+08} elucidated a degeneracy between molecular abundances and total atmospheric pressure (i.e. a ``normalization degeneracy''). This same degeneracy was explored in \citet{BennekeSeager12} in the context of super-Earths, where they determine that the degeneracy can be broken by the inclusion of optical data covering the molecular Rayleigh scattering signature, assuming that \ch{N2} and \ch{H2}+\ch{He} are the only spectrally inactive species. Further work from \citet{BennekeSeager13} showed how a degeneracy between a cloudy hydrogen-dominated atmosphere and a cloud-free high mean molecular weight atmosphere can be broken by precisely measuring the wings and depths of absorption features, given sufficiently small uncertainties on the data points. \citet{LineParmentier16} showed how this degeneracy means that, in wavelength regions with a single absorber (such as WFC3), one is only able to constrain very high or low molecular abundances, and are not sensitive to changes between these extremes. 

Additional studies have also considered the planet's bulk properties, such as mass and radius. \citet{deWitSeager13} investigated how the planet mass could be retrieved from transmission spectra, although they noted a degeneracy with high-altitude clouds. This is further explored in \citet{Batalha17}, who found this degeneracy to be detrimental when attempting to retrieve the masses of super-Earths and sub-Neptunes, in particular. In addition, \citet{Griffith14} found that small uncertainties in the planet's radius can cause uncertainties in the gas mixing ratios of several orders of magnitude -- another variation of the normalization degeneracy. In agreement with previous studies, they identify how the inclusion of optical data can mitigate this issue. \citet{Betremieux16} also investigate how refraction signatures can be used to break this degeneracy, and further studies explore the effects of accounting for a surface or optically thick clouds \citep{BetremieuxSwain17, BetremieuxSwain18}. More recently, \citet{WelbanksMadhusudhan19} performed an in-depth study of retrieval degeneracies and found that the combination of optical and infrared spectra can break the degeneracy between clouds and chemical abundances. They also highlighted the limitations of the isobaric assumption in analytical models \citep[e.g.][]{HengKitzmann17}, and discussed the importance of accurately accounting for collision-induced absorption (CIA) in correctly determining molecular abundances. The impact of CIA on transmission spectra was also explored in some of the previous studies \citep{deWitSeager13, LineParmentier16, BetremieuxSwain17, BetremieuxSwain18}. In \citet{WelbanksMadhusudhan19}, they performed a case study of optical and infrared data of HD 209458b by implementing retrievals with increasingly complex physical models, and analysed the resulting degeneracies. However, for low-resolution data with limited wavelength coverage, higher model complexity is not always warranted, and can lead to over-fitting.

\subsection{\textit{HST} in the \textit{JWST} era}

With the launch of the \textit{James Webb Space Telescope} (\textit{JWST}), we are seeing a drastic improvement in the wavelength coverage and precision of exoplanet spectra over \textit{HST}. \textit{JWST} spectra of previously observed exoplanets are bringing to light the limitations of \textit{HST}, with an array of new molecular detections and a deeper understanding of their atmospheric properties \citep[e.g.][]{Ahrer+23, Feinstein+23, Rustamkulov+23, Alderson+23, Tsai+23, Hammond+24}. Despite the number of studies investigating degeneracies in transmission spectra retrievals, there are several cases where \textit{JWST} measurements are demonstrating that our conclusions from \textit{HST} data were incorrect, due to degenerate solutions within the \textit{HST} wavelength range \citep{Madhusudhan+23}. In addition, \citet{Nixon+24} showed that uncertainties in the retrieved parameters from \textit{HST} data have likely been underestimated, due to uncertainties on the correct model to use.

With \textit{JWST} instruments such as NIRISS encompassing the entire WFC3 wavelength range at a higher resolution, and expanding further into the near infrared, future observations are likely to continue to challenge our interpretations from previous WFC3 exoplanet spectra. \citet{Lueber+24} using WASP-39b data from different \textit{JWST} instruments (NIRCam, NIRISS, NIRSpec G395H, and NIRSpec PRISM) from 0.5 to 5.5 $\upmu$m have shown that both the increase in data resolution as well as the wide wavelength coverage were able to break the degeneracy between \ch{H2O} abundances and normalization parameters. Moreover, a comparison study between \textit{HST} WFC3 and \textit{JWST} NIRISS data for the same object by \citet{Fisher+24} has further demonstrated that WFC3 retrievals alone are not able to accurately constrain \ch{H2O} abundances, due to the difficulty to determine the continuum in such a narrow wavelength range.

However, even in the era of \textit{JWST}, there is still a strong argument for using \textit{HST} data, especially in the optical range, which is not covered by \textit{JWST}. A recent study by \citet{Fairman+24} demonstrated the importance of optical data from STIS for constraining cloud properties in exoplanet atmospheres, and \cite{Fisher+24} explored the benefits of combining this with data from \textit{JWST}'s NIRISS instrument. Furthermore, despite the anticipated removal of WFC3 from Hubble proposal cycles as of October 2025, there is an abundance of available WFC3 exoplanet spectra. This archival data still has much to offer for population studies, and understanding the challenges associated with its interpretations -- both individually and collectively -- provides a powerful lesson for our analysis of \textit{JWST} spectra.

\begin{table*}
\centering
\caption{Prior ranges assumed in our retrievals. For $R_{\rm star}$ and log $g_{\rm p}$, $\mu$ and $\sigma$ represent the mean and standard deviation of the Gaussian distribution, respectively, assumed to be the mean and uncertainty values listed in Table \ref{table:input_parameters}.}
\label{table:priors}
\resizebox{0.9\textwidth}{!}{
\begin{tabular}{lcccc}
\hline
Parameter & Symbol & Range & Distribution & Units \\ \hline
Pressure & $P$ & [10$^{-9}$, 10$^{1}$] & Log-uniform & bar \\
Temperature & $T$ & [200, 3100] & Uniform & K \\
Water abundance & $X_{\ch{H2O}}$ & [10$^{-12}$, 10$^{-1}$] & Log-uniform & $-$ \\
Cloud-top pressure & $P_{\rm cloud\mbox{-}top}$ & [10$^{-4}$, 10] & Log-uniform & bar \\
Cloud composition parameter & $Q_0$ & [1, 100] & Uniform & $-$ \\
Cloud slope index & $a_0$ & [2, 13] & Uniform & $-$ \\
Spherical cloud particle radius & $r_{\rm cloud}$ & [10$^{-9}$, 10$^{-3}$] & Log-uniform & cm \\
Reference optical depth & $\tau_{\rm ref}$ & [10$^{-5}$, 10$^{2}$] & Log-uniform & $-$ \\
Cloud bottom pressure & $P_{\rm cloud\mbox{-}bottom}$ & [10$^{0}$, 10$^{2}$] & Log-uniform & bar \\
Reference transit radius & $R_{\rm p}$ & [$R_{\rm p}$ $-$ $\sigma_{R_{\rm p}}$, $R_{\rm p}$ $+$ $\sigma_{R_{\rm p}}$] & Uniform & R$_{\rm Jup}$ \\
Stellar radius & $R_{\rm star}$ & $\mu = R_{\rm star}, \sigma = \sigma_{R_{\rm star}}$ & Gaussian & R$_\odot$ \\
Planetary surface gravity & log $g_{\rm p}$ & $\mu = {\rm log\,} g_{\rm p}, \sigma = \sigma_{{\rm log\,} g_{\rm p}}$ & Gaussian & cm s$^{-2}$ \\ \hline
\end{tabular}
}
\end{table*}

\begin{table*}
\centering
\caption{Free parameters retrieved by each model in our study.}
\label{table:free_parameters}
\resizebox{\textwidth}{!}{
\begin{tabular}{lccccccccccccc}
\hline
Model & Number of parameters & $T$ & $X_{\ch{H2O}}$ & $P_{\rm cloud\mbox{-}top}$ & $Q_0$ & $a_0$ & $r_{\rm cloud}$ & $\tau_{\rm ref}$ & $P_{\rm cloud\mbox{-}bottom}$ & $R_{\rm p}$ & $R_{\rm star}$ & log $g_{\rm p}$ & line \\ \hline
cloud free & 5 & $\checkmark$ & $\checkmark$ & $-$ & $-$ & $-$ & $-$ & $-$ & $-$ & $\checkmark$ & $\checkmark$ & $\checkmark$ & $-$ \\
grey clouds & 6 & $\checkmark$ & $\checkmark$ & $\checkmark$ & $-$ & $-$ & $-$ & $-$ & $-$ & $\checkmark$ & $\checkmark$ & $\checkmark$ & $-$ \\
non-grey clouds & 11 & $\checkmark$ & $\checkmark$ & $\checkmark$ & $\checkmark$ & $\checkmark$ & $\checkmark$ & $\checkmark$ & $\checkmark$ & $\checkmark$ & $\checkmark$ & $\checkmark$ & $-$ \\
flat line & 1 & $-$ & $-$ & $-$ & $-$ & $-$ & $-$ & $-$ & $-$ & $-$ & $-$ & $-$ & $\checkmark$ \\ \hline
\end{tabular}
}
\end{table*}

\subsection{Structure of the study}

The present study is a follow-up from \citet{FisherHeng18}, analysing the same sample of 38 transmission spectra observed by \textit{HST} WFC3, focusing on degeneracies and challenges associated with determining atmospheric parameters from WFC3 transmission spectra. Section \ref{sec:methodology} explains the retrieval code used in this work, the chosen atmospheric models, as well as the method of comparison between them. In Section \ref{sec:degeneracies} we investigate the effect of accounting for uncertainties in the stellar radius and planetary surface gravity, which are commonly considered fixed parameters. We analyse the individual influence of model properties (e.g. temperature, pressure, chemical abundance, cloud parameterization) and degeneracies between them in retrieval outcomes from WFC3 data. In Section \ref{sec:population_study} we compare our current \ch{H2O} abundance and temperatures with \citet{FisherHeng18} outcomes. In particular, we examine the temperature and its influence in chemical abundances, which are seen to be artificially low in retrievals from transmission spectra due to day-night asymmetries \citep{MacDonald+20}. In Section \ref{sec:case_studies} we perform tests using three case studies (HD 209458b, WASP-12b, and WASP-39b) to further investigate the low temperature ``problem''. In Section \ref{sec:discussion} we compare our findings with previous literature results, highlighting how retrieved temperatures and chemical abundances are dependent on the choice of model. Finally, we perform retrievals using data reduced by two different works, in order to verify if model parameters are dependent on the reduction. Section \ref{sec:summary} summarises the conclusions of our study.

\section{Methodology}
\label{sec:methodology}

We perform atmospheric retrievals using the open-source, GPU-accelerated code \texttt{BeAR} \citep[Bern Atmospheric Retrieval code\footnote{\texttt{BeAR} is available at the GitHub repository \url{https://github.com/newstrangeworlds/bear}.},][]{Kitzmann+20}. \texttt{BeAR} uses the nested-sampling algorithm MultiNest \citep{Skilling06, Feroz+09, Buchner+14, Buchner21} to fit exoplanet transmission or occultation spectra, as well as emission spectra of self-luminous objects. We refer the reader to \citet{Kitzmann+20}, \citet{Lueber+22}, and references therein for a full description of the \texttt{BeAR} framework, including temperature-pressure ($T$-$P$) profiles. Detailed descriptions of \texttt{BeAR}'s capabilities can also be found in its documentation.

\begin{figure*}
\centering
\includegraphics[width=0.9\linewidth]{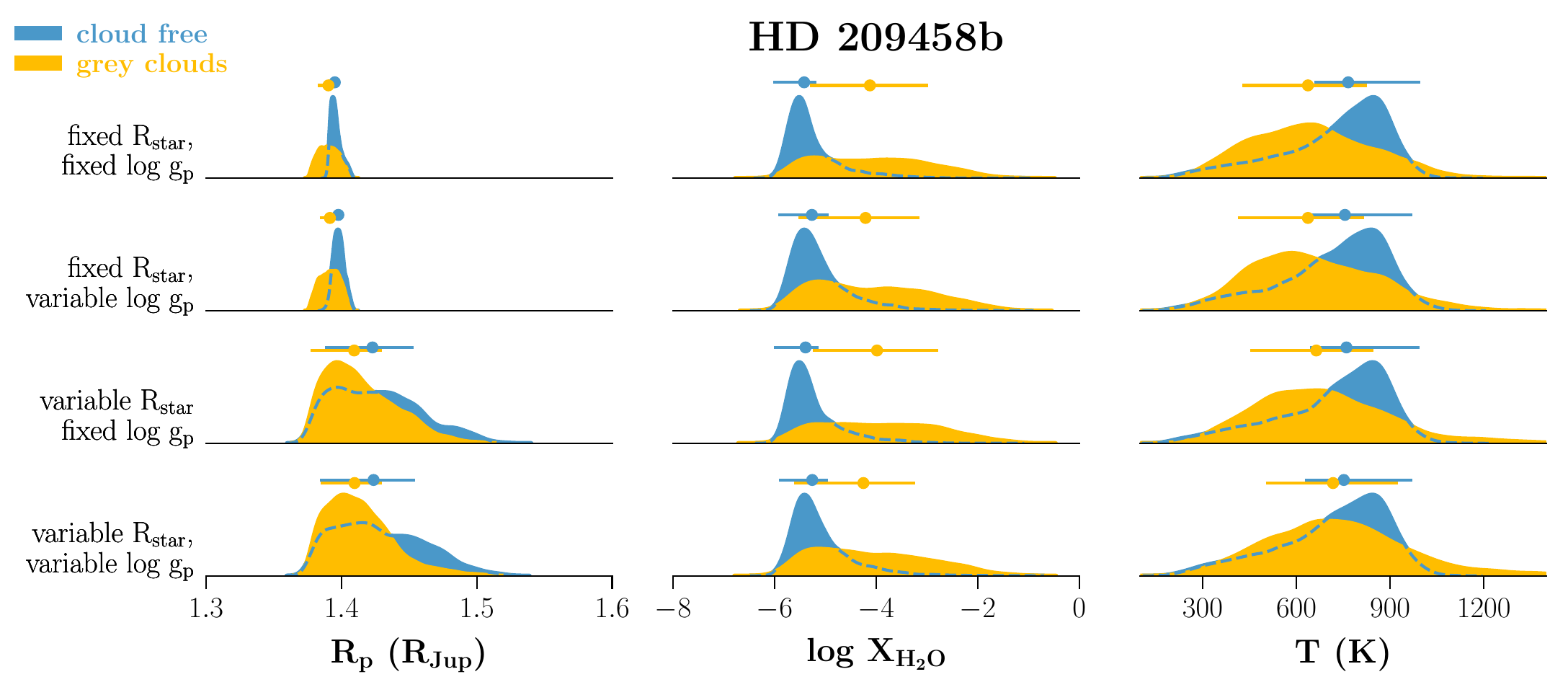}
\vspace{-0.3cm}
\caption{Posterior distributions of the reference planet radius $R_{\rm p}$, water abundance $X_{\rm \ch{H2O}}$, and temperature $T$ retrieved using fixed or variable values of the stellar radius ($R_{\rm p}$) and planetary surface gravity (log $g_{\rm p}$) for HD 209458b. Cloud free (blue) and grey cloud (yellow) models were tested.}
\label{fig:fixed_variable_rstar_logg}
\end{figure*}

\begin{figure*}
\centering
\includegraphics[width=0.9\textwidth]{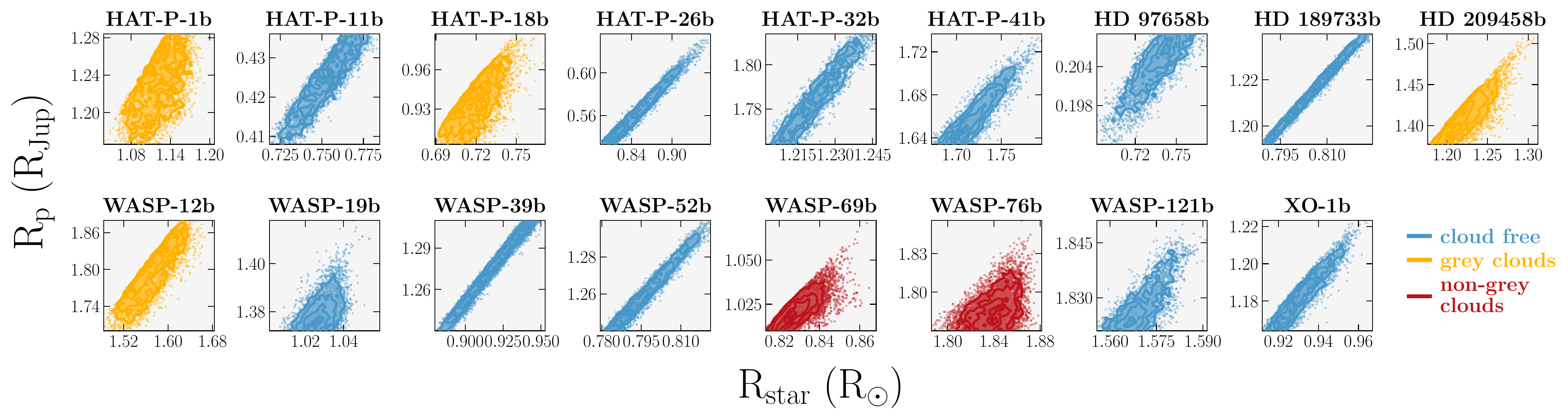}
\vspace{-0.25cm}
\caption{Correlation between retrieved stellar radii $R_{\rm star}$ and reference transit radii $R_{\rm p}$ for all non-flat-line objects using their cloud free (blue), grey cloud (yellow), or non-grey cloud (red) best-fit model.}
\label{fig:Rstar_Rp_degeneracy_bestfit}
\end{figure*}

\subsection{\textit{HST} WFC3 transmission spectra}

To provide continuity with the study of \citet{FisherHeng18}, we perform retrieval on the same set of 38 \textit{HST} WFC3 transmission spectra. Spectra for GJ 1214b were obtained from \citet{Kreidberg+14}, WASP-12b from \citet{Kreidberg+15}, WASP-17b from \citet{Mandell+13}, WASP-19b from \citet{Huitson+13}, HD 97658b from \citet{Knutson+14}, and TRAPPIST-1 objects from \citet{deWit+18}. The remaining WFC3 data were provided by \citet{Tsiaras+18}.

\subsection{Opacity sources}
\label{sec:opacities}

As most of our objects are presented in the literature as Jupiter-sized or Saturn-sized planets, our models assume atmospheres dominated by \ch{H2} and \ch{He}. Besides these background gases, \ch{H2O} is considered the main molecular opacity source, as the \ch{H2O} spectral feature at 1.4 $\upmu$m is the strongest molecular signature in the \textit{HST} WFC3 wavelength range. \ch{H2} and \ch{He} are initially set to solar abundances, which will vary according to the additional opacity sources included in the model (i.e. \ch{H2O}, CIA, Rayleigh scattering, and clouds).

For completeness, our work also contemplates four TRAPPIST-1 planets, for which we roughly define an Earth-like, \ch{N2}-dominated atmosphere. Therefore, we adopt the treatment of \citet{FisherHeng18}, submitting the TRAPPIST-1 planets to two sets of retrievals, separately representing an \ch{H2}-dominated and an \ch{N2}-dominated atmosphere.

We use \ch{H2O} opacities from the \texttt{DACE} database \citep{Grimm+21}, which converts spectroscopic line lists from \citet{Polyansky+18} into opacities using the open-source opacity calculator \texttt{HELIOS-K} \citep{GrimmHeng15, Grimm+21}. Our models also include collision-induced absorption (CIA) opacities of \ch{H2}--\ch{H2} from \citet{Abel_jp109441f} and \ch{H2}--\ch{He} from \citet{Abel2012JChPh.136d4319A}, available within the \texttt{HITRAN} database \citep{Karman+19}, and Rayleigh scattering by \ch{H2} \citep{Allen2000asqu.book.....C} and \ch{He} \citep{Sneep2005JQSRT..92..293S, Thalman2014JQSRT.147..171T} in the total atmospheric opacity. We adopted an opacity resolution of 0.1 cm$^{-1}$ for all runs.

Several of the planets in our sample do have detections of other species in their atmosphere. For example, the ultra-hot Jupiter WASP-121b shows evidence of \ch{VO} from its STIS data \citep{Evans+18}, and \ch{H-} from the WFC3 thermal phase curve \citep{MikalEvans+22}. \citet{MacDonaldMadhusudhan17b} also found tentative evidence of \ch{NH3} in the WFC3 transmission spectra of HD 209458b and WASP-31b, and \ch{HCN} in WASP-63b (see also \citet{Kilpatrick+18}). However, to keep our analysis homogeneous across the sample and focus on degeneracies present with a single spectral feature, we choose to only include \ch{H2O} in our retrievals.

\begin{figure*}
\centering
\vspace{-0.1cm}
\includegraphics[width=0.9\textwidth]{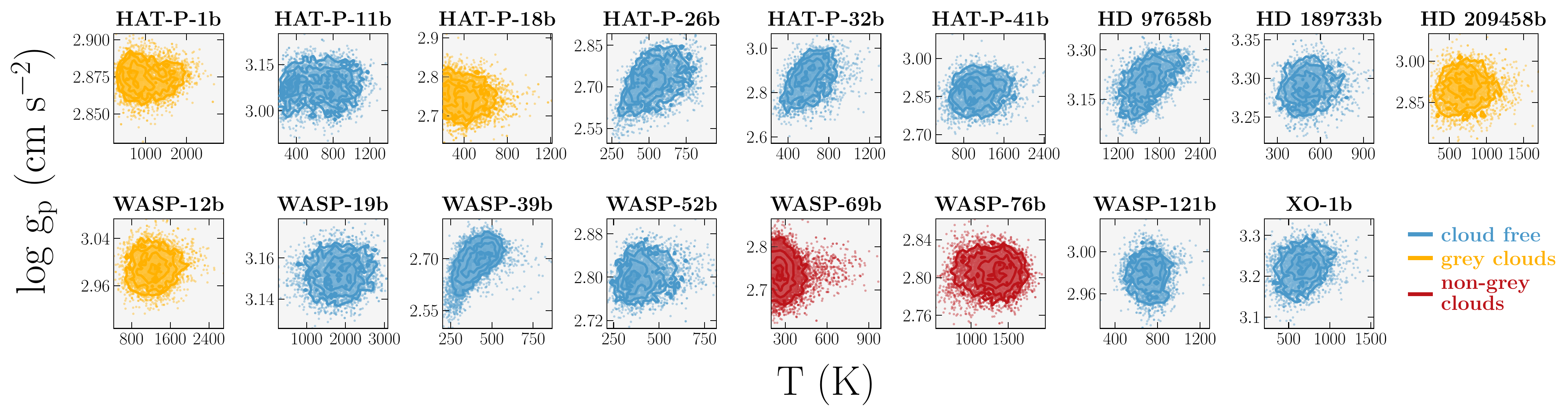}
\vspace{-0.25cm}
\caption{Same as Figure \ref{fig:Rstar_Rp_degeneracy_bestfit}, but between temperatures $T$ and planetary surface gravities log $g_{\rm p}$.}
\label{fig:T_log_g_degeneracy_bestfit}
\end{figure*}

\begin{figure*}
\centering
\vspace{-0.1cm}
\includegraphics[width=0.9\textwidth]{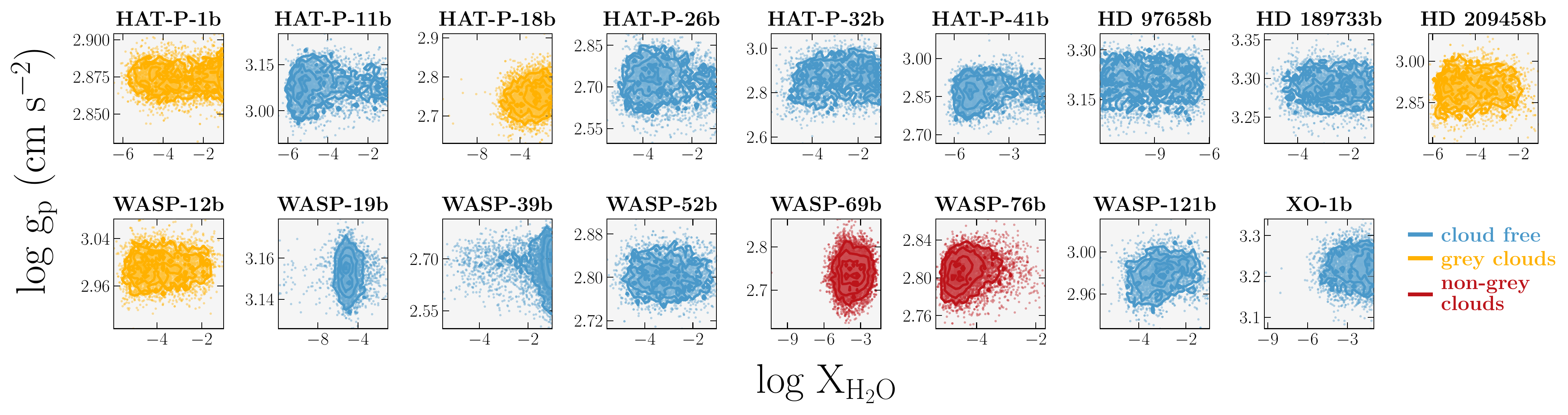}
\vspace{-0.25cm}
\caption{Same as Figure \ref{fig:Rstar_Rp_degeneracy_bestfit}, but between \ch{H2O} abundances and planetary surface gravities log $g_{\rm p}$.}
\label{fig:H2O_log_g_degeneracy_bestfit}
\end{figure*}

\begin{figure*}
\centering
\vspace{-0.1cm}
\includegraphics[width=0.9\textwidth]{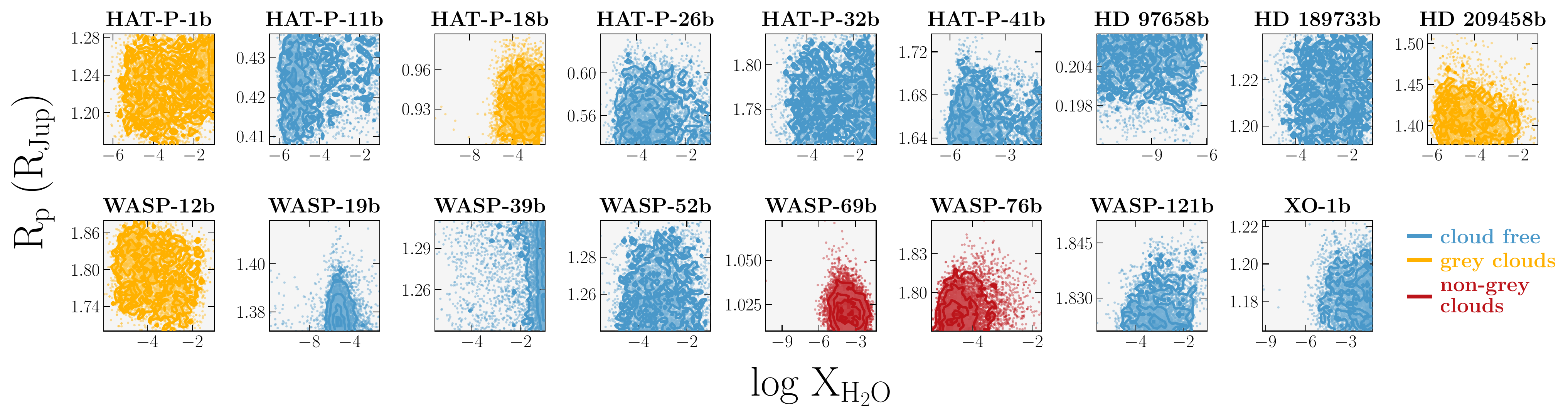}
\vspace{-0.25cm}
\caption{Same as Figure \ref{fig:Rstar_Rp_degeneracy_bestfit}, but between \ch{H2O} abundances and reference transit radii $R_{\rm p}$.}
\label{fig:H2O_Rp_degeneracy_bestfit}
\end{figure*}

\begin{figure*}
\centering
\vspace{-0.1cm}
\includegraphics[width=0.9\textwidth]{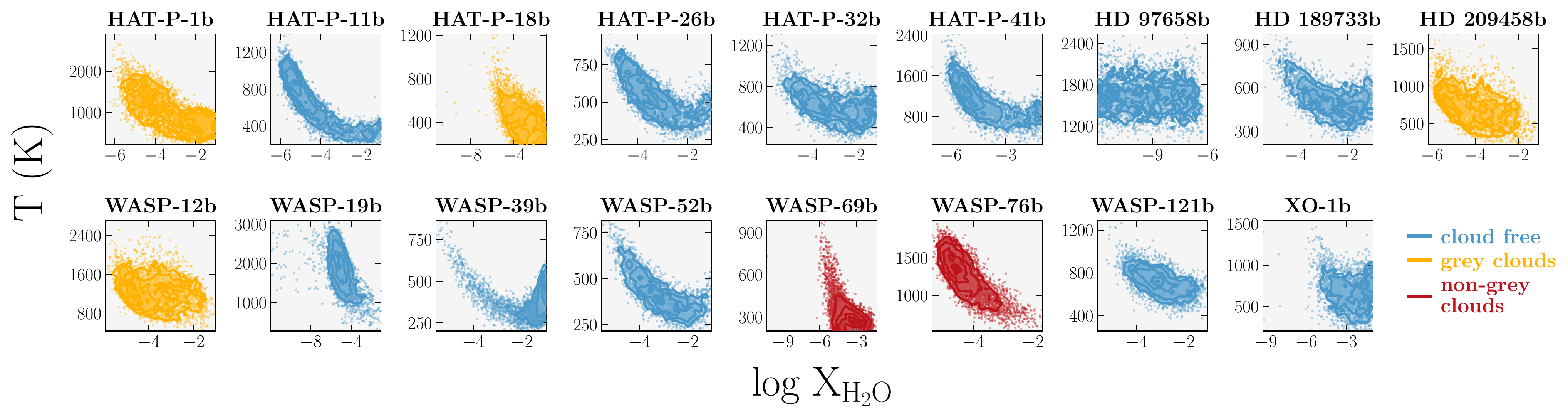}
\vspace{-0.25cm}
\caption{Same as Figure \ref{fig:Rstar_Rp_degeneracy_bestfit}, but between \ch{H2O} abundances and temperatures $T$.}
\label{fig:H2O_T_degeneracy_bestfit}
\end{figure*}

\begin{figure*}
\centering
\vspace{-0.1cm}
\includegraphics[width=0.72\textwidth]{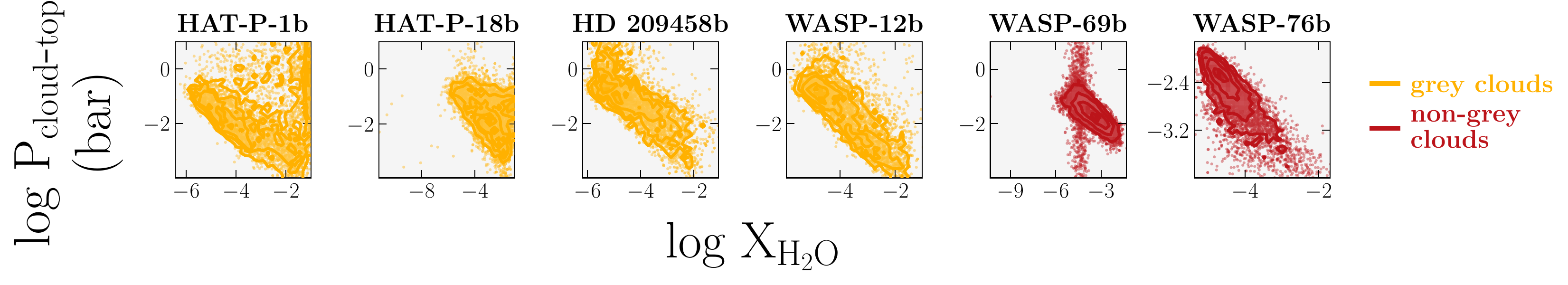}
\vspace{-0.25cm}
\caption{Same as Figure \ref{fig:Rstar_Rp_degeneracy_bestfit}, but between \ch{H2O} abundances and cloud-top pressures $P_{\rm cloud\mbox{-}top}$.}
\label{fig:H2O_Pcloudtop_degeneracy_bestfit}
\end{figure*}

\subsection{Retrieval code}

The atmospheric retrieval code used by \citet{FisherHeng18} assumes isobaric transit chords, i.e. where opacities are assumed not to vary with pressure. \citet{HengKitzmann17} showed that the isobaric approximation was sufficiently accurate in many cases for modelling WFC3 data, given its resolution and precision, but that inaccuracies could occur at lower temperatures in particular. In this work we adopt a non-isobaric treatment using the retrieval code \texttt{BeAR}, to consider pressure-dependent transit chords. By assuming an isothermal atmosphere, some numerical approximations can be made to speed up the calculations. We consider an atmospheric pressure range from $10^{-9}$ to 10 bar, divided into 199 equal layers (200 levels) in log space. For comparison, isobaric retrievals by \citet{FisherHeng18} use a fixed pressure of $10^{-2}$ bar, assumed to be the WFC3 photospheric pressure.

Additionally, \citet{FisherHeng18} do not include Rayleigh scattering in their retrievals, which is included in this work. Further differences include the stellar radius $R_{\rm star}$ and the planetary surface gravity log $g_{\rm p}$, which were previously considered fixed input parameters, and are now allowed to vary within a range of priors. The surface gravity is assumed to vary with pressure. Table \ref{table:input_parameters} summarises the input values for these parameters for all objects analysed in this study. The uncertainty ranges of the reference transit radius $R_{\rm p}$, $R_{\rm star}$, and log $g_{\rm p}$ from Table \ref{table:input_parameters} are considered the prior ranges for each of these parameters, as listed in Table \ref{table:priors} along with their considered distributions.

\begin{table*}
\centering
\caption{Values used to test the spectral influence of each free parameter, as shown in Figure \ref{fig:param_effect}.}
\label{table:param_effect}
\resizebox{0.9\textwidth}{!}{
\begin{tabular}{lccccccc}
\hline
Planet & Value & $T$ (K) & log $g_{\rm p}$ (cm s$^{-2}$) & log $X_{\ch{H2O}}$ & $R_{\rm p}$ (R$_{\rm Jup}$) & $R_{\rm star}$ (R$_\odot$) & log $P_{\rm cloud\mbox{-}top}$ (bar) \\ \hline
\multirow{3}{*}{HD 209458b} & low & 700 & 2.68 & $-5.0$ & 1.40 & 1.15 & $-6.0$ \\
 & medium & 1450 & 2.88 & $-3.0$ & 1.45 & 1.20 & $-3.0$ \\
 & high & 2100 & 3.08 & $-1.0$ & 1.50 & 1.25 & 0.0 \\ \hline
\end{tabular}
}
\end{table*}

\begin{figure*}
\centering
\vspace{-0.2cm}
\includegraphics[width=0.95\linewidth]{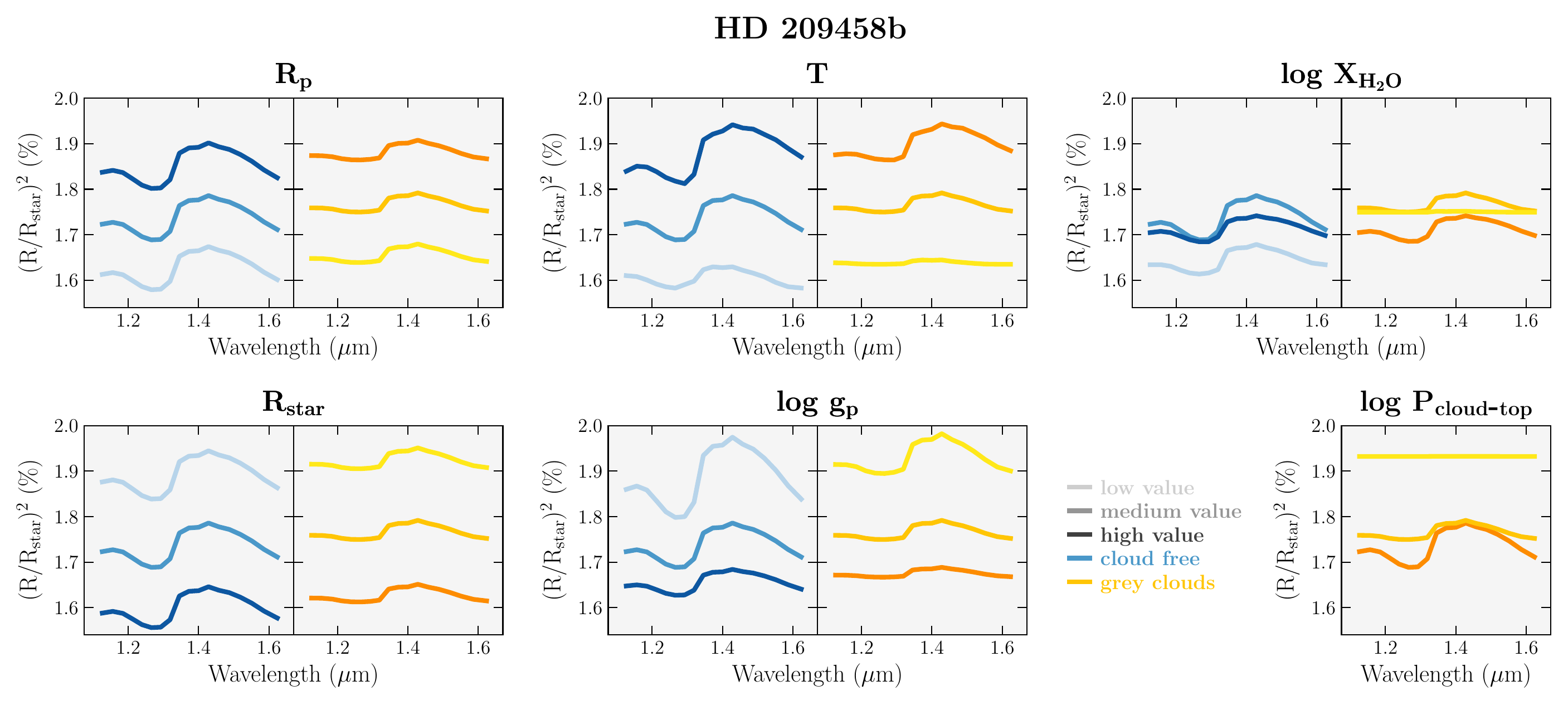}
\vspace{-0.2cm}
\caption{Forward models for HD 209458b data showing the difference when a low (light shades), medium (medium shades), or high (dark shades) value is considered for each free parameter. Cloud free (blue) and grey cloud (yellow) models were tested. For each panel, all values are fixed to the medium value except for the tested parameter (e.g. if the tested parameter is temperature, all other parameters are fixed to their medium values). Values used for each parameter (low, medium, high) are listed in Table \ref{table:param_effect}.}
\label{fig:param_effect}
\end{figure*}

\subsection{Atmospheric models}
\label{sec:models}

Each planet is examined by a range of models, which consider different atmospheric parameters in the retrieval code. These parameters include the temperature, treated as constant, a reference transit radius, and opacity sources, such as water, CIA, Rayleigh scattering, and clouds. While \ch{H2O}, CIA, and Rayleigh scattering are included in all models, the inclusion of clouds is optional. We test the presence of clouds using four independent models: with no clouds (``cloud free''), with grey clouds, with non-grey clouds, and a flat line model. ``Grey'' clouds consist of particles with sizes greater than the wavelengths probed and result in a constant cloud opacity. Non-grey clouds consist of particles with sizes similar than the wavelengths probed and are associated with a variable opacity across these same wavelengths. Each model has a different number of free parameters, according to their complexity, as listed in Table \ref{table:free_parameters}. A ``flat line'' model is also tested, representing a spectrum with a flat continuum (i.e. no spectral features).

For our grey cloud model, we define a cloud-top pressure parameter P$_{\rm cloud\mbox{-}top}$ which will establish a minimum altitude below which the atmosphere becomes 100\% opaque. Accordingly, for pressures higher than the cloud-top pressure (i.e. altitudes below the level where $P = P_{\rm cloud\mbox{-}top}$), the cloud optical depth is set to infinity.

For non-grey clouds, we follow Equation 5 of \citet{Lueber+22}, motivated by Equation 32 of \citet{KitzmannHeng18}:

\begin{equation}
\tau_{\rm nongrey} = \tau_{\rm ref} \frac{Q_0 x_{{\uplambda}_{\rm ref}}^{-a_0} + x_{\uplambda_{\rm ref}}^{0.2}}{Q_0 x_\uplambda^{-a_0} + x_\uplambda^{0.2}} \text{ ,}
\end{equation}

\noindent
where $x_\uplambda = 2 \pi r_{\rm cloud} / \uplambda$ and $x_{\uplambda_{\rm ref}} = 2 \pi r_{\rm cloud} / \uplambda_{\rm ref}$, $\uplambda$ is the wavelength, $\uplambda_{\rm ref}$ is the reference wavelength fixed at $1$ $\upmu$m, and $\tau_{\rm ref}$ is the vertical optical depth at the reference wavelength. $Q_0$, $a_0$, and $r_{\rm cloud}$ are the composition parameter, small particle slope, and particle radius, respectively. The cloud-top pressure is a free parameter, and the cloud bottom pressure is fixed to the bottom of the atmosphere, i.e. at 10 bar, because this choice has no consequences for transmission spectra as long as the atmosphere becomes optically thick already at lower pressures. Consequently, $\tau_{\rm nongrey}$ is added to the total atmospheric optical depth for pressures higher than the cloud-top pressure, extending the cloud from the cloud-top pressure to the assumed bottom of the atmosphere.

\subsection{Non-isothermal retrievals}
\label{sec:nonisothermal_methodology}

In order to explore temperature variations with altitude, we also implement non-isothermal retrievals. \texttt{BeAR} has the option to include a variety of different $T$-$P$ profiles, and in this work we select the polynomial profile. For this, an $n$-parameter profile sets the boundaries of the $n - 1$ atmospheric layers, and a polynomial is fit between them. In our case, we use linear profiles between the boundaries (see \citealt{Kitzmann+20} for details on the $T$-$P$ profile parametrisation). All non-isothermal runs in this work utilise a 4-parameter $T$-$P$ profile, which is enough to show the differences between the upper and lower layers, yet not as complex to require long calculations \citep{Kitzmann+20, Schleich+24}.

\begin{figure*}
\centering
\includegraphics[width=0.95\textwidth]{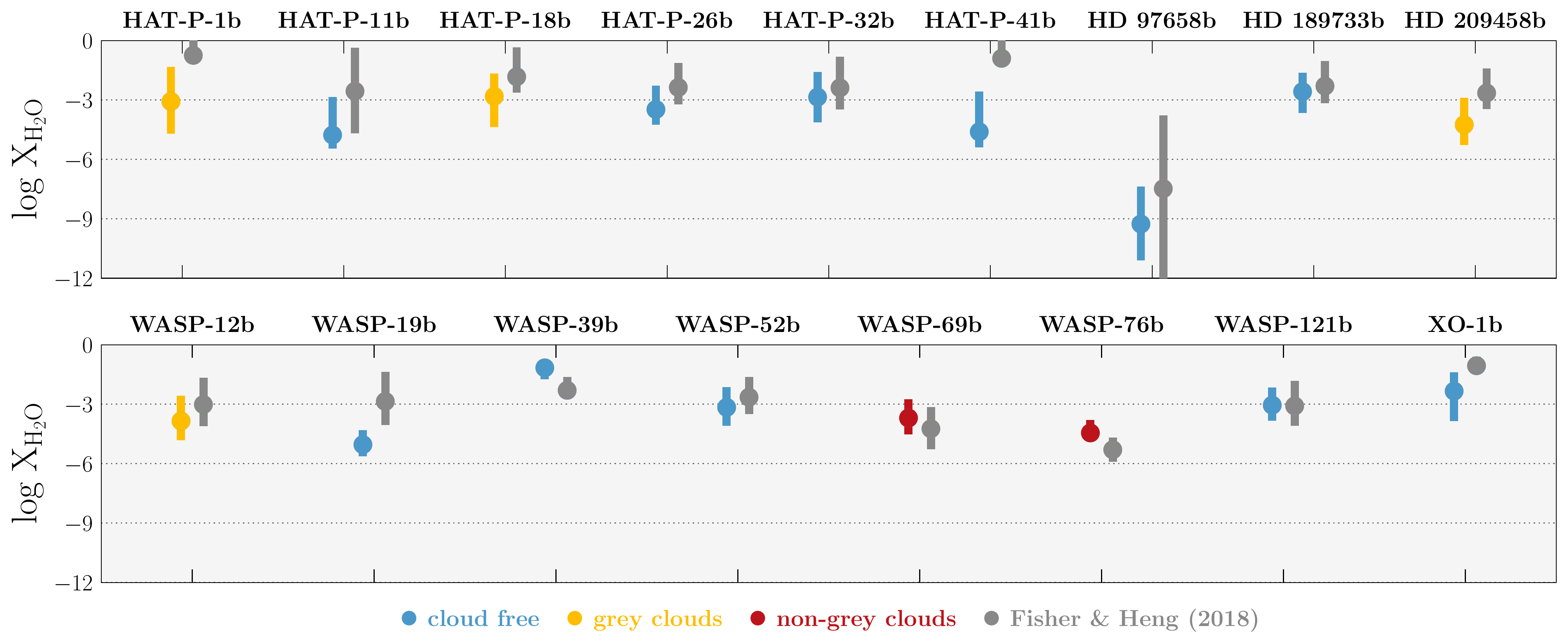}
\vspace{-0.15cm}
\caption{Comparison between retrieved \ch{H2O} abundances for all non-flat-line object using the best-fit model for each object. Cloud-free, grey-cloud, and non-grey-cloud models are represented in blue, yellow, and red, respectively. Grey circles correspond to isobaric values found by \citet{FisherHeng18}.}
\label{fig:H2O_abundance_fav}
\end{figure*}

\begin{figure*}
\centering
\includegraphics[width=0.95\textwidth]{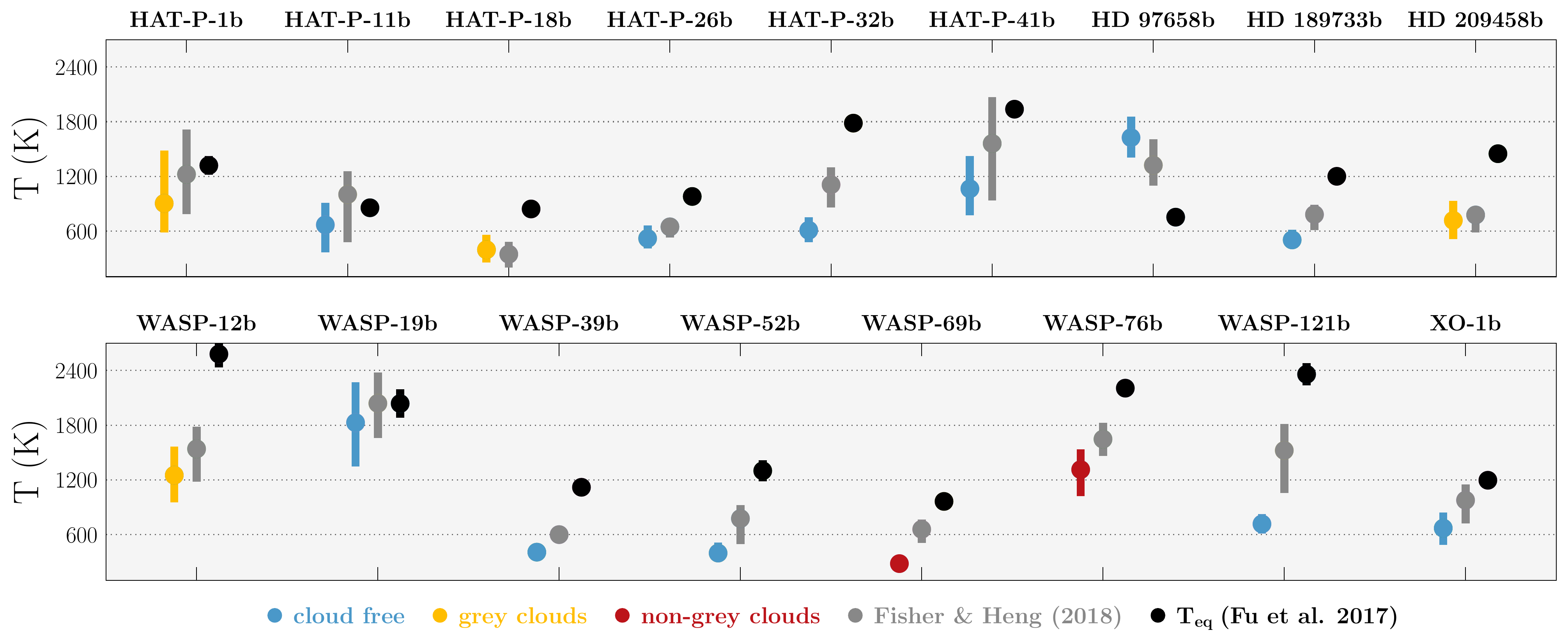}
\vspace{-0.15cm}
\caption{Same as Figure \ref{fig:H2O_abundance_fav}, but for retrieved temperatures. Equilibrium temperatures calculated by \citet{Fu+17} are represented in black.}
\label{fig:temperature_fav}
\end{figure*}

\subsection{Bayesian model comparison}
\label{sec:bayesian_comparison}

To effectively explore the multi-dimensional parameter space, we use the method of nested sampling \citep{Skilling06} as implemented via the \texttt{MULTINEST} software package \citep{Feroz+09}. Nested sampling allows the practitioner to formally implement Occam's Razor via Bayesian model comparison by calculating the Bayesian evidence (marginalized likelihood) associated with a pair of models \citep{Trotta08}. Table 1 of \citet{Trotta08}, which is reproduced in Table 2 of \citet{BennekeSeager13}, lists the correspondence between the so-called Bayes factor (ratio of Bayesian evidences) and number of standard deviations ($\sigma$) for which a model is disfavoured relative to the best-fit model.

The first step after running retrievals using our four non-isobaric models (cloud free, grey clouds, non-grey clouds, and flat line) is to compare these models by their corresponding Bayes factors. Bayesian comparison for all 38 objects (42 sets of retrievals) is displayed in Figures \ref{fig:bayesian_comparison_all1} and \ref{fig:bayesian_comparison_all2}, where the natural logarithm of the Bayes factor is calculated relative to the model with the highest Bayesian evidence. Hence, the highest Bayesian evidence model is denoted by a null Bayes factor, and all models with Bayes factors below unity are considered favoured.

According to Figures \ref{fig:bayesian_comparison_all1} and \ref{fig:bayesian_comparison_all2}, a large number of the objects can be sufficiently fit by a flat line model, i.e. the flat-line model has $\log$ Bayes factor of less than unity: GJ 436b, GJ 1214b, GJ 3470b, HAT-P-3b, HAT-P-12b, HAT-P-17b, HAT-P-38b, HD 149026b, WASP-17b, WASP-29b, WASP-31b, WASP-43b, WASP-63b, WASP-67b, WASP-74b, WASP-80b, WASP-101b, TRAPPIST-1d, TRAPPIST-1e, TRAPPIST-1f, and TRAPPIST-g. We continue our study excluding these objects from our analyses, as no further model complexity is required to fit the data. The next sections include analyses of the remaining 17 planets.

We find that most of the non-flat-line objects present at least two models with $\log$ Bayes factor < 1. The cloud complexity defines which are the best-fit models of each object, meaning the simplest model with $\log$ Bayes factor < 1 is considered the most favoured model. For example, if the cloud-free and one of the cloudy models both have $\log$ Bayes factor < 1 (e.g. HAT-P-11b, HAT-P-26b, HAT-P-32b, HAT-P-41b, HD 97658b, HD 189733b, WASP-19b, WASP-39b, WASP-52b, XO-1b), it means that clouds are not required to fit the data. If only both cloudy models have $\log$ Bayes factor < 1 (e.g. HAT-P-1b, HAT-P-18b, HD 209458b, WASP-12b), then the simpler, grey-cloud model is favoured, as the extra cloud parameters are not required to fit the spectrum. Tables \ref{table:retrieval_results_all1} and \ref{table:retrieval_results_all2} show our retrieval outcomes for all models, except flat lines (due to the lack of spectral features). Posterior distributions for each object using all models (except flat lines) are provided online (see Data Availability).

\section{Parameter degeneracies}
\label{sec:degeneracies}

Degeneracies between multiple parameters are commonly seen in retrievals from low-resolution data. These degeneracies have been previously investigated by a number of studies \citep[e.g.][]{BennekeSeager12, Griffith14, HengKitzmann17, FisherHeng18, WelbanksMadhusudhan19}. In this section we explore the specific degeneracies in our own retrievals, and particularly focus on the effects of our additional parameters, the stellar radius and planet gravity.

\subsection{Correlations between parameters}

Figure \ref{fig:fixed_variable_rstar_logg} illustrates the posterior distributions of \ch{H2O} abundance, temperature, and reference transit radius (for an example case -- HD 209458b) when $R_{\rm star}$ and/or log $g_{\rm p}$ are fixed or allowed to vary in the retrievals. We demonstrate that $X_{\ch{H2O}}$ and $T$ are barely affected by variable $R_{\rm star}$ and log $g_{\rm p}$ assumption, which mainly affects $R_{\rm p}$ posteriors. A strong degeneracy arises between $R_{\rm star}$ and $R_{\rm p}$ as seen for all non-flat-line objects in Figure \ref{fig:Rstar_Rp_degeneracy_bestfit}, which is expected from the calculation of the transit depth. Figure \ref{fig:T_log_g_degeneracy_bestfit} shows the relationship between the retrieved surface gravity and temperature values. We expect a degeneracy between these parameters due to their opposing effects on the atmospheric scale height, and this can be seen more clearly in some planets than others, such as WASP-39b. However, no clear degeneracy is observed between log $g_{\rm p}$ and water abundance for any of the planets, as shown in Figure \ref{fig:H2O_log_g_degeneracy_bestfit}.

The degeneracy between $R_{\rm star}$ and $R_{\rm p}$ likely dominates the spectral normalization, suppressing the so-called normalization degeneracy expected between $R_{\rm p}$ and \ch{H2O} abundance, as seen in \citet{FisherHeng18}. Therefore, Figure \ref{fig:H2O_Rp_degeneracy_bestfit} shows $R_{\rm p}$ and $X_{\ch{H2O}}$ are not correlated in our retrievals. However, note that the water abundance is quite widely constrained across several orders of magnitude, a result of the underlying normalization degeneracy \citep{LineParmentier16}. The lack of impact of log $g_{\rm p}$ and $R_{\rm star}$ on the water abundance is reassuring, as these parameters have regularly been fixed in previous retrieval studies, despite large uncertainties on their values.

Figure \ref{fig:H2O_T_degeneracy_bestfit} shows an additional degeneracy between \ch{H2O} abundance and temperature in most cases. This often presents as a ``banana-shaped'' joint posterior distribution -- at lower \ch{H2O} abundances, we see temperature decreasing with increasing \ch{H2O} abundance, whereas at higher \ch{H2O} abundances, the temperature increases with increasing abundance (e.g. HAT-P-11b, HAT-P-26b, HAT-P-32b, HAT-P-41b, HD 189733b, WASP-52b). This behaviour can be explained by the relationship between the spectral feature and the atmospheric scale height. At lower abundances, only the temperature is affecting the scale height, causing an increase in the amplitude of the feature at higher temperatures due to the larger scale height. At very high abundances, the increasing \ch{H2O} abundance starts to increase the mean molecular weight of the atmosphere, causing a reduction in the scale height when log $X_{\ch{H2O}}$ $\gtrsim$ $-2$, and thus muting the \ch{H2O} spectral features. To compensate this effect, the temperature increases to increase the scale height again.

In retrievals where grey or non-grey clouds are included, there is a further degeneracy between cloud-top pressure and \ch{H2O} abundance (Figure \ref{fig:H2O_Pcloudtop_degeneracy_bestfit}). This is because the cloud-top pressure can be shifted up and down to mute or extend the spectral feature, balancing the change in abundance.

\begin{figure}
\centering
\includegraphics[width=0.9\columnwidth]{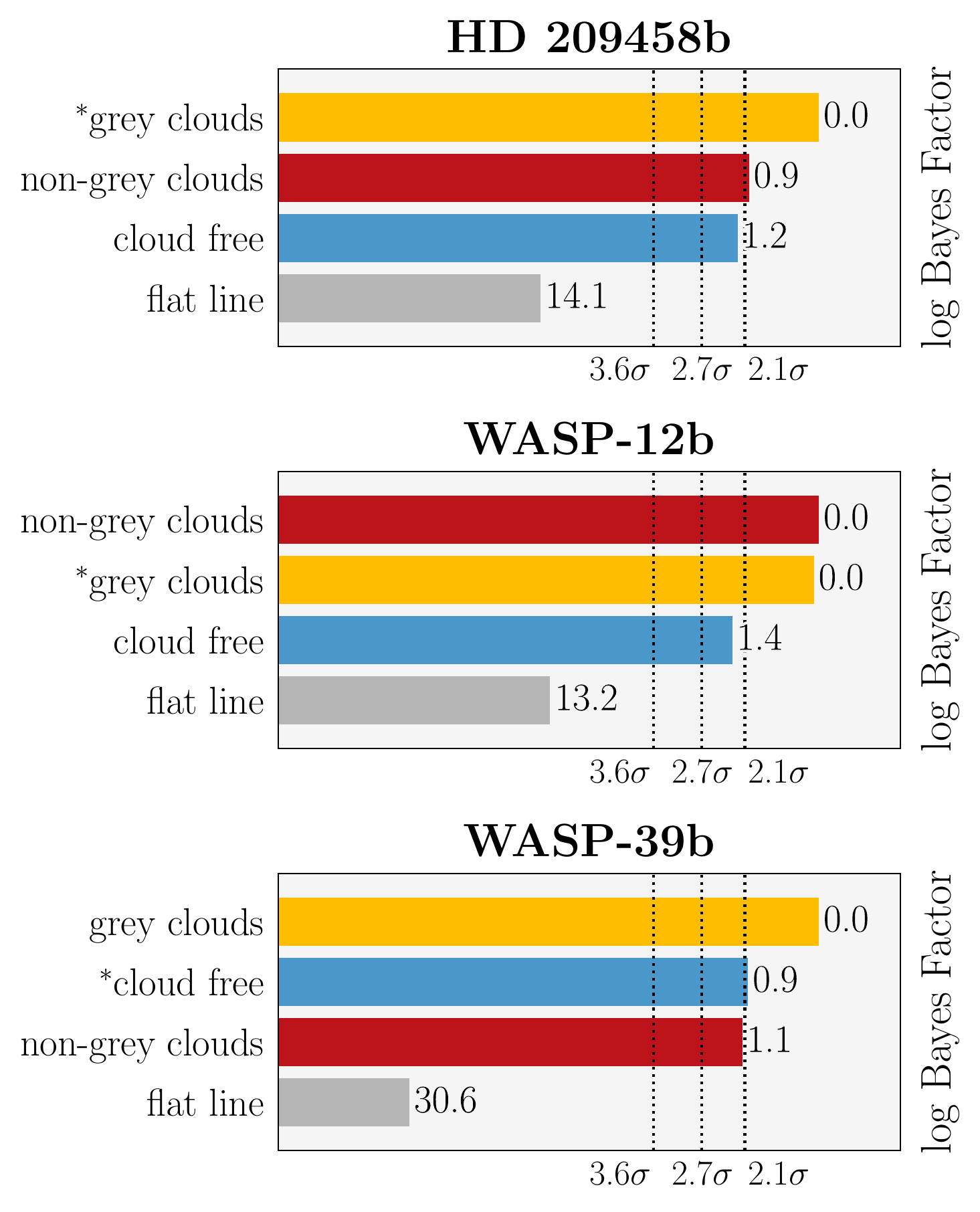}
\vspace{-0.1cm}
\caption{Bayesian comparison between cloud-free (blue), grey-cloud (yellow), non-grey cloud (red), and flat-line (grey) models for HD 209458b, WASP-12b, and WASP-39b. For each object, models are ordered from highest (top) to lowest (bottom) Bayesian evidence, and consequently from most favoured (highest evidence) to least favoured in comparison to the highest evidence model. Values on the right side of each bar show the corresponding natural logarithm of the Bayes factor relative to the highest-evidence model. Vertical dotted lines show significances of 3.6$\sigma$, 2.7$\sigma$, and 2.1$\sigma$, corresponding to $\log$ Bayes factors of 5.0, 2.5, and 1.0, considered ``strong'', ``moderate'', and ``weak'' evidences compared to the highest-evidence model, according to \citet{Trotta08}. Therefore, models with $\log$ Bayes factor < 1 have inconclusive evidence compared to the highest-evidence model, and cannot be ruled out as favoured, while models with $\log$ Bayes factor $\geq$ 1 are not favoured in comparison to the highest-evidence model. All runs are calculated at a spectral resolution of 0.1 cm$^{-1}$.}
\label{fig:bayesian_comparison_case_study}
\end{figure}

\subsection{Influence of each free parameter}
\label{sec:param_influence}

Due to the ambiguity caused by degeneracies, in order to reliably fit the spectral data, retrieval models may artificially compensate the lack of one (or more) parameter(s) with the lack or excess of other(s). In other words, when one limits a parameter to higher values, the model compensates by adjusting other free parameters to higher (or lower) values. Therefore, it is useful to understand the effect of each individual parameter on the transmission spectra. We note that several studies have investigated the effects of atmospheric parameters on the shape of transmission spectra \citep[e.g.][]{LineParmentier16}, but in this work we perform additional tests for our particular model setup. Figure \ref{fig:param_effect} demonstrates how our cloud free, grey cloud, and non-grey cloud modelled spectra behave when parameters ($T$, $X_{\rm \ch{H2O}}$, $R_{\rm p}$, $R_{\rm star}$, $P_{\rm cloud\mbox{-}top}$, and log $g_{\rm p}$) are fixed to low, medium, or high values, listed in Table \ref{table:param_effect}. We used spectra of HD 209458b as an example, considering input parameters listed in Table \ref{table:input_parameters}.

Trivially, the reference transit radius and stellar radius have a normalization effect of the spectrum, shifting it upwards or downwards. The main impact of both temperature and surface gravity is on the atmospheric scale height, with higher temperature and lower gravity increasing it. Larger scale heights cause both a continuum shift and a stretching of the spectrum, due to the larger spectral features.

The influence of \ch{H2O} abundance and cloud-top pressure in the spectra is not as straightforward. The previously mentioned degeneracy between low \ch{H2O} abundance and very high \ch{H2O} abundance, causing a high mean molecular weight, can be seen in the top right panels of Figure \ref{fig:param_effect}. In the cloud free case in particular, we can see how increasing \ch{H2O} abundance initially increases the spectral features, before reducing them again at very high abundance. When clouds are present, the cloud-top pressure sets an altitude limit where one can no longer probe below, and the spectrum becomes very flat at low \ch{H2O} abundances. The bottom right panel of Figure \ref{fig:param_effect} shows how decreasing the cloud-top pressure causes the continuum to shift upwards without raising the spectral features, muting them from below and flattening the spectrum. Several other parameters can then adjust to shift this overall spectrum upwards or downwards. This explains how degeneracies may conspire to fit low-resolution data, regardless of their physical plausibility. This is particularly the case for the temperature, as we have seen previously, and we will investigate this next in more detail.

\begin{figure}
\centering
\includegraphics[width=\linewidth]{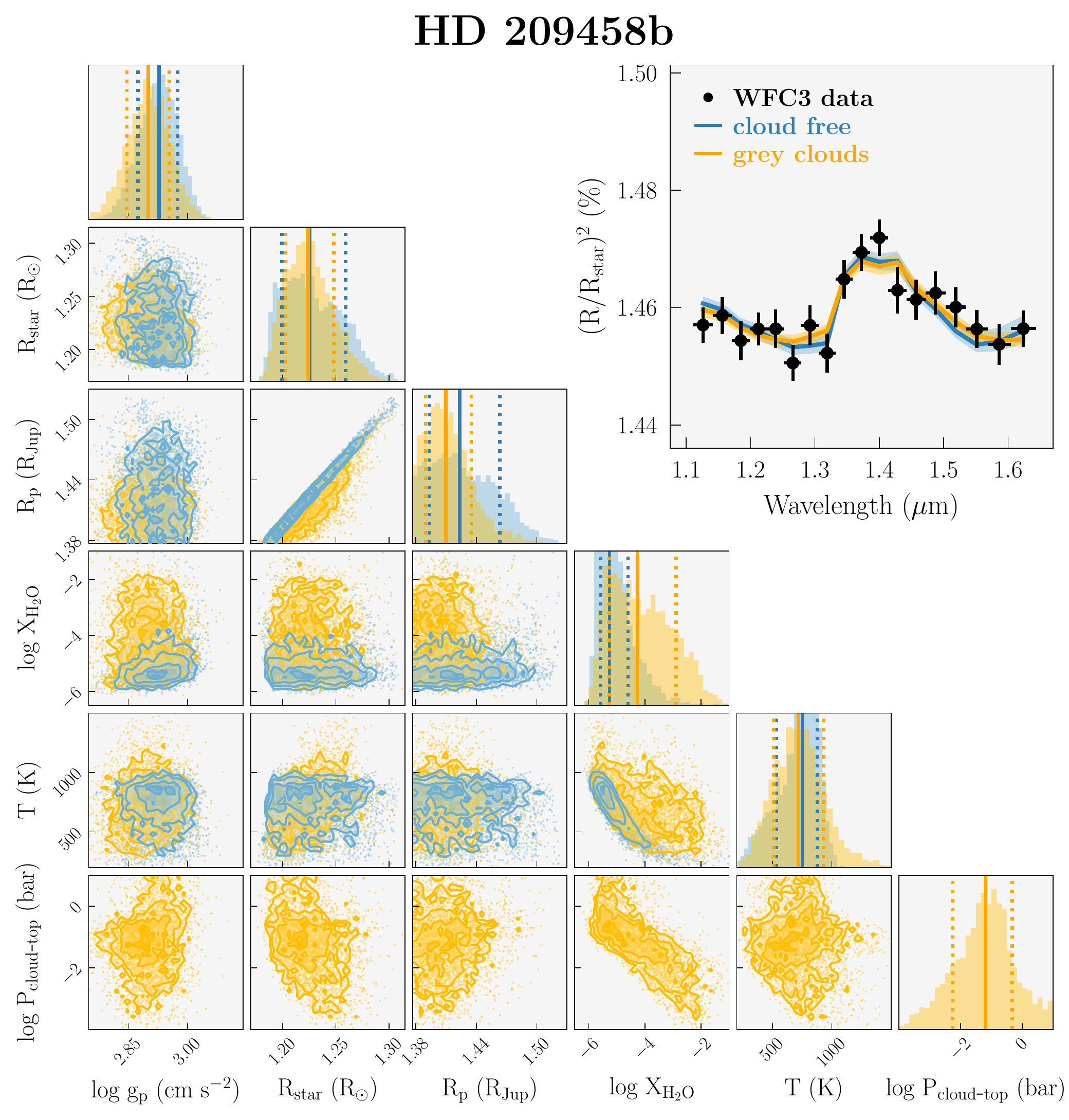}
\vspace{-0.5cm}
\caption{Retrieval outcomes for HD 209458b using our cloud free (blue) and grey cloud (yellow) models. Histograms represent the posterior distribution for each free parameter, with their median value (solid lines) and 1$\sigma$ uncertainty range (dashed lines). Top-right panel in each figure shows the fitted spectrum and its associated 1$\sigma$ uncertainty region, calculated at a resolution of 0.1 cm$^{-1}$ and binned to the data resolution for better visualisation. Black points correspond to data observed by \textit{HST} WFC3.}
\label{fig:corner_cloudfree_greycloud_HD209458b}
\end{figure}

\begin{figure}
\centering
\includegraphics[width=\linewidth]{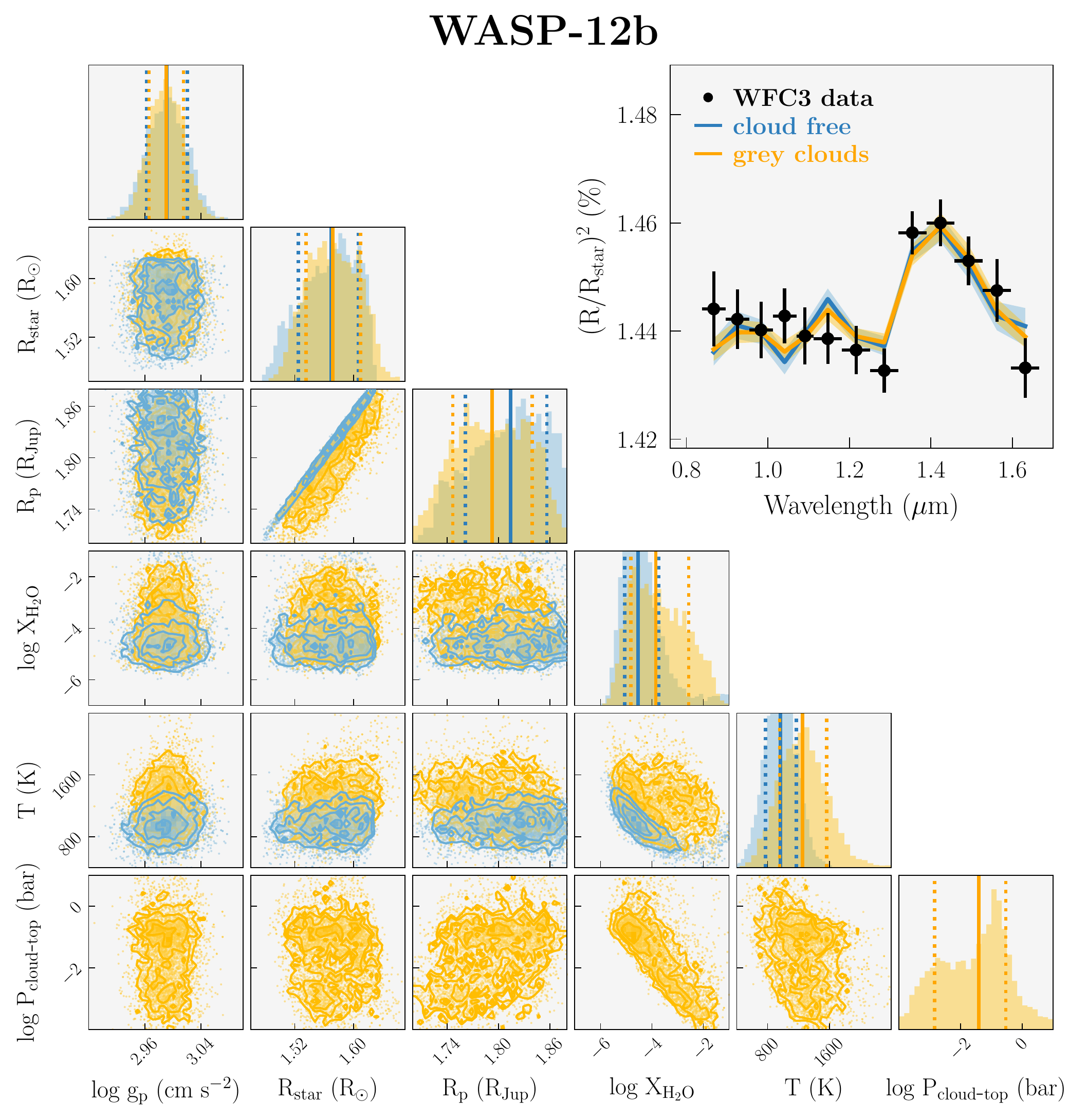}
\vspace{-0.5cm}
\caption{Same as Figure \ref{fig:corner_cloudfree_greycloud_HD209458b} but for WASP-12b.}
\label{fig:corner_cloudfree_greycloud_WASP-12b}
\end{figure}

\begin{figure}
\centering
\includegraphics[width=\linewidth]{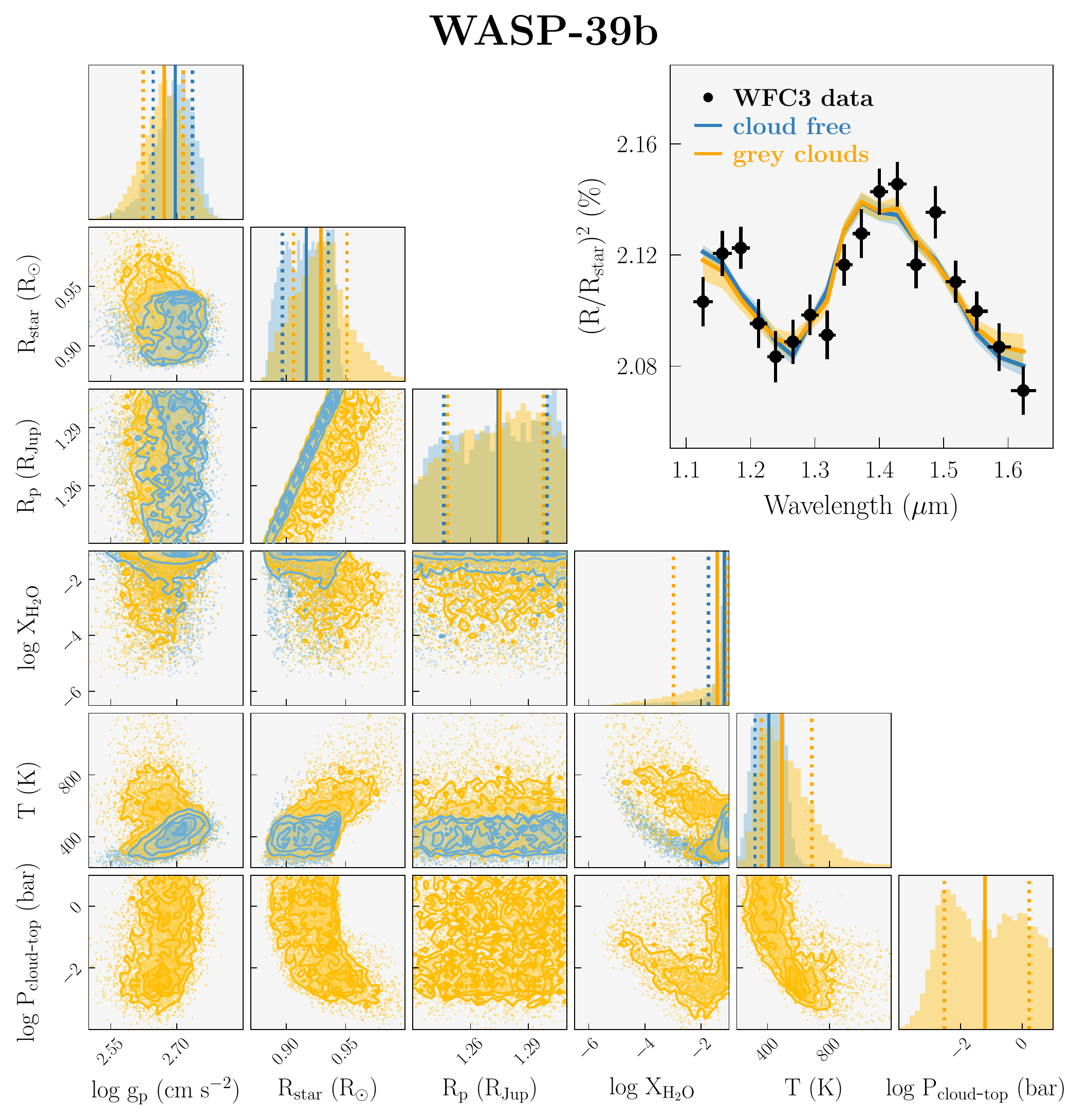}
\vspace{-0.5cm}
\caption{Same as Figures \ref{fig:corner_cloudfree_greycloud_HD209458b} and \ref{fig:corner_cloudfree_greycloud_WASP-12b} but for WASP-39b.}
\label{fig:corner_cloudfree_greycloud_WASP-39b}
\end{figure}

\section{Population study: full sample retrievals}
\label{sec:population_study}

\subsection{Water abundances}
\label{sec:H2O_abundances}

As \ch{H2O} is the main absorber in the \textit{HST} WFC3 wavelength range, it is also the most sensitive parameter in our models. Therefore, we examine the \ch{H2O} abundances retrieved for the favoured models (cloud free, grey clouds, or non-grey clouds) for each object. These values are represented in Figure \ref{fig:H2O_abundance_fav}, along with values retrieved by \citet{FisherHeng18} using an isobaric model.

\begin{figure*}
\centering
\begin{minipage}{\columnwidth}
 \centering
 \includegraphics[width=\columnwidth]{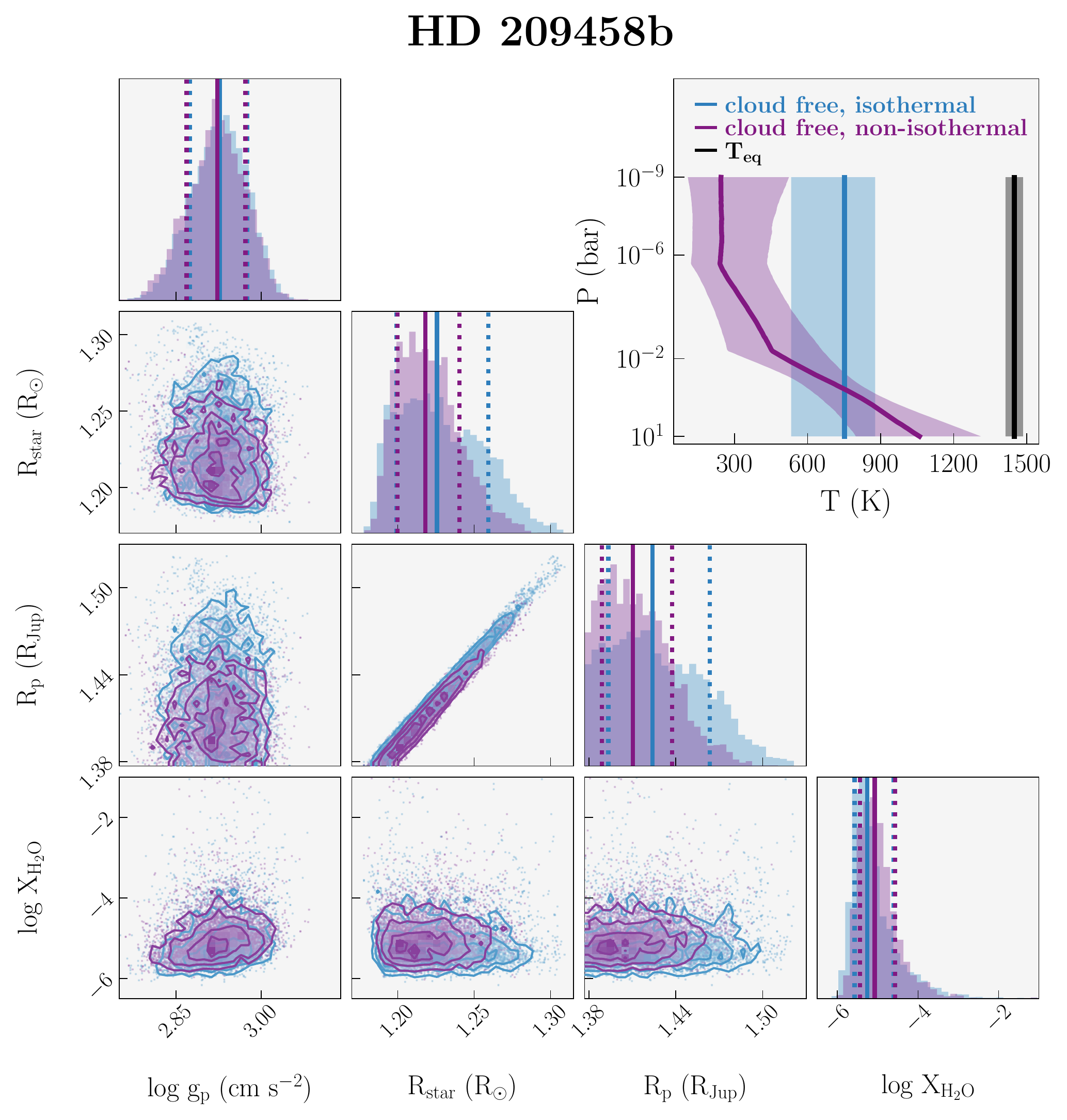}
\end{minipage}
\hfill
\begin{minipage}{\columnwidth}
 \centering
 \includegraphics[width=\columnwidth]{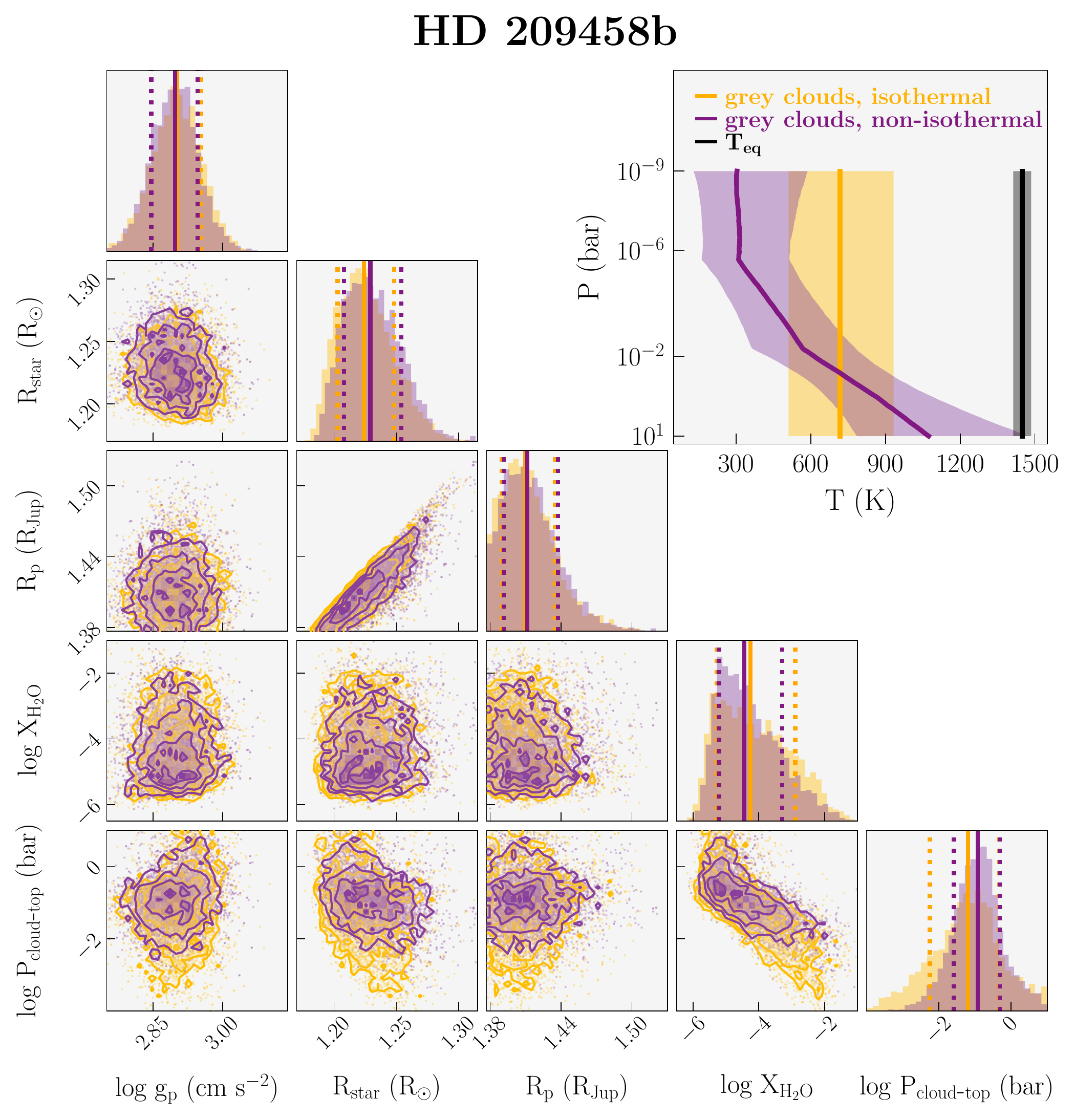}
\end{minipage}
\vspace{0.1cm}
\begin{minipage}{\columnwidth}
 \centering
 \includegraphics[width=\columnwidth]{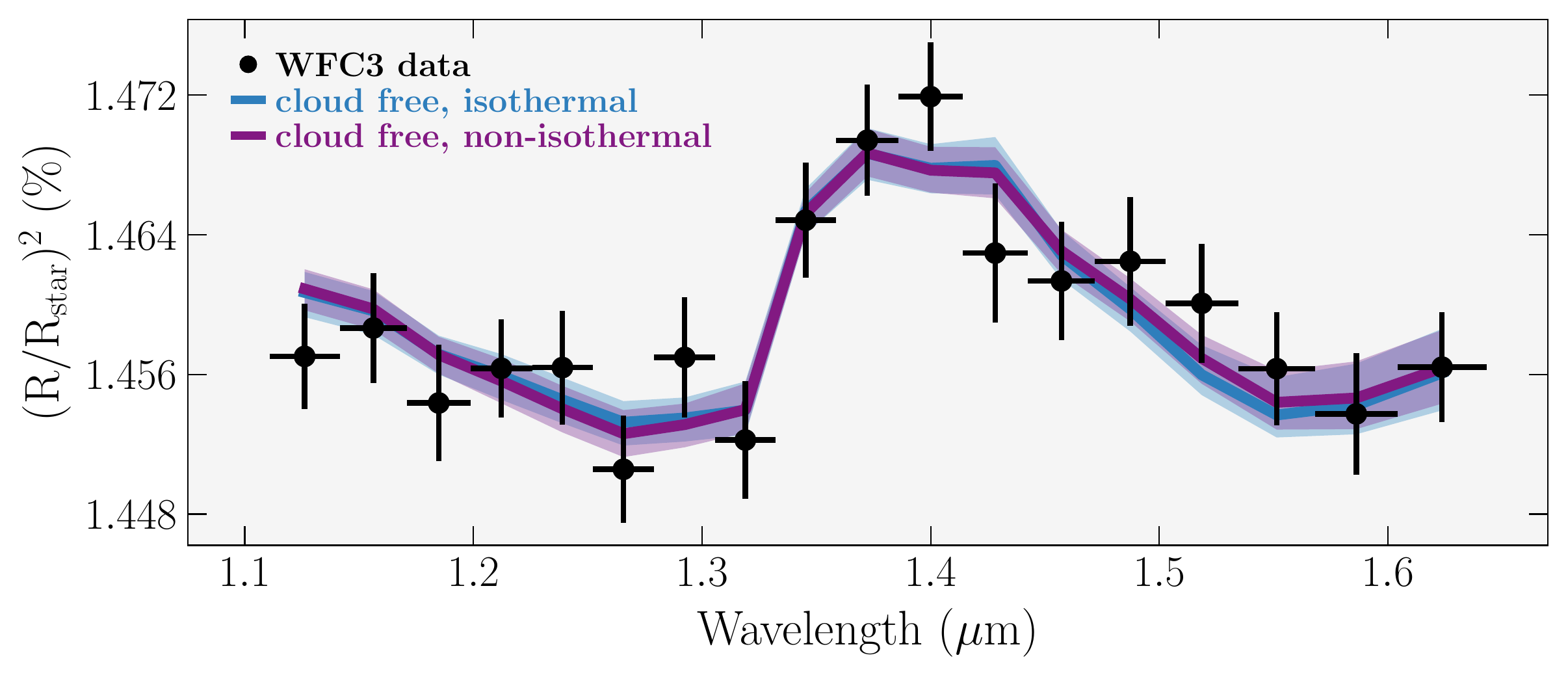}
\end{minipage}
\hfill
\begin{minipage}{\columnwidth}
 \centering
 \includegraphics[width=\columnwidth]{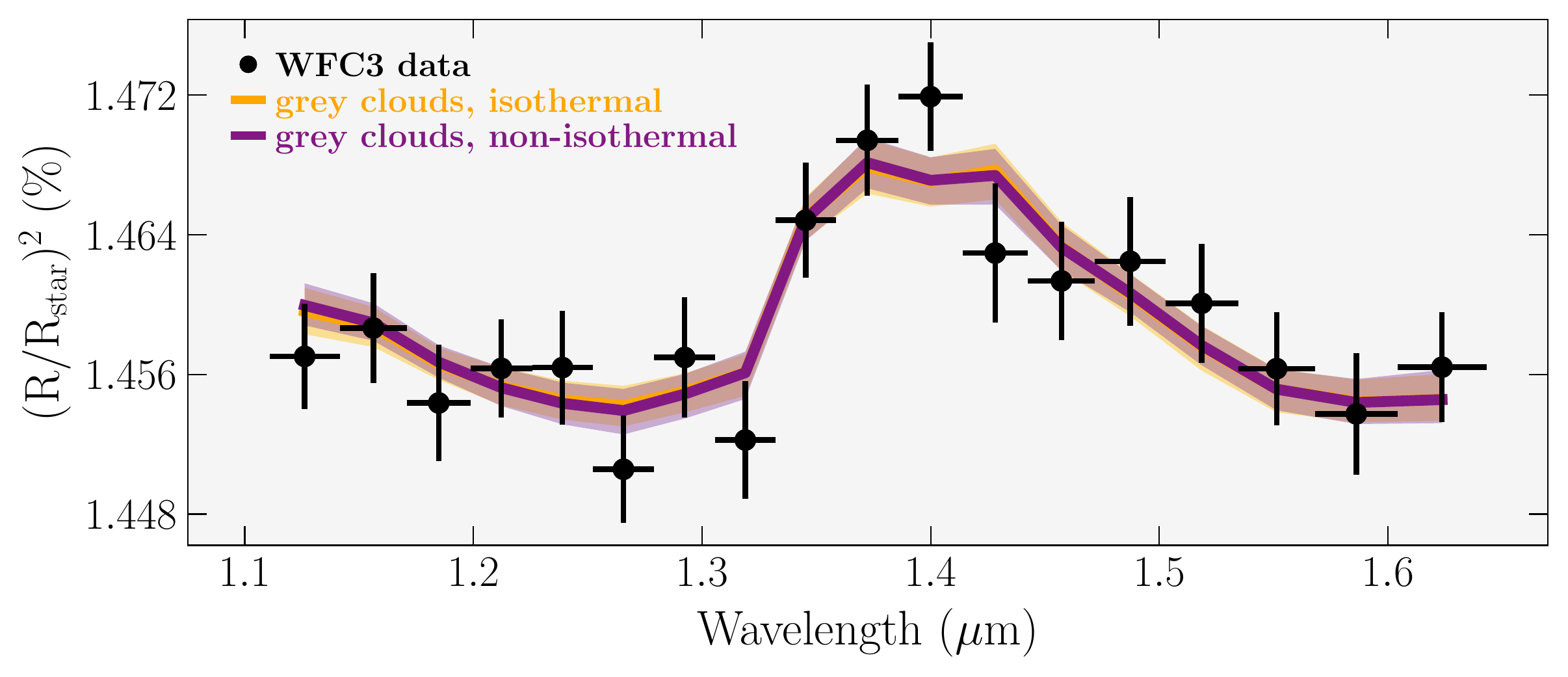}
\end{minipage}
\caption{Top figures: HD 209458b isothermal full cloud-free retrieval (blue, left) and grey-cloud retrieval (yellow, right) vs. non-isothermal cloud-free (purple, left) and grey-cloud (purple, right) outcomes. Top-right panel in each figure shows the temperature-pressure profiles, with the equilibrium temperature ($T_{\rm eq}$) and its 1$\sigma$ uncertainty calculated by \citet[][see Table \ref{table:input_parameters} for each value]{Fu+17}. Bottom figures: best-fit spectra and its associated 1$\sigma$ uncertainty region for the same object, binned to the data resolution. Black points correspond to data observed by \textit{HST} WFC3.}
\label{fig:corner_non_isothermal_HD209458b}
\end{figure*}

\begin{figure*}
\centering
\begin{minipage}{\columnwidth}
 \centering
 \includegraphics[width=\columnwidth]{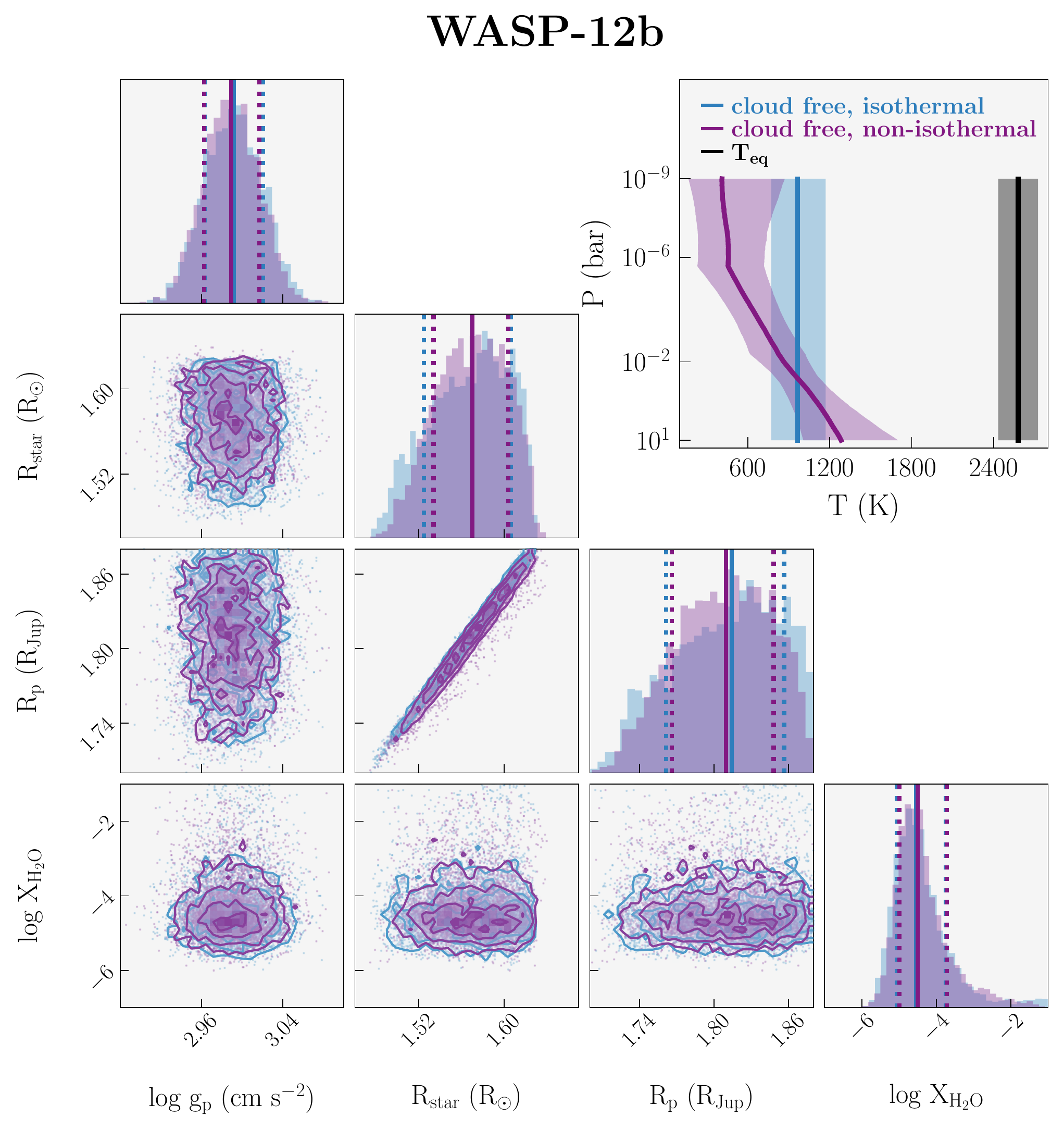}
\end{minipage}
\hfill
\begin{minipage}{\columnwidth}
 \centering
 \includegraphics[width=\columnwidth]{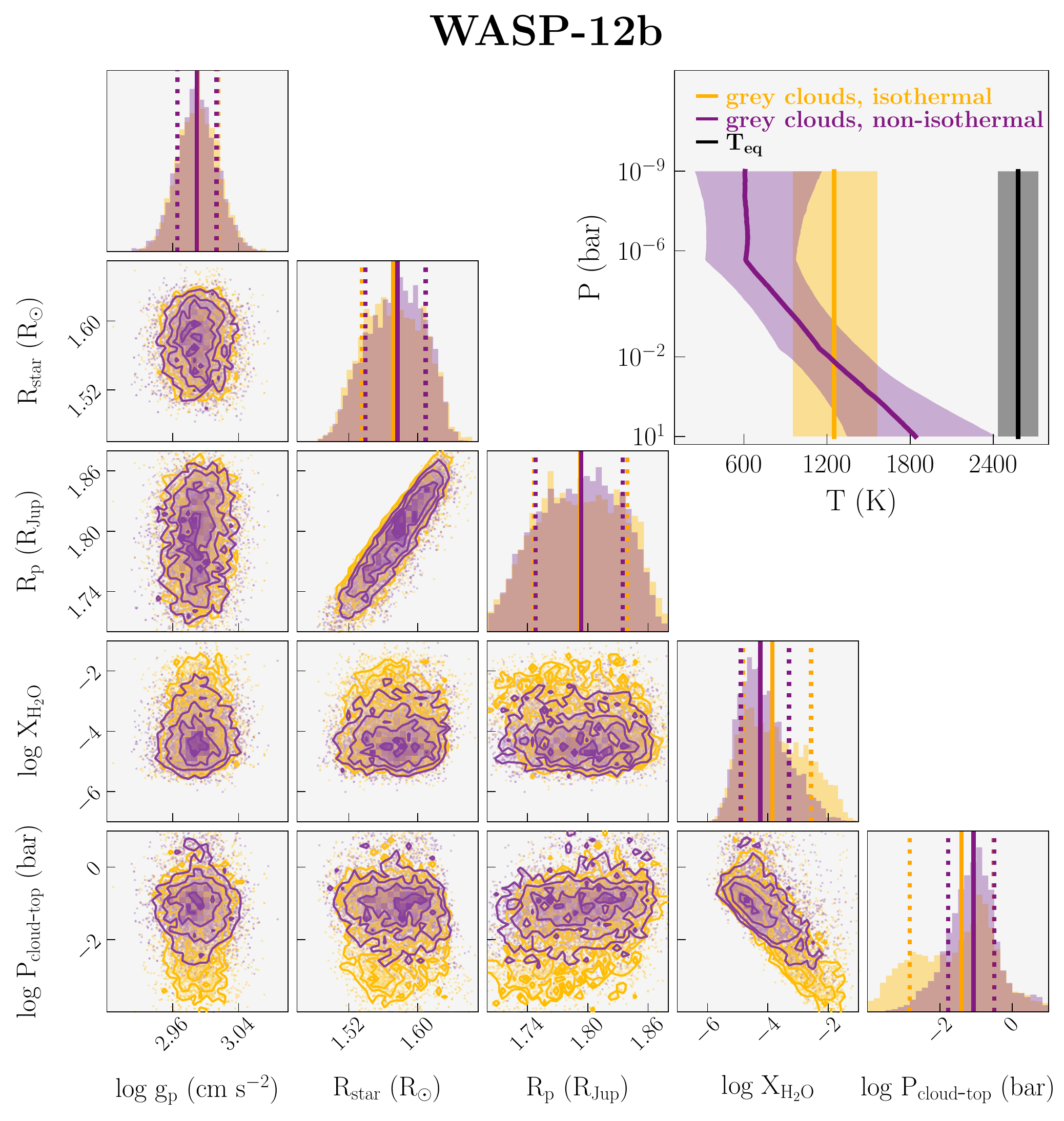}
\end{minipage}
\vspace{0.1cm}
\begin{minipage}{\columnwidth}
 \centering
 \includegraphics[width=\columnwidth]{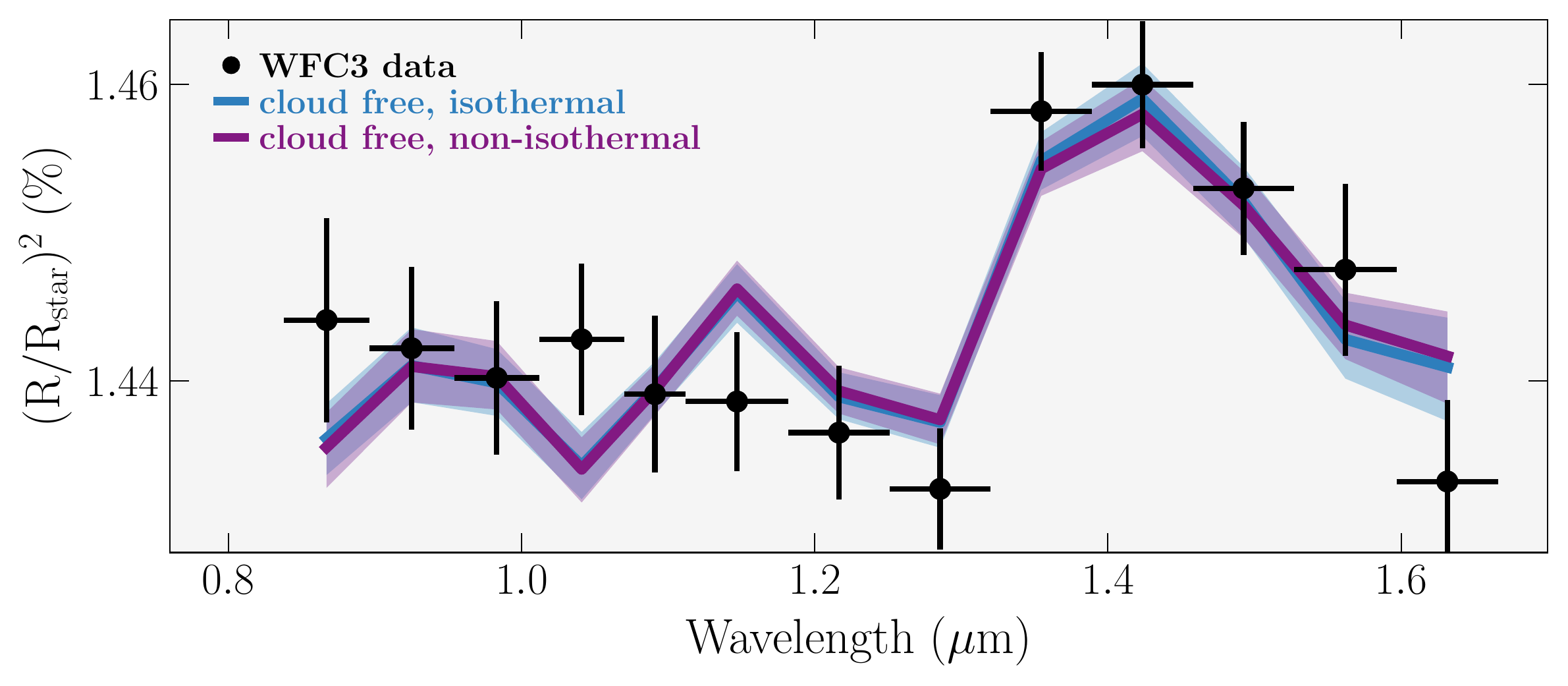}
\end{minipage}
\hfill
\begin{minipage}{\columnwidth}
 \centering
 \includegraphics[width=\columnwidth]{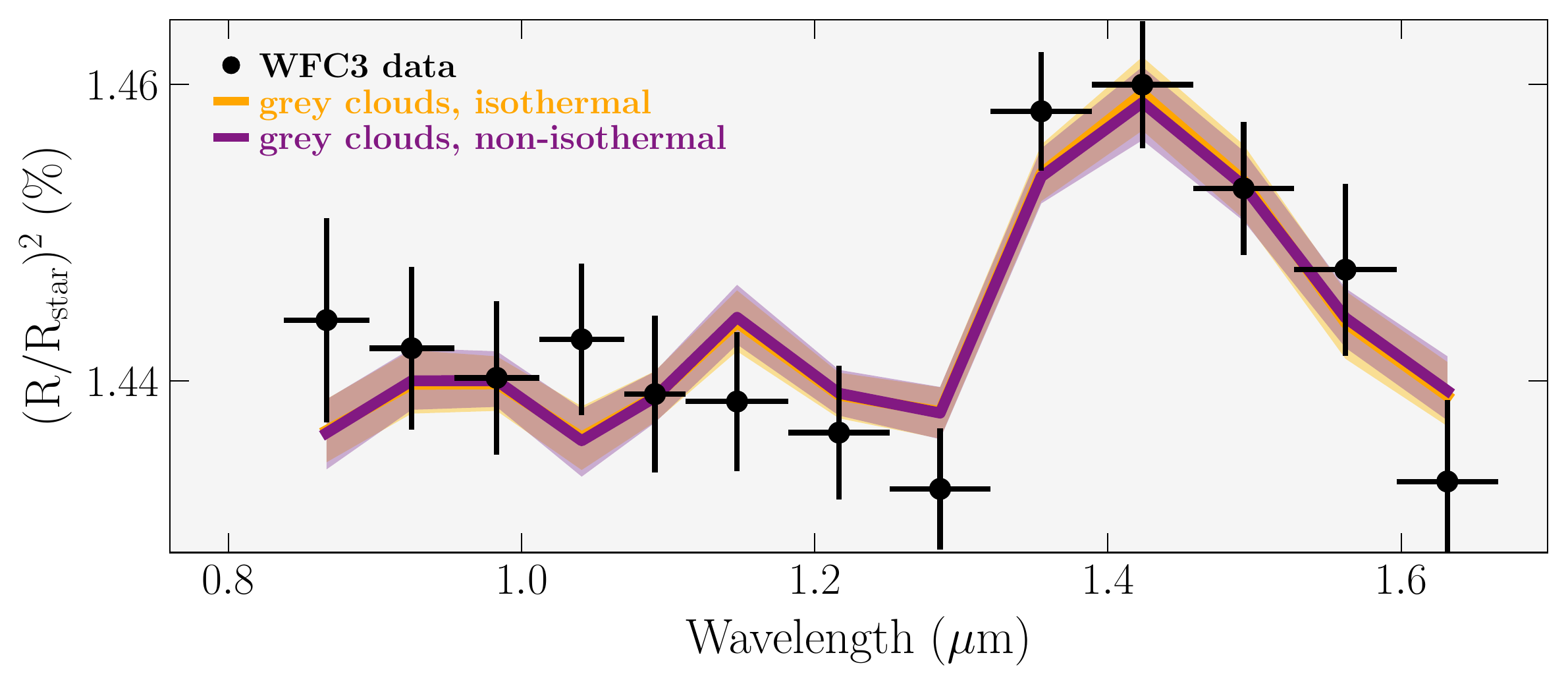}
\end{minipage}
\caption{Same as Figure \ref{fig:corner_non_isothermal_HD209458b} but for WASP-12b.}
\label{fig:corner_non_isothermal_WASP-12b}
\end{figure*}

\begin{figure*}
\centering
\begin{minipage}{\columnwidth}
 \centering
 \includegraphics[width=\columnwidth]{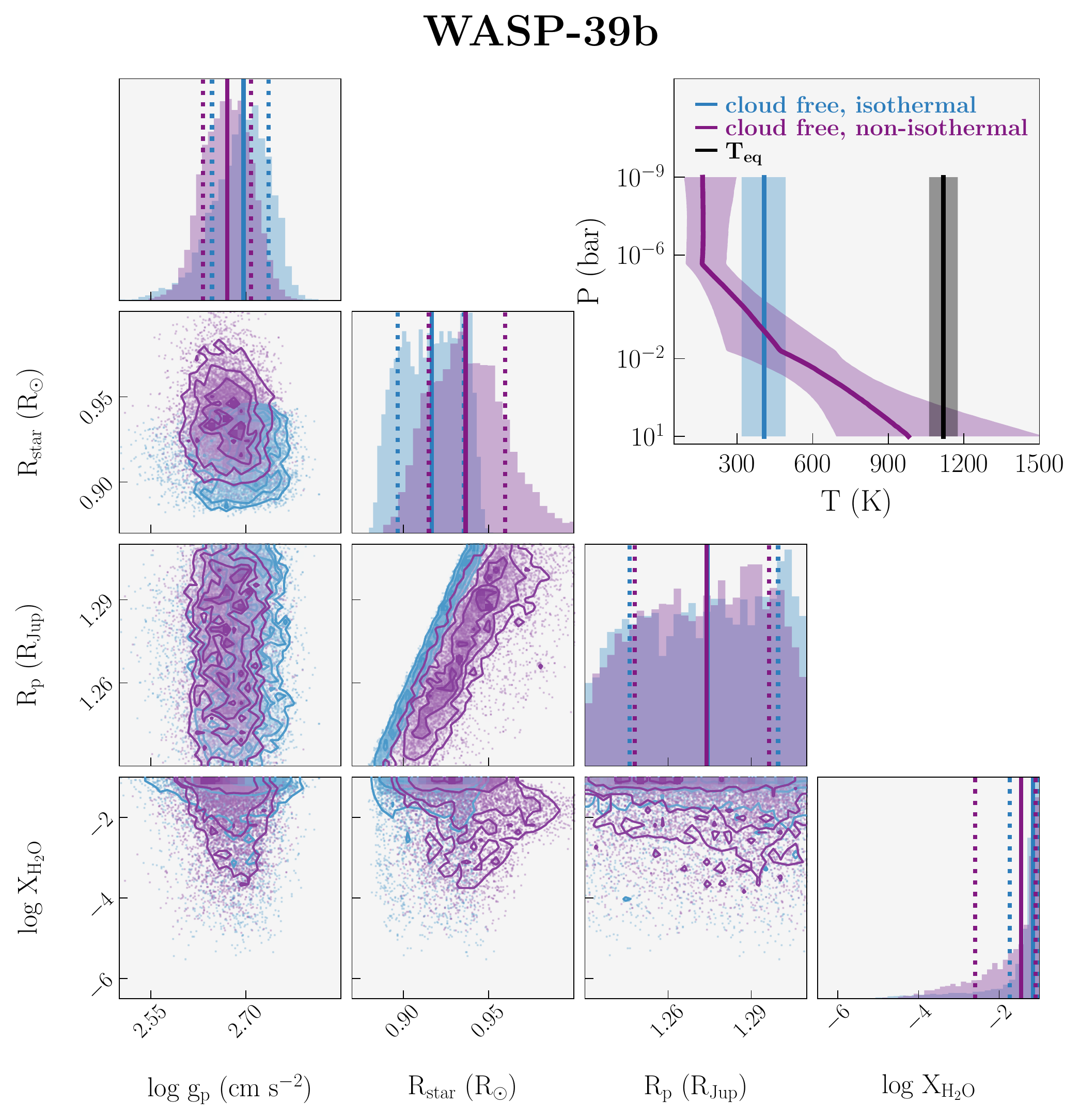}
\end{minipage}
\hfill
\begin{minipage}{\columnwidth}
 \centering
 \includegraphics[width=\columnwidth]{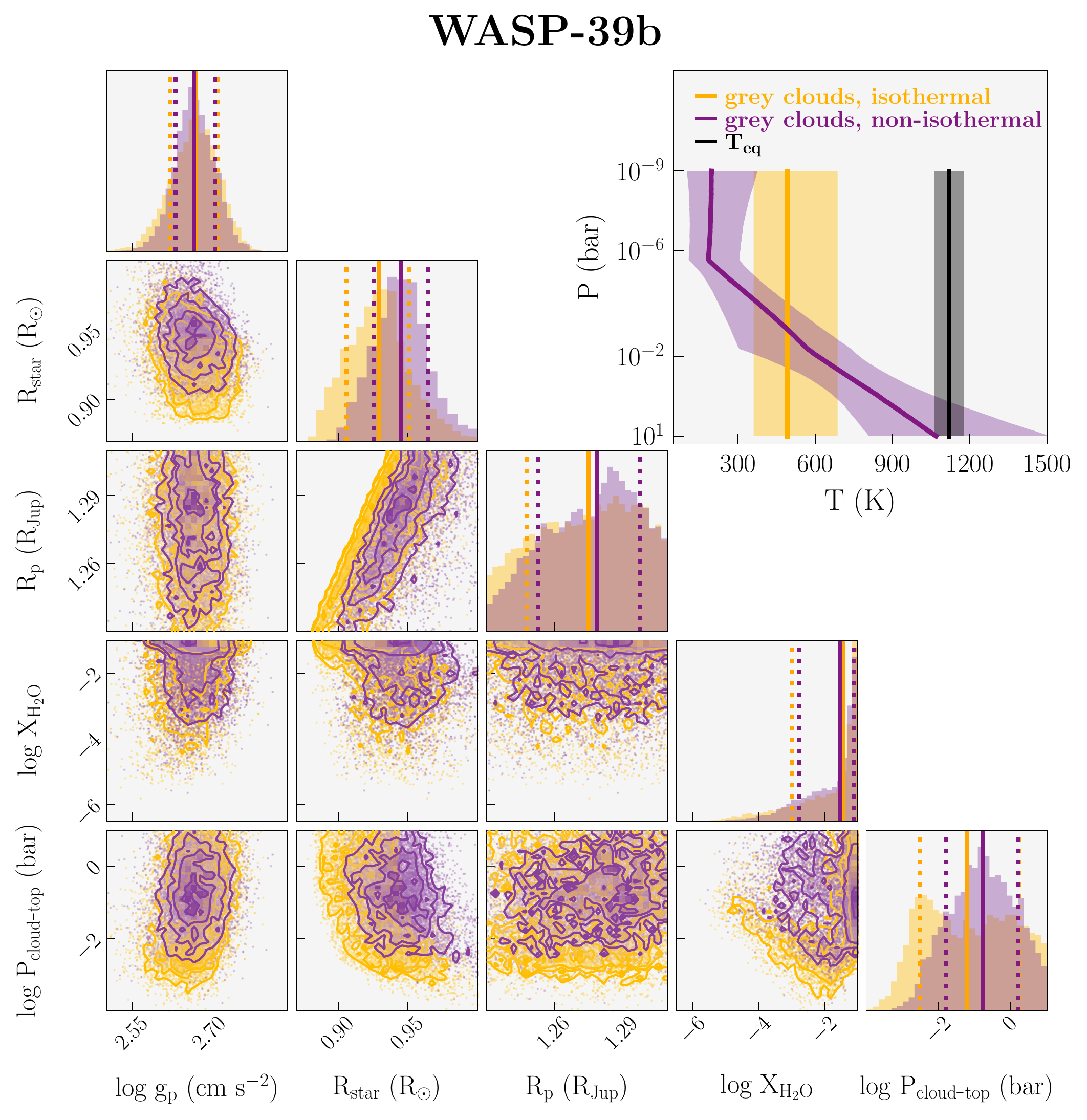}
\end{minipage}
\vspace{0.1cm}
\begin{minipage}{\columnwidth}
 \centering
 \includegraphics[width=\columnwidth]{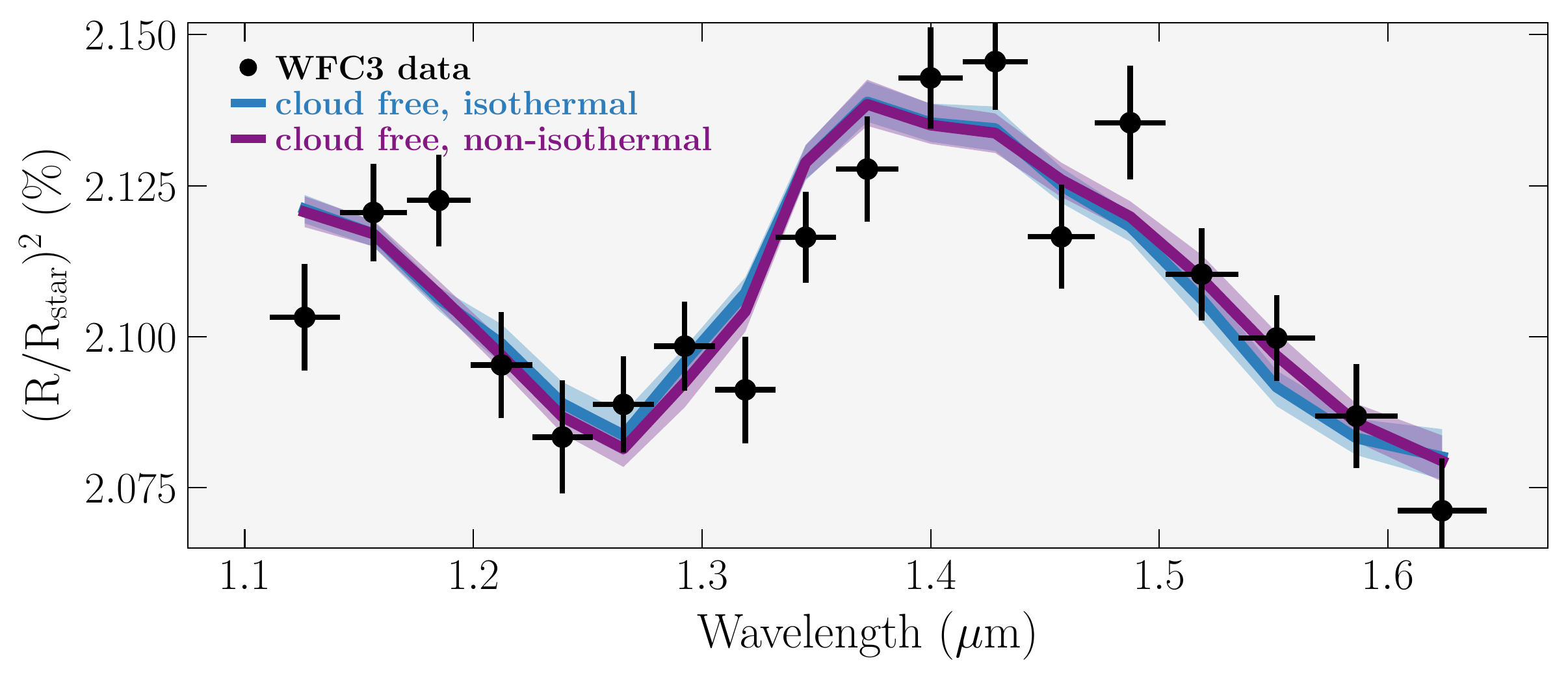}
\end{minipage}
\hfill
\begin{minipage}{\columnwidth}
 \centering
 \includegraphics[width=\columnwidth]{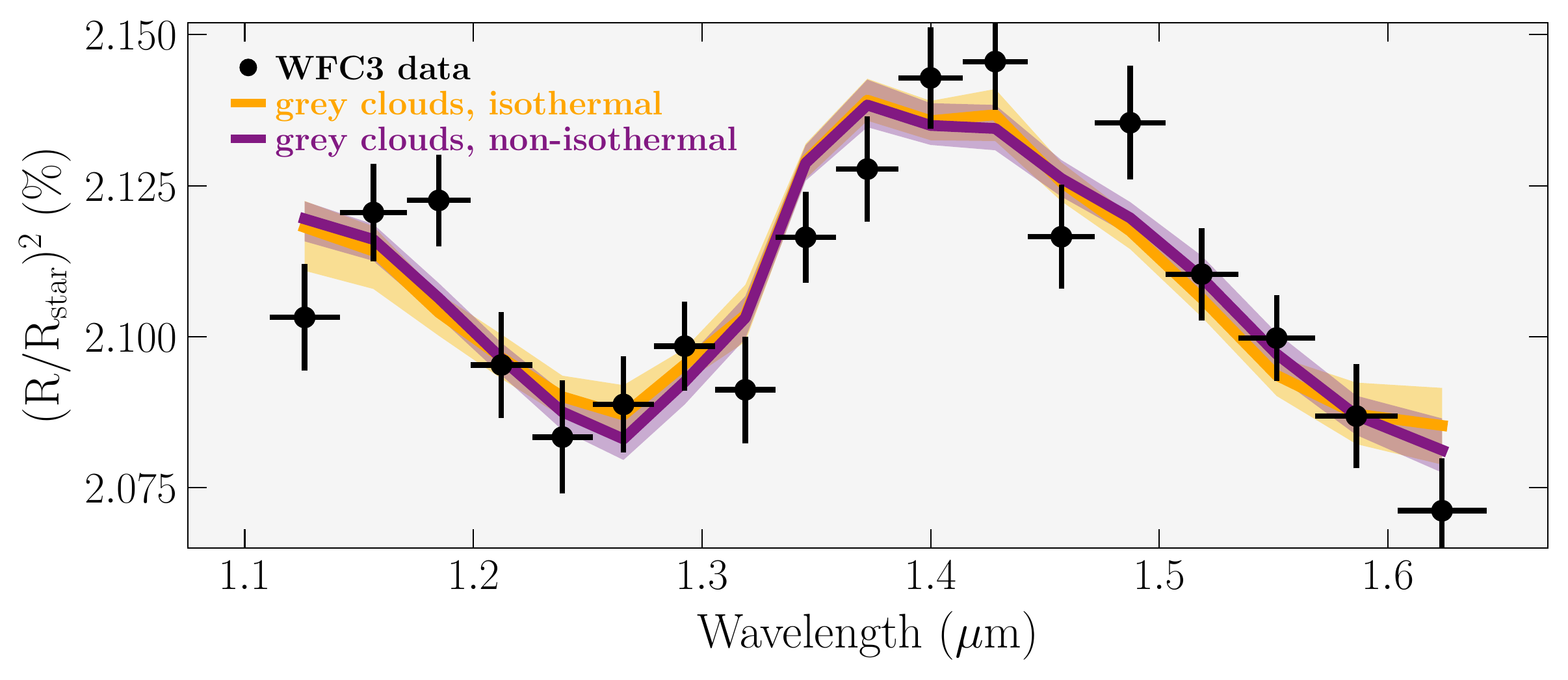}
\end{minipage}
\caption{Same as Figures \ref{fig:corner_non_isothermal_HD209458b} and \ref{fig:corner_non_isothermal_WASP-12b} but for WASP-39b.}
\label{fig:corner_non_isothermal_WASP-39b}
\end{figure*}

Isobaric \ch{H2O} abundances retrieved by \citet{FisherHeng18} often present larger uncertainties, and are sometimes inconsistent with our non-isobaric values (e.g. HAT-P-1b and HAT-P-41b). An explanation for these large uncertainties and inconsistencies is the absolute spectral continuum sourced by CIA. In \citet{FisherHeng18}, even though CIA is included in all models, their isobaric approach neglects the effects of CIA in their retrievals. As CIA strongly depends on pressure, our pressure-gradient models correctly calculate CIA in each pressure layer, providing more accurate \ch{H2O} abundances (as shown in \citealt{WelbanksMadhusudhan19}) and highlighting the need for a non-isobaric assumption.

Abnormally high \ch{H2O} abundances close to 10\% (i.e. $\log X_{\ch{H2O}} \simeq -1$) are retrieved for WASP-39b, which favours a cloud free model. This is because WFC3 mainly identifies \ch{H2O} features, but fails to cover spectral features in wider wavelength ranges, such as molecular and cloud signatures. With the lack of additional opacity features, accurately constraining the water abundance is challenging (as shown in \citealt{LineParmentier16}) and our outcomes may result in unphysical values \citep{BarstowHeng20}. This will be further examined in the next section. Furthermore, \ch{H2O} abundances that are too high (or too low) could be compensated within the retrievals by higher or lower values for the other parameters, such as temperature or clouds.

\subsection{Temperatures}
\label{sec:temperatures}

After water abundance, temperature is the most sensitive parameter in our models. Analogous to Figure \ref{fig:H2O_abundance_fav}, Figure \ref{fig:temperature_fav} examines temperature values retrieved for the favoured models (cloud free, grey clouds, or non-grey clouds) for each object. Isobaric values retrieved by \citet{FisherHeng18} and equilibrium temperatures calculated by \citet{Fu+17} are also shown.

Isobaric temperatures retrieved by \citet{FisherHeng18} are overall comparable or higher than this work's values. As explained in the previous section, our models sets the CIA continuum in each pressure layer, as opposed to \citet{FisherHeng18}, which affects the retrieved \ch{H2O} abundances, and consequently the scale height, and thus the retrieved temperatures. This effect is also demonstrated in Figure \ref{fig:param_effect}.

Low temperatures are a well-known puzzle in atmospheric retrievals of transmission spectra. \citet{MacDonald+20} showed that 1D retrievals often obtain temperatures $\sim$1000 K cooler than the planet's equilibrium temperature. They attribute these unphysically low temperatures to the 1D model's inability to reproduce chemical differences between the morning and evening terminators. This effect is found to be strongest for ultra-hot Jupiters. Figure \ref{fig:temperature_fav} shows that, for nearly all objects, we retrieve a temperature below the equilibrium value of \citet{Fu+17}, with differences of sometimes 1000 K or higher. 

In addition to limb asymmetries, there are other issues that could cause erroneous temperatures. Firstly, our model assumes an isothermal atmosphere, so not only do we assume longitudinal symmetry, but also vertical symmetry. Since transmission spectra probe a relatively high-altitude region, this could account for some retrieved temperatures being cooler than expected. Secondly, the temperature only arises in two computations in the model -- in the calculation of the opacities and the atmospheric scale height. Since the effect of temperature on the opacities is fairly small with respect to the resolution of \textit{HST} data, this is unlikely to be constraining the temperature. Therefore, the main effect of temperature is on the scale height of the atmosphere. With only two spectral features covered by WFC3 (a strong one at $\sim$1.4 $\upmu$m and a weaker one at $\sim$1.2 $\upmu$m), and little baseline for the continuum, it can be challenging to retrieve an accurate scale height from this data alone. For example, one of the biggest temperature discrepancies occurs for the ultra-hot Jupiter WASP-121b, which, as previously mentioned, has detections of additional species in its atmosphere that could be affecting this wavelength range \citep{Evans+18,MikalEvans+22}. This is a potential case where neglecting other molecular species could be affecting the retrieval results.

These issues of low retrieved temperatures will be further investigated in Section \ref{sec:case_studies}.

\section{Case studies: Investigating the low temperature ``problem''}
\label{sec:case_studies}

As discussed in Section \ref{sec:temperatures}, our retrievals obtain substantially lower temperatures than expected from the equilibrium values, in most cases. We speculated that this could be caused by limb asymmetries (as investigated in \citealt{MacDonald+20}), the narrow and relatively high-altitude pressure region probed by transmission, or challenges retrieving an accurate scale height from WFC3 data. The latter causes further issues in retrieving the abundance of \ch{H2O}, demonstrated by the abnormally high abundance values retrieved for WASP-39b. 

In order to further investigate the low temperature problem, we choose three planets to study in more detail: HD 209458b, WASP-12b, and WASP-39b. These objects were selected to represent our full dataset due to being three of the best-studied giant exoplanets in the literature. Additionally, WASP-12b and WASP-39b represent some extremes of the previously mentioned cases of low temperature and high \ch{H2O} abundance, respectively. HD 209458b provides a more intermediate case study. The next sections provide additional tests on these three objects.

Figure \ref{fig:bayesian_comparison_case_study} shows that, amongst the models with $\log$ Bayes factor < 1, grey clouds are the simplest model for HD 209458b and WASP-12b, while the cloud-free model is the simplest for WASP-39b. Thus following Occam's razor, we consider these the best-fit models for each planet (see \ref{sec:bayesian_comparison} for definition of ``best-fit model''). In order to compare the presence or absence of clouds in our additional analyses, we perform tests using both the cloud-free and grey-cloud models.

\subsection{HD 209458b}

Figure \ref{fig:corner_cloudfree_greycloud_HD209458b} highlights the differences between HD 209458b cloud-free and grey-cloud retrieval outcomes. The inclusion of clouds leads to a higher \ch{H2O} abundance with a wider posterior, due to the degeneracy between $X_{\ch{H2O}}$ and the cloud-top pressure $P_{\rm cloud\mbox{-}top}$. As previously discussed in Section \ref{sec:degeneracies}, secondary normalization degeneracies are also present between $R_{\rm p}$ and $R_{\rm star}$ and between \ch{H2O} and temperature. With the inclusion of clouds, the \ch{H2O} abundance is allowed to reach higher values, as expected.

\subsection{WASP-12b}

As expected, Figure \ref{fig:corner_cloudfree_greycloud_WASP-12b} of WASP-12b displays a similar behaviour for $X_{\ch{H2O}}$ and $R_{\rm p}$ posteriors as for HD 209458b, where the addition of clouds leads to a wider $X_{\ch{H2O}}$ posterior. $X_{\ch{H2O}}$--$P_{\rm cloud\mbox{-}top}$, $X_{\ch{H2O}}$--$T$, and $R_{\rm p}$--$R_{\rm star}$ degeneracies are also present. In particular, although the median and 1$\sigma$ cloud-free and grey-cloud retrieved temperatures remain substantially lower than the $T_{\rm eq}$ value of 2580 K, the grey-cloud model retrieves higher temperatures, and the grey-cloud posterior distribution encompasses higher temperatures in its posterior tail.

\subsection{WASP-39b}

Figure \ref{fig:corner_cloudfree_greycloud_WASP-39b} shows cloud-free and grey-cloud retrieval outcomes for WASP-39b provide very high \ch{H2O} abundances and very low temperatures. As the cloud-free model is favoured, and therefore clouds are not required to fit the data, the inclusion of clouds in the case of WASP-39b has a more modest effect in comparison to HD 209458b and WASP-12b. The \ch{H2O} posterior has a stronger left tail tending to slightly lower values when clouds are included. The cloud-top pressure is fairly unconstrained, demonstrating its lack of effect on the spectrum due to the extremely high retrieved water abundance. The $X_{\ch{H2O}}$--$T$ degeneracy for the cloud-free model displayed here is representative of the high abundance section of the banana shape discussed in Section \ref{sec:degeneracies}, where the high \ch{H2O} abundance increases the mean molecular weight, which is balanced by an increase in temperature.

\begin{figure*}
\centering
\begin{minipage}{\columnwidth}
 \centering
 \includegraphics[width=\columnwidth]{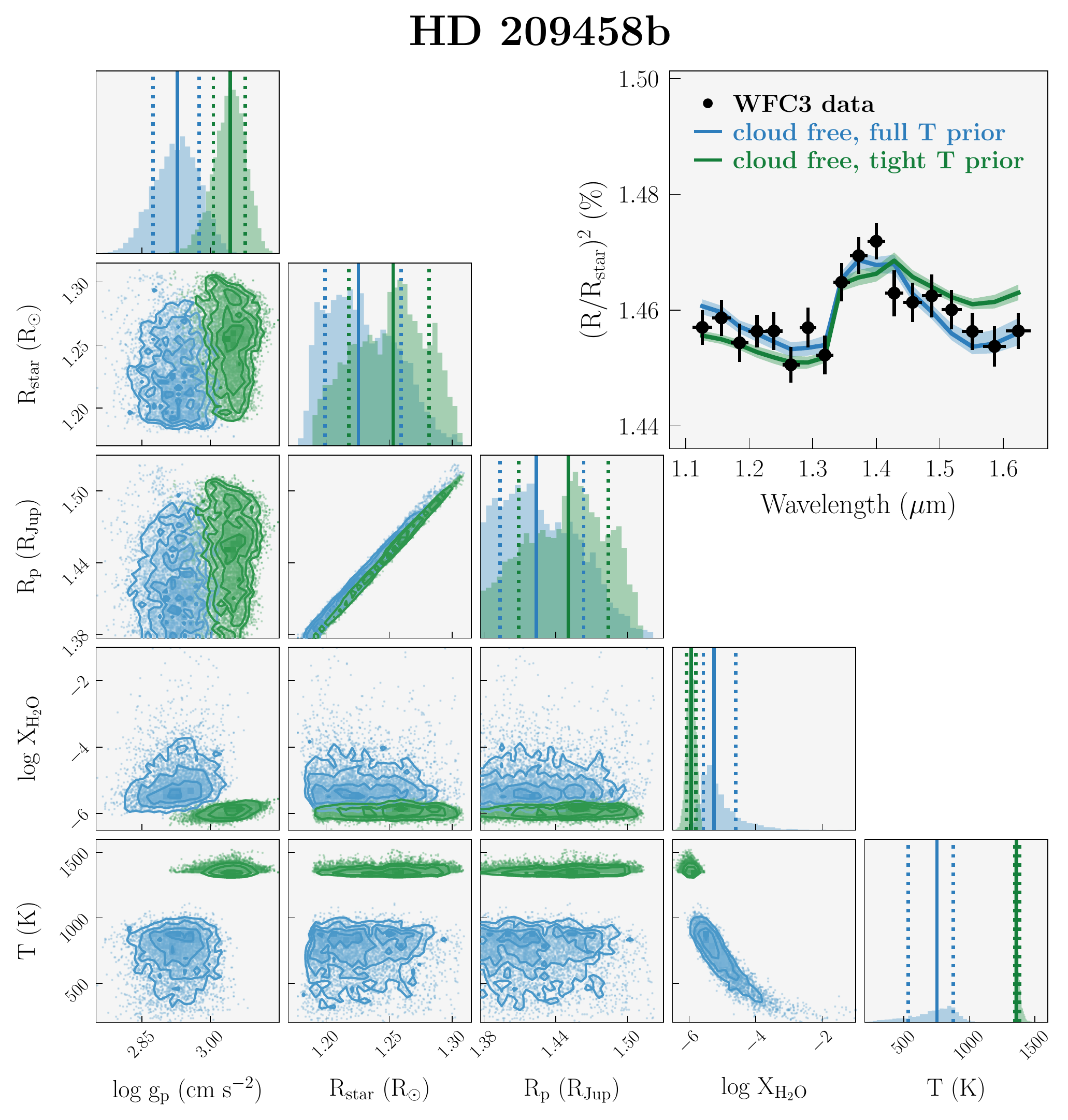}
\end{minipage}
\hfill
\begin{minipage}{\columnwidth}
 \centering
 \includegraphics[width=\columnwidth]{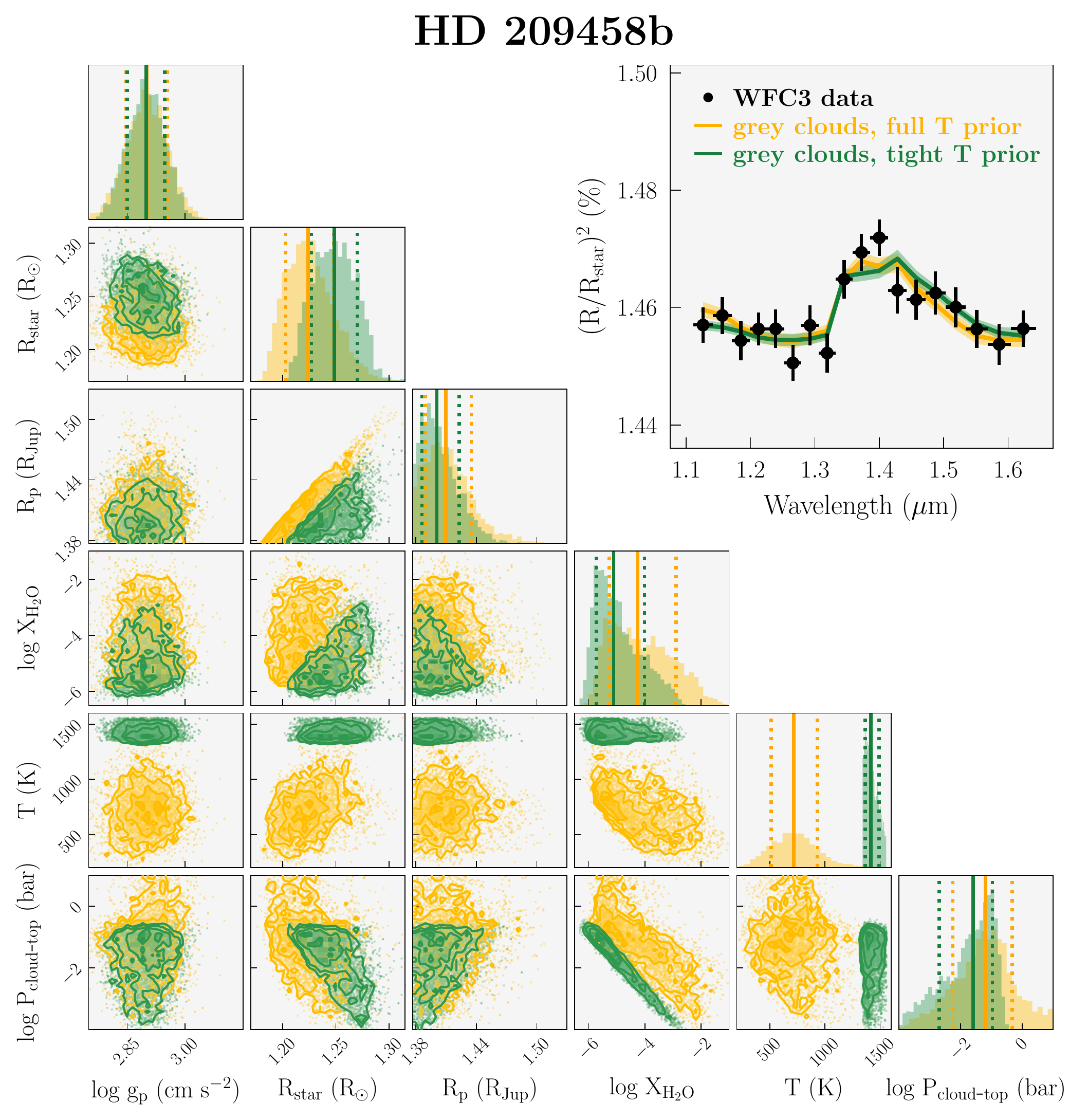}
\end{minipage}
\caption{Retrievals for HD 209458b using a full temperature prior range (200--3100 K) vs. using a tight prior range around the equilibrium temperature (1341--1557 K, i.e. $T_{\rm eq}$ $\pm$ 3$\sigma$). Blue and yellow represent cloud free and grey cloud models using full priors, while green represents the same models but using tight priors. $T_{\rm eq}$ and $\sigma$ values are listed in Table \ref{table:input_parameters}. Top-right panel in each corner plot shows the fitted spectrum and its associated 1$\sigma$ uncertainty region, binned to the data resolution. Black points correspond to data observed by \textit{HST} WFC3.}
\label{fig:corner_tight_temp_HD209458b}
\end{figure*}

\begin{figure*}
\centering
\begin{minipage}{\columnwidth}
 \centering
 \includegraphics[width=\columnwidth]{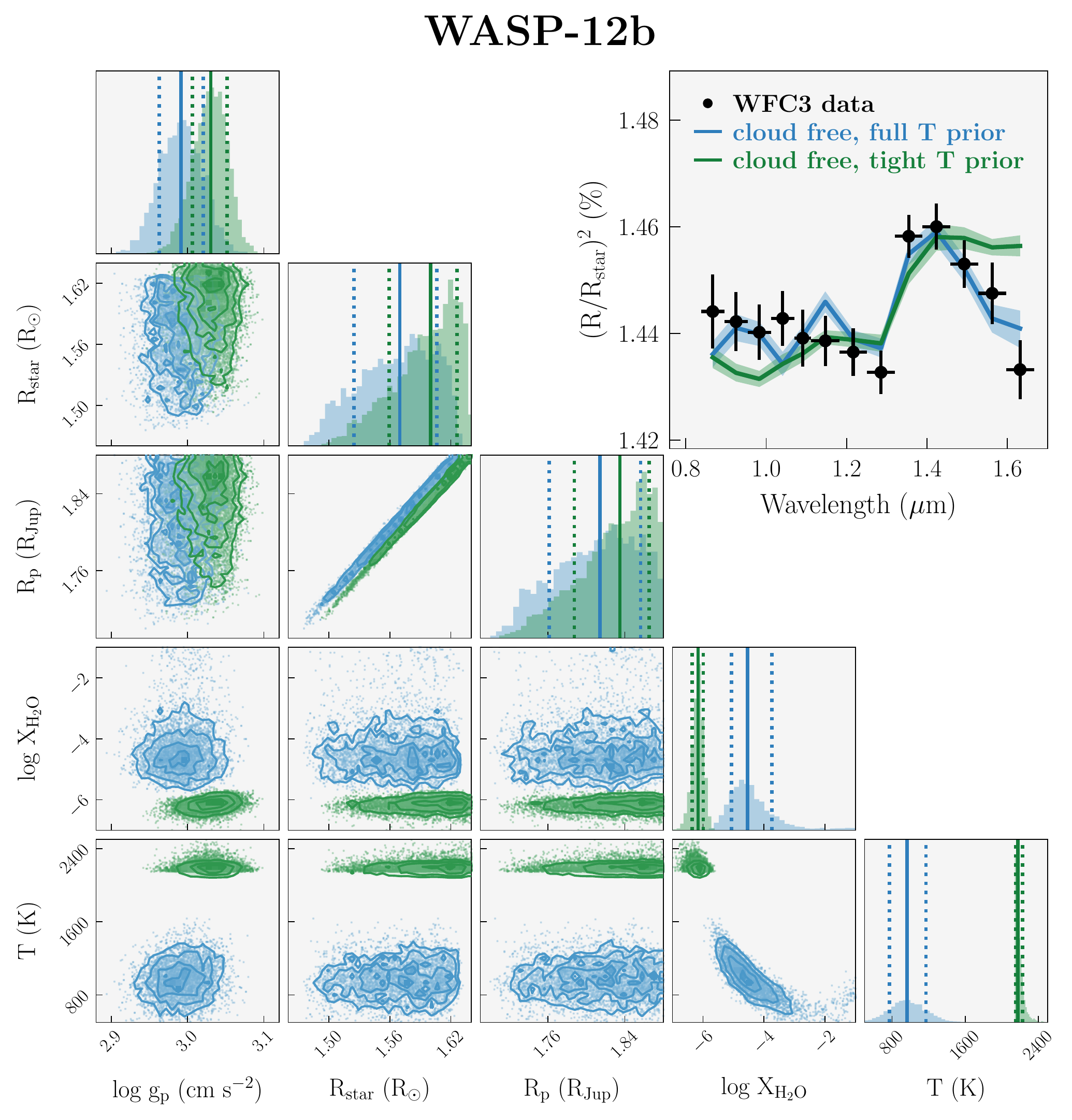}
\end{minipage}
\hfill
\begin{minipage}{\columnwidth}
 \centering
 \includegraphics[width=\columnwidth]{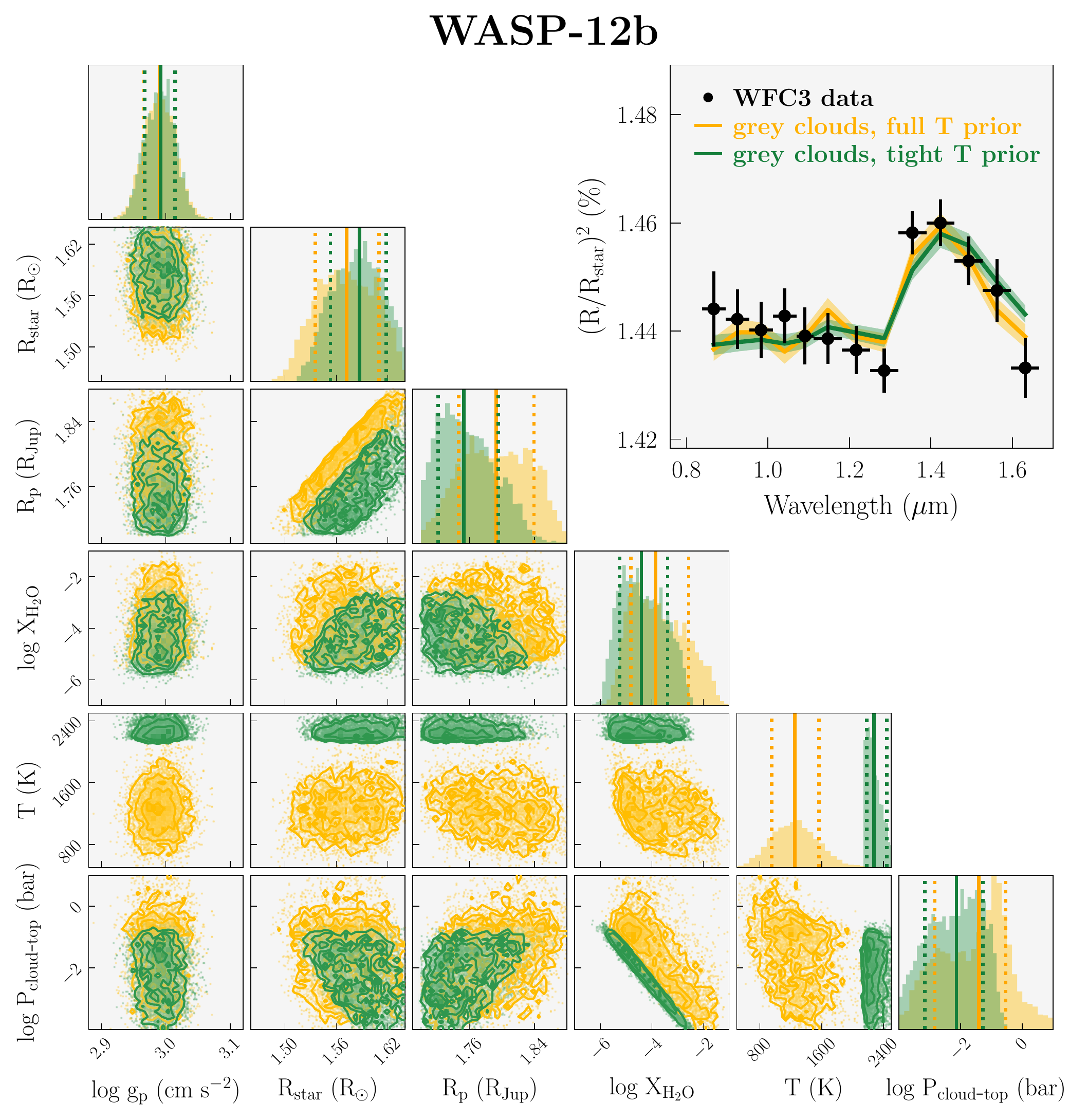}
\end{minipage}
\caption{Same as Figure \ref{fig:corner_tight_temp_HD209458b} but for WASP-12b, using a tight temperature prior range of 2142--3018 K (i.e. $T_{\rm eq}$ $\pm$ 3$\sigma$).}
\label{fig:corner_tight_temp_WASP-12b}
\end{figure*}

\subsection{Temperature tests}

As previously mentioned, most spectra in our sample retrieve temperatures much lower than their equilibrium values. In this sub-section we explore some of the possible explanations for these low temperatures. Firstly, we test the effect of the isothermal assumption on our case studies, to investigate if the narrow and high-altitude pressure region probed by transmission could explain the low temperatures. Secondly, we repeat our retrievals with constrained prior on temperature around the equilibrium value, forcing the temperature posteriors to higher values, and compare the results with the wider prior retrievals.

\begin{figure*}
\centering
\begin{minipage}{\columnwidth}
 \centering
 \includegraphics[width=\columnwidth]{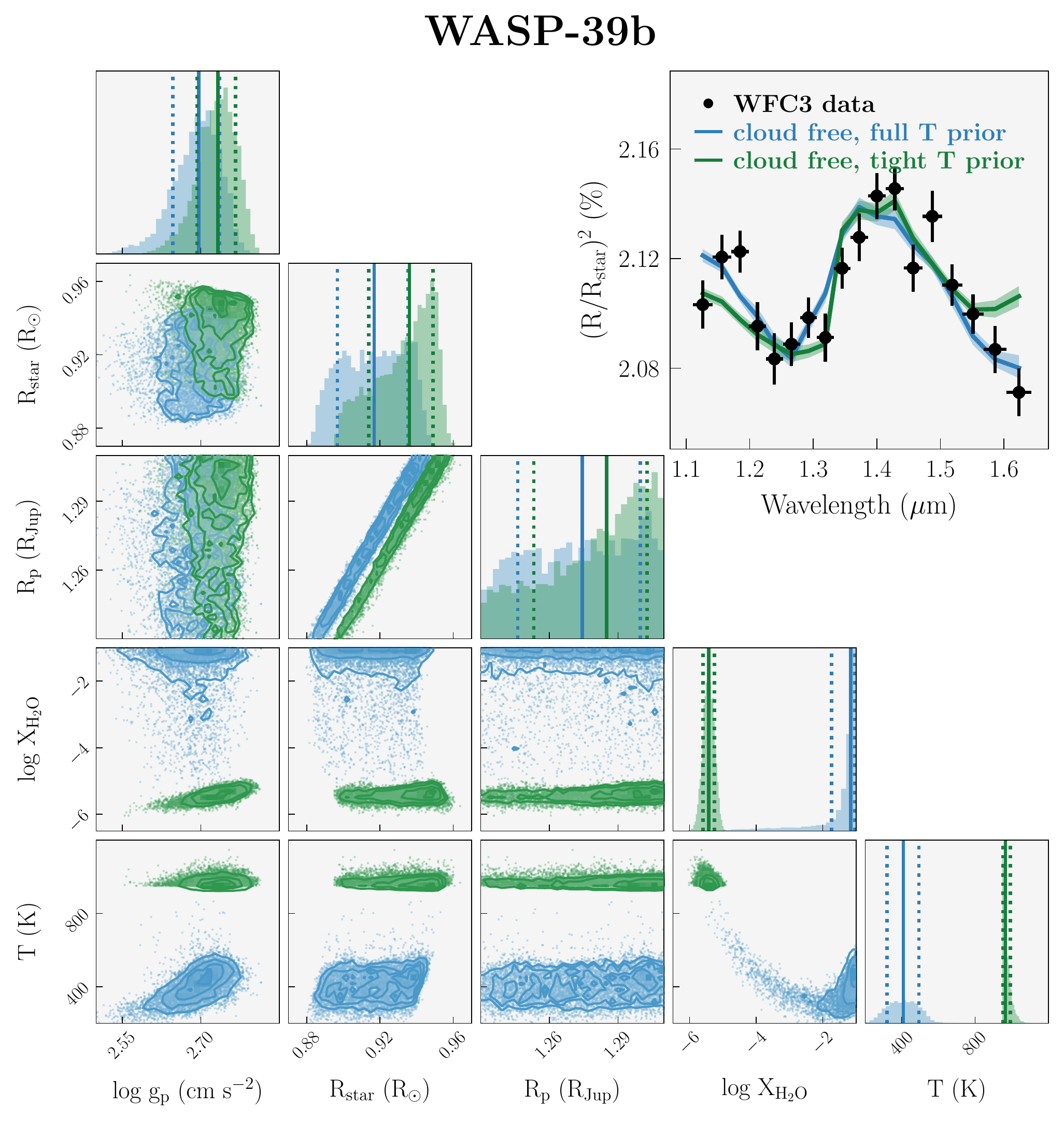}
\end{minipage}
\hfill
\begin{minipage}{\columnwidth}
 \centering
 \includegraphics[width=\columnwidth]{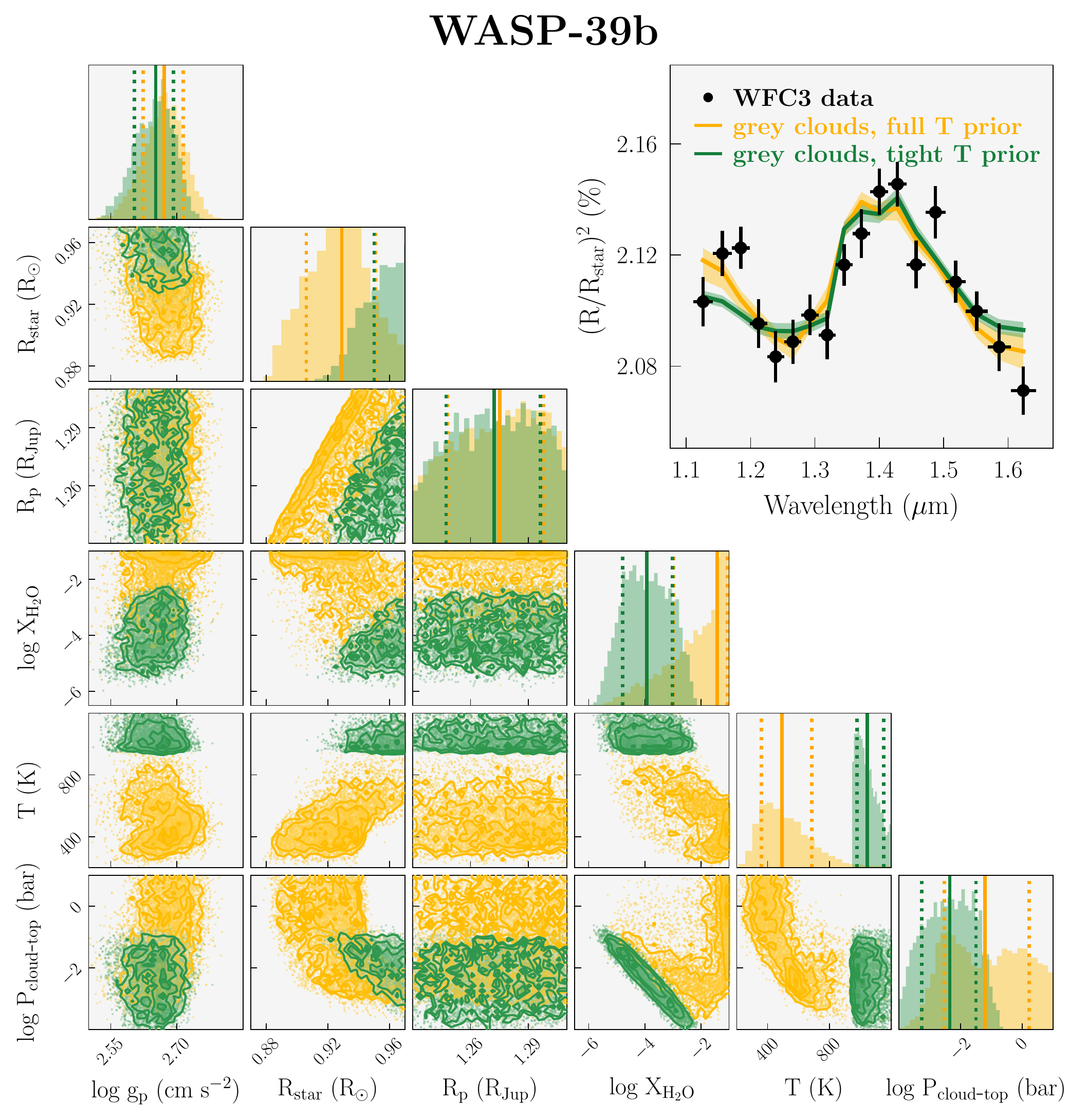}
\end{minipage}
\caption{Same as Figures \ref{fig:corner_tight_temp_HD209458b} and \ref{fig:corner_tight_temp_WASP-12b} but for WASP-39b, using a tight temperature prior range of 948--1290 K (i.e. $T_{\rm eq}$ $\pm$ 3$\sigma$).}
\label{fig:corner_tight_temp_WASP-39b}
\end{figure*}

\subsubsection{Isothermal vs. non-isothermal retrievals}
\label{sec:nonisothermal_tests}

In order to test the effect of the isothermal assumption, we performed retrievals using cloud-free and grey-cloud non-isothermal models, with the setup described in Section \ref{sec:nonisothermal_methodology}. Temperature-pressure profiles, posterior distributions, and modelled spectra are displayed in Figures \ref{fig:corner_non_isothermal_HD209458b}, \ref{fig:corner_non_isothermal_WASP-12b}, and \ref{fig:corner_non_isothermal_WASP-39b} for HD 209458b, WASP-12b, and WASP-39b, respectively. Retrieved parameter values are listed in Table \ref{table:nonisothermal_results}. The isothermal retrievals are the same as previously presented (Figures \ref{fig:corner_cloudfree_greycloud_HD209458b} to \ref{fig:corner_cloudfree_greycloud_WASP-39b}).

\begin{table}
 \centering
 \begin{tabular}{c|c|c|c|c}
 \hline
 \multirow{ 2}{*}{Planet} & \multicolumn{2}{c}{Cloud Free} & \multicolumn{2}{c}{Grey Cloud}\\
 & Full Prior & Tight Prior & Full Prior & Tight Prior \\ \hline
 HD 209458b & $1.07^{+0.24}_{-0.26}$ & $2.34^{+0.33}_{-0.27}$ & $1.05^{+0.22}_{-0.11}$ & $1.14^{+0.19}_{-0.10}$ \\
 \rule{0pt}{4ex} 
 WASP-12b & $1.71^{+0.39}_{-0.20}$ & $4.8^{+0.49}_{-0.37}$ & $1.54^{+0.42}_{-0.22}$ & $1.98^{+0.39}_{-0.22}$ \\
 \rule{0pt}{4ex} 
 WASP-39b & $2.27^{+0.37}_{-0.24}$ & $3.85^{+0.32}_{-0.25}$ & $2.58^{+0.32}_{-0.28}$ & $2.94^{+0.21}_{-0.14}$ \\
 \hline
 \end{tabular}
 \caption{Reduced $\chi^2$ values for the retrievals of HD 209458b, WASP-12b, and WASP-39b using the two different temperature priors -- the full prior and the tight prior around the $T_{\rm eq}$ value.}
 \label{tab:chi_sq}
\end{table}

The first aspect to notice is the temperature profile on the top-right corner of each figure (\ref{fig:corner_non_isothermal_HD209458b}, \ref{fig:corner_non_isothermal_WASP-12b}, and \ref{fig:corner_non_isothermal_WASP-39b}). For all three planets, the temperature profiles show a steep decrease from 10 to 10$^{-5}$--10$^{-6}$ bar, and remain approximately isothermal above this region. At altitudes around 10$^{-1}$--10$^{-3}$ bar, where the \textit{HST} WFC3 transmission spectra are expected to probe, the non-isothermal temperature values are slightly lower than isothermal, yet they are generally consistent within the uncertainty range. All isothermal and non-isothermal profiles are substantially lower than the equilibrium temperatures. For WASP-39b, the non-isothermal profile does overlap with the $T_{\rm eq}$ at the very bottom of the atmosphere ($\sim$10 bar), however it is highly unconstrained as we do not expect transmission spectra to probe these deep pressures.

We call attention to the sharp posteriors and elevated values for \ch{H2O} abundances retrieved for WASP-39b, which confirms high \ch{H2O} abundances are not caused by our isothermal assumption. The agreement between the isothermal and non-isothermal temperatures at the WFC3 photospheric region also suggests that the isothermal assumption is sufficient for interpreting these low-resolution spectra. In fact, despite the wider wavelength coverage and increased precision of \textit{JWST}, \citet{Lueber+24} demonstrated that the isothermal model is sometimes favoured for their retrievals on \textit{JWST} transmission spectra of WASP-39b, in particular for the NIRISS and PRISM instruments.

In contrast, for WASP-12b even the deep atmosphere is inconsistent with the equilibrium temperature. However, as investigated in \citet{MacDonald+20}, this ultra-hot Jupiter is a likely candidate for strong limb asymmetries, which can cause substantially lower temperatures in 1D retrievals.

Whilst we use these non-isothermal retrievals to test for variable temperature structures in the atmospheres, it is worth noting that the molecular abundances are kept constant with altitude. This could introduce variations in the temperature profile that compensate for the underlying water abundance profile. Although testing these non-constant molecular profiles is out of the scope of this paper, it is an important caveat to bear in mind.

\begin{figure}
\centering
\includegraphics[width=\linewidth]{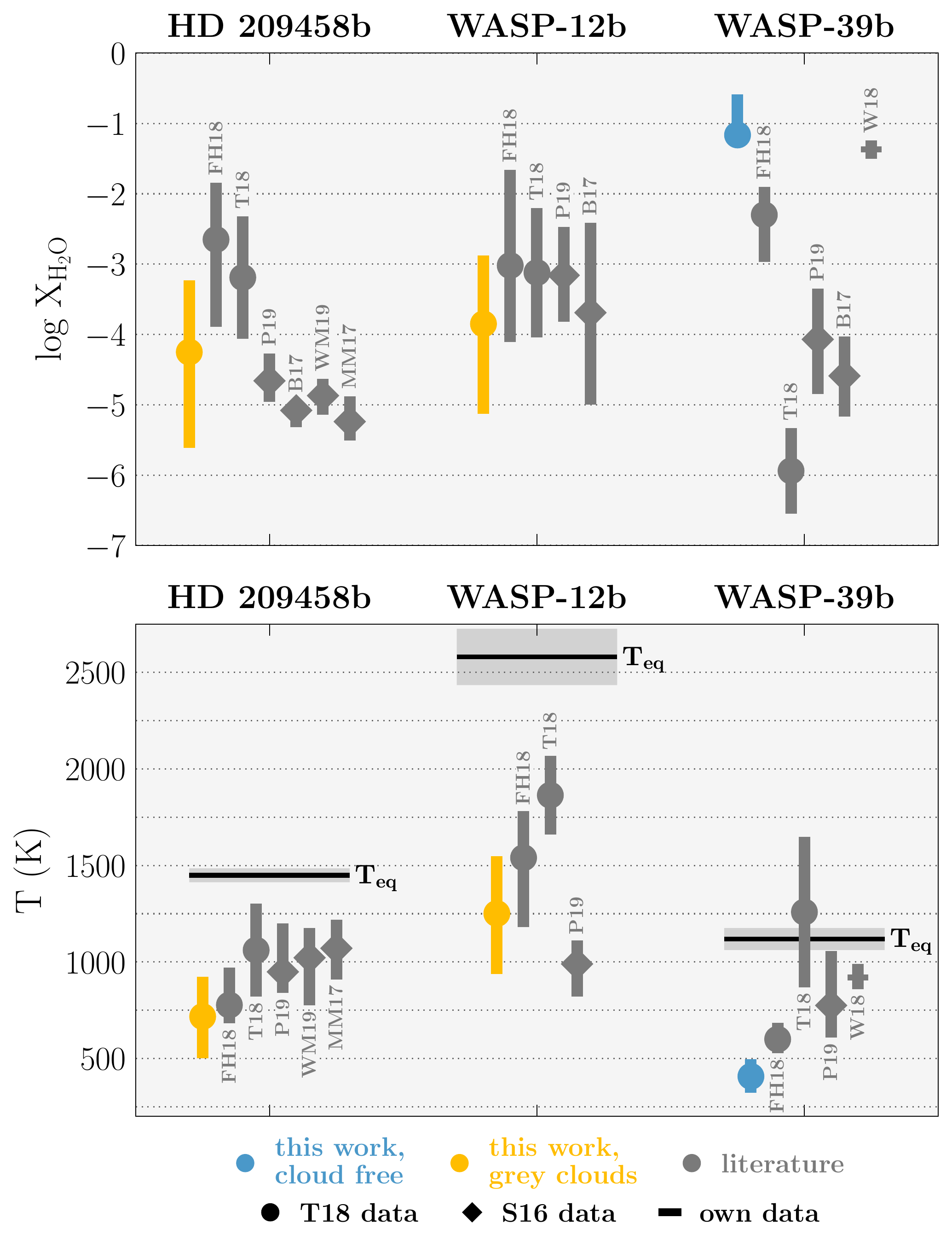}
\vspace{-0.5cm}
\caption{Comparison between retrieved \ch{H2O} abundances (top) and temperatures (bottom) for the best-fit cloud-free (blue) and grey-cloud (yellow) models using \texttt{BeAR}, and for best-fit models found by different works (grey). Abbreviations in the plot correspond to the works of \citealt{Barstow+17} (B17), \citealt{FisherHeng18} (FH18), \citealt{Tsiaras+18} (T18), \citealt{Wakeford+18} (W18), \citealt{Pinhas+19} (P19), \citealt{WelbanksMadhusudhan19} (WM19), and \citealt{MacDonaldMadhusudhan17a} (MM17). Values retrieved using data reduced by \citet{Tsiaras+18} and \citealt{Sing+16} (S16) are represented by circles and diamonds, respectively. Works that use their own data are denoted by a horizontal line. For non-isothermal retrievals (e.g. P19, WM19, MM17), the temperatures at the bottom panel represent the temperature calculated at the top of the atmosphere. $T_{\rm eq}$ represents equilibrium temperatures by \citet{Fu+17}.}
\label{fig:H2O_T_literature_models}
\end{figure}

\subsubsection{Tight temperature priors}
\label{sec:tight_temp}

So far, all our retrievals used a temperature prior range of 200--3100 K. We now perform a test in which we narrow the uniform temperature prior range to values around the equilibrium temperature for each object, forcing the retrieved temperature to higher values. Our priors for this test range from $T_{\rm eq}-3\sigma$ to $T_{\rm eq}+3\sigma$, where $T_{\rm eq}$ are the equilibrium temperatures calculated by \citet[][assuming zero albedo and uniform re-distribution of heat]{Fu+17} and $\sigma$ is the standard deviation. These values are listed in Table \ref{table:input_parameters}.

Figures \ref{fig:corner_tight_temp_HD209458b}, \ref{fig:corner_tight_temp_WASP-12b}, and \ref{fig:corner_tight_temp_WASP-39b} show how our cloud-free and grey-cloud retrievals are affected by this test. Retrieved parameter values are listed in Table \ref{table:tight_T_prior_results}. Full retrievals are the same as in Figures \ref{fig:corner_cloudfree_greycloud_HD209458b} to \ref{fig:corner_cloudfree_greycloud_WASP-39b}. Full retrievals show a degeneracy between temperature and \ch{H2O} abundance, and tight prior retrievals exhibit the same behaviour, presenting lower $X_{\ch{H2O}}$ due to the forced high temperatures. In all cases, the temperature posterior is up against the lower limit of the prior, as expected.

When tight priors are used, the cloud-free best-fit spectra in Figures \ref{fig:corner_tight_temp_HD209458b}, \ref{fig:corner_tight_temp_WASP-12b}, and \ref{fig:corner_tight_temp_WASP-39b} (left panels) reveal a unusually high CIA continuum in comparison to the tight prior, grey-cloud fits (right panels). This is likely due to the tight prior retrievals being forced to extremely low \ch{H2O} abundances, to compensate for the higher temperatures, and the lack of additional absorbers in our models. This results in a visibly worse fit to the data, particularly at the reddest wavelength points.

WASP-39b is an extreme case, as the very high $X_{\ch{H2O}}$ median value decreases by about 6 orders of magnitude in the cloud-free tight-prior results, and 3 orders of magnitude in the grey-cloud tight-prior results. In the grey-cloud tight temperature prior retrievals (right panels of Figure \ref{fig:corner_tight_temp_WASP-39b}), the increased scale height from the higher temperature values can be compensated for by a combination of a lower $X_{\ch{H2O}}$, decreasing the mean molecular weight, and a lower $P_{\rm cloud\mbox{-}top}$, bringing up the continuum level to reduce the size of the spectral feature. As for the cloud-free tight-prior retrievals (left panels of Figure \ref{fig:corner_tight_temp_WASP-39b}), a different effect occurs. Besides lower $X_{\ch{H2O}}$ values, the high scale height from the high temperatures are additionally compensated by an increase in the planetary surface gravity, which is not seen in the grey-cloud retrievals.

This tight temperature priors test supplements our previous conclusions: as clouds are an extra free parameter to normalize the spectral continuum, they are necessary to compensate for too high or too low $X_{\ch{H2O}}$ and temperature values, providing a more accurate fit to WFC3 spectra. When clouds are not included, the continuum can be compensated for by a variable, pressure-dependent planetary surface gravity. Whilst this tight temperature prior test demonstrated an alternative to accurately fit WFC3 transmission spectra, a careful consideration of model degeneracies and data limitations is still indispensable. Additionally, these parameters are not enough to cover missing molecular features, accentuating the necessity of a broader wavelength coverage.

Although these tight prior tests offer an alternative fit to the data that is perhaps more physically realistic, their fits remain worse than the cooler-temperature solutions found by the wide prior retrievals. A reduced $\chi^2$ test reveals values consistently closer to 1 for the wide prior retrievals for HD 2095458b and WASP-12b (see Table \ref{tab:chi_sq}), as expected since the higher temperatures were rejected in the wide prior retrievals. For WASP-39b, the reduced $\chi^2$ values reveal that the fit was already poor in the wide prior cases, indicating that our model struggles to fit this dataset. This could be due to missing physics, or additional noise in the data.

Whilst Section \ref{sec:nonisothermal_tests} demonstrated that these low temperatures could be explained by the high altitudes probed by transmission, it could also be due to the quality of the spectra and the data-driven nature of atmospheric retrievals that could favour an incorrect solution. For example, additional noise effects that are unaccounted for in the error bars could lead to biased retrieval results. This serves as another warning for relying on the results from WFC3 spectra without careful consideration of model degeneracies and data limitations.

\section{Comparison with other works}
\label{sec:discussion}

As \ch{H2O} is the main absorber in the WFC3 wavelength range, it is also the most sensitive parameter in our models. Whilst the error bars are wide, several planets do get quite different \ch{H2O} abundances for different models. Moreover, the number of degeneracies regarding water in our models imply the need for additional data to accurately retrieve abundance values, in agreement with previous studies \citep[e.g.][]{BennekeSeager12, Griffith14, LineParmentier16, WelbanksMadhusudhan19, Fairman+24}.

Although the WFC3 range encompasses two water features, it fails to cover additional opacity features in wider spectral ranges, such as molecular and haze signatures, which may result in unreliable retrieved \ch{H2O} abundances \citep{BarstowHeng20}. Figure \ref{fig:H2O_T_literature_models} presents our cloud-free and grey-cloud \ch{H2O} abundance and temperature values found for HD 209458b, WASP-12b, and WASP-39b, along with values retrieved by a number of studies \citep{Barstow+17, FisherHeng18, Tsiaras+18, Wakeford+18, Pinhas+19, WelbanksMadhusudhan19, MacDonaldMadhusudhan17a}.

\begin{figure*}
\centering
\vspace{-0.2cm}
\begin{minipage}{\columnwidth}
 \centering
 \includegraphics[width=\columnwidth]{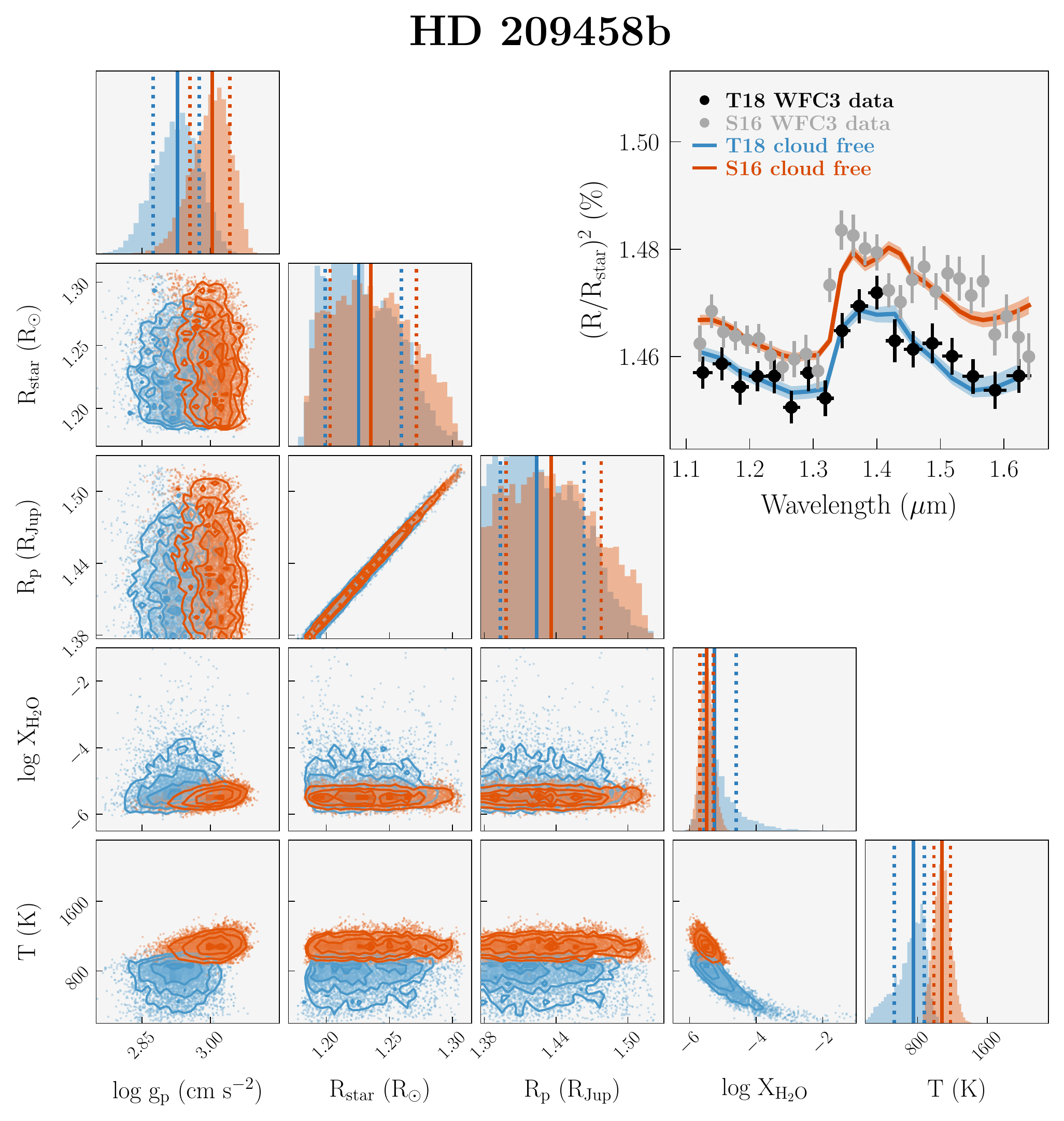}
\end{minipage}
\hfill
\begin{minipage}{\columnwidth}
 \centering
 \includegraphics[width=\columnwidth]{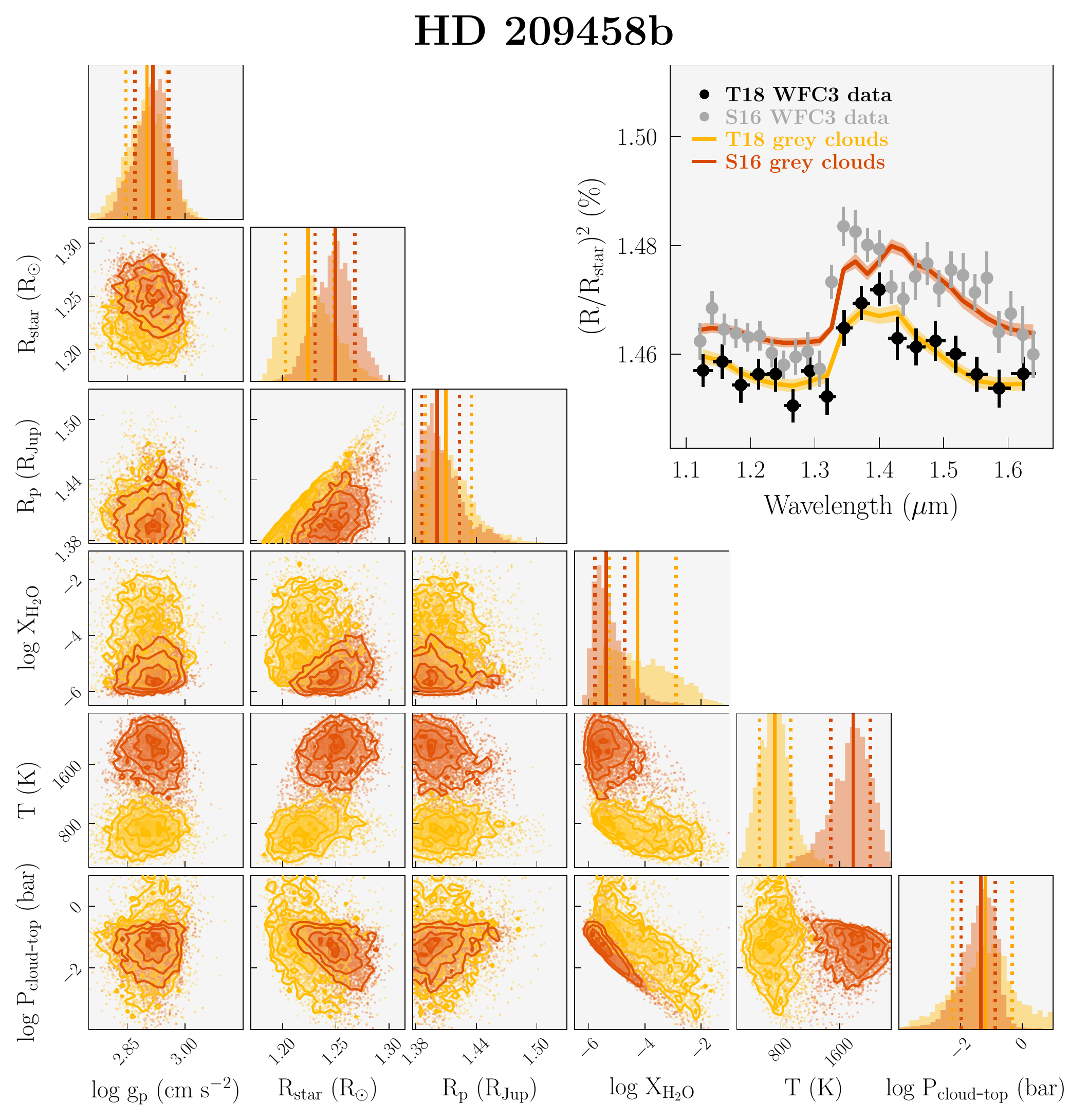}
\end{minipage}
\vspace{0.01cm}
\begin{minipage}{\columnwidth}
 \centering
 \includegraphics[width=\columnwidth]{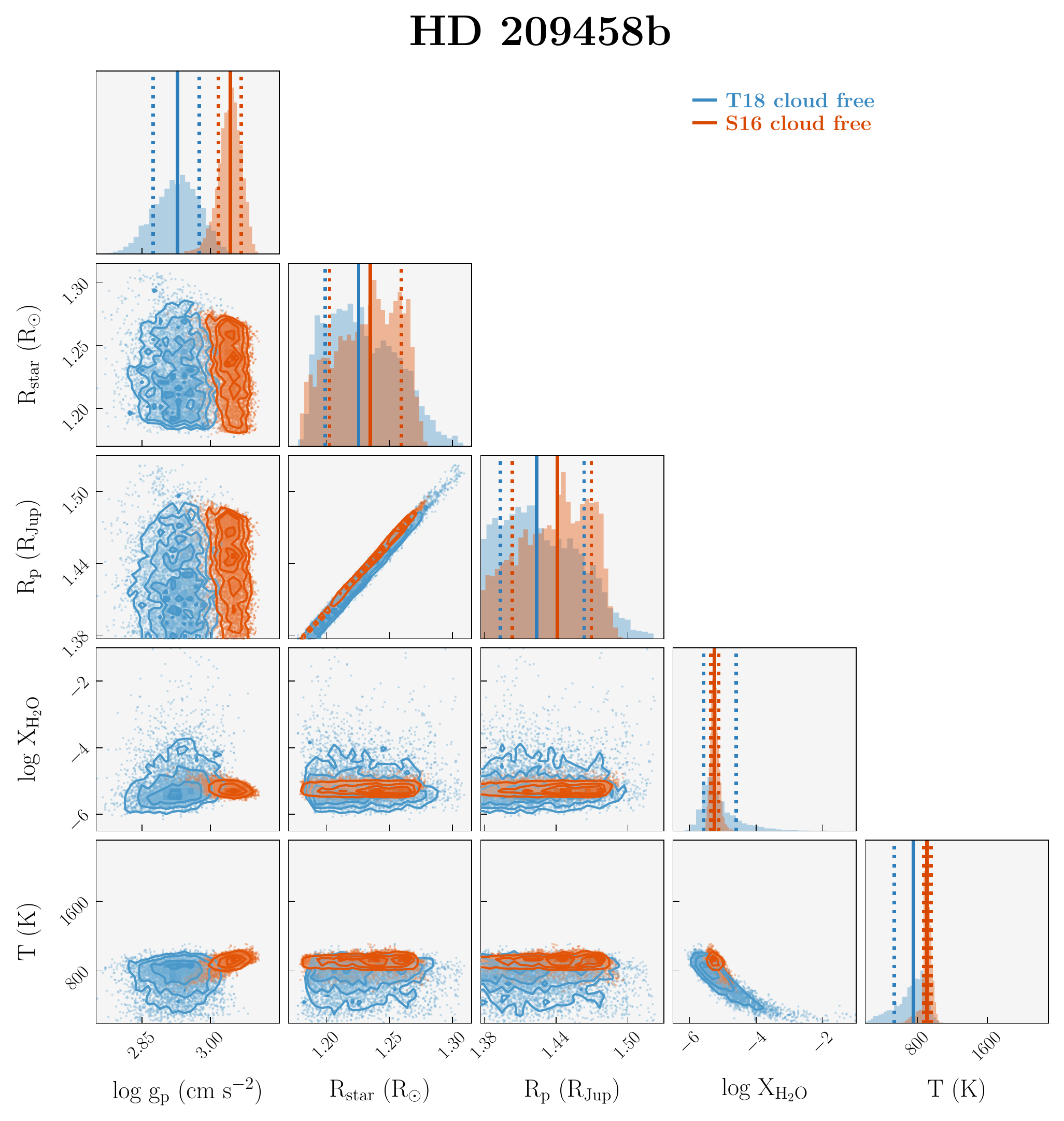}
\end{minipage}
\hfill
\begin{minipage}{\columnwidth}
 \centering
 \includegraphics[width=\columnwidth]{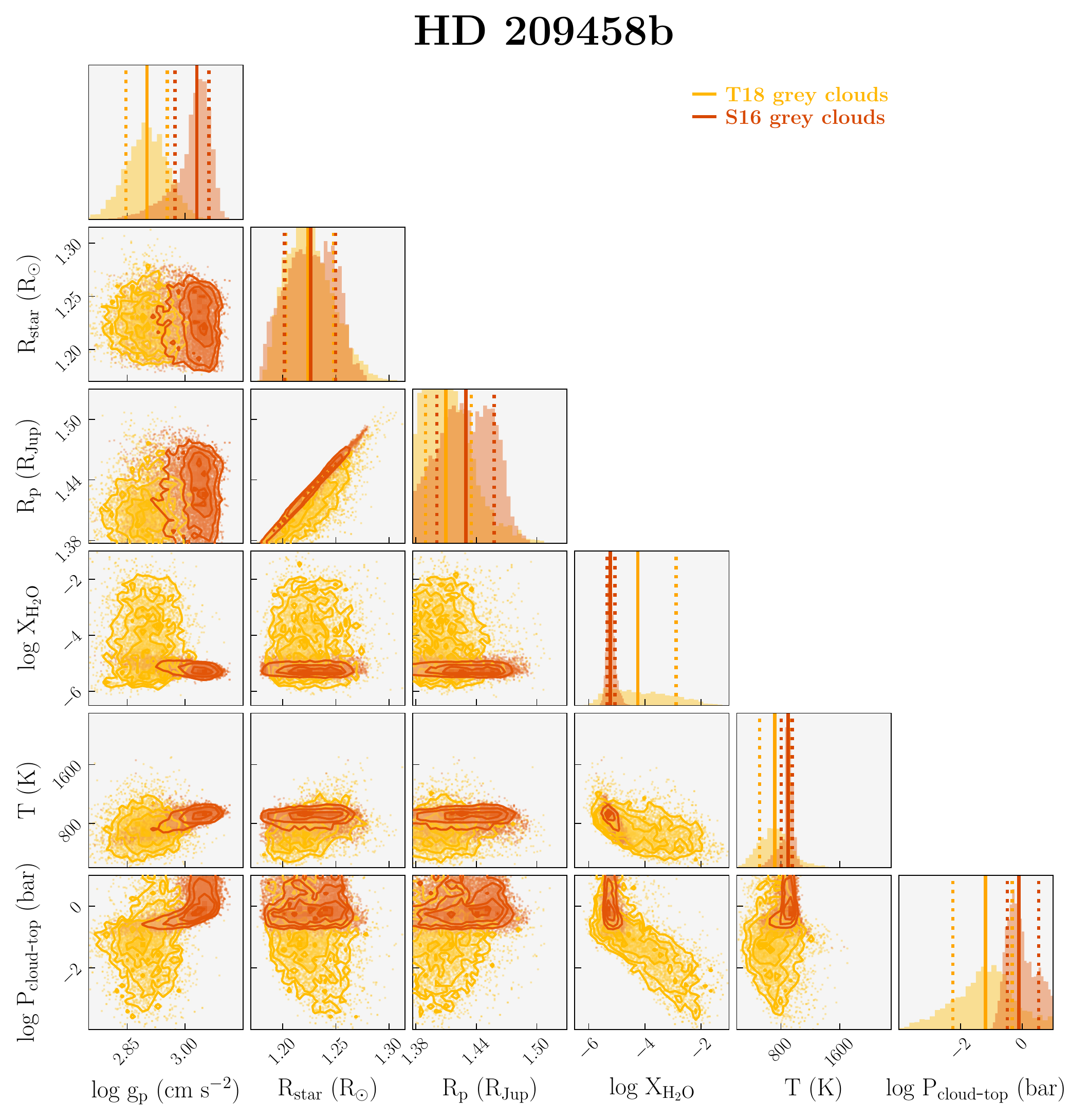}
\end{minipage}
\vspace{-0.1cm}
\begin{minipage}{\columnwidth}
 \centering
 \includegraphics[width=\columnwidth]{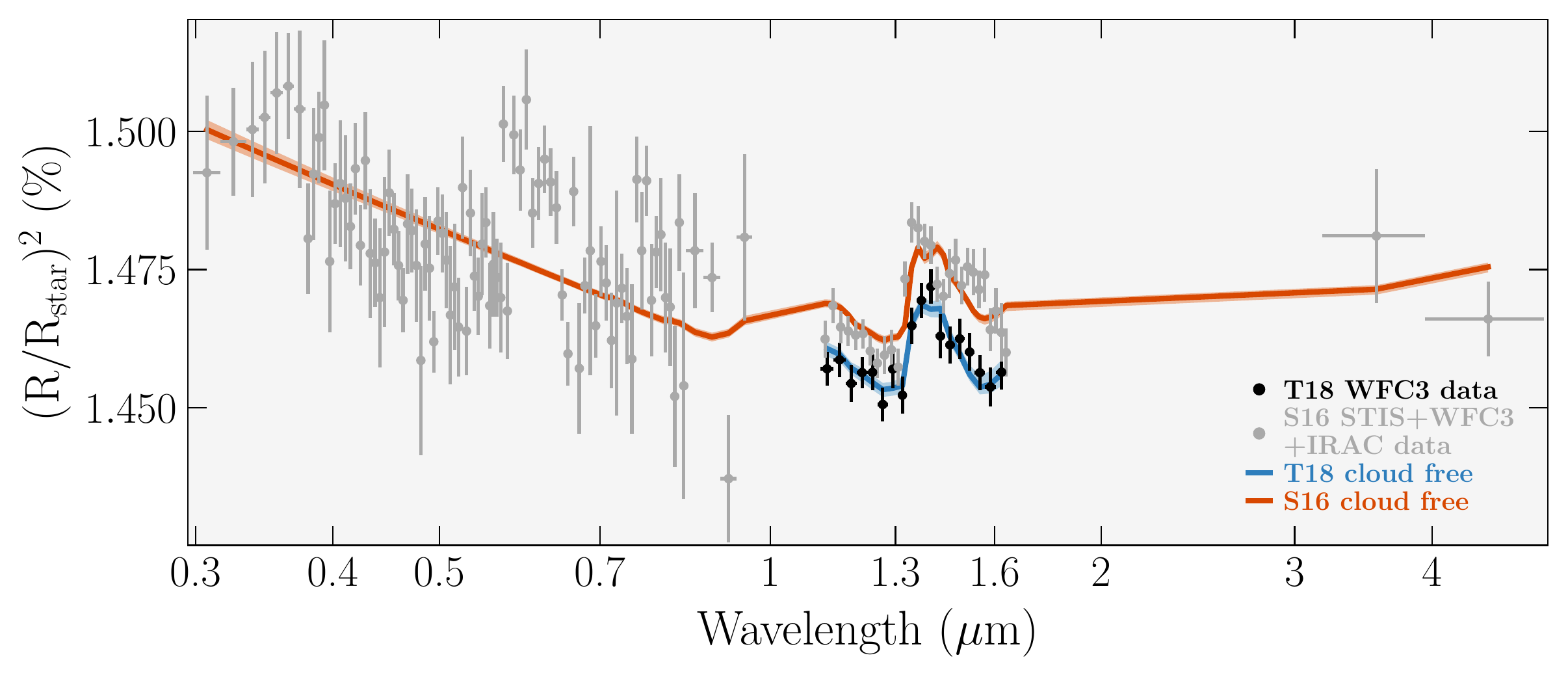}
\end{minipage}
\hfill
\begin{minipage}{\columnwidth}
 \centering
 \includegraphics[width=\columnwidth]{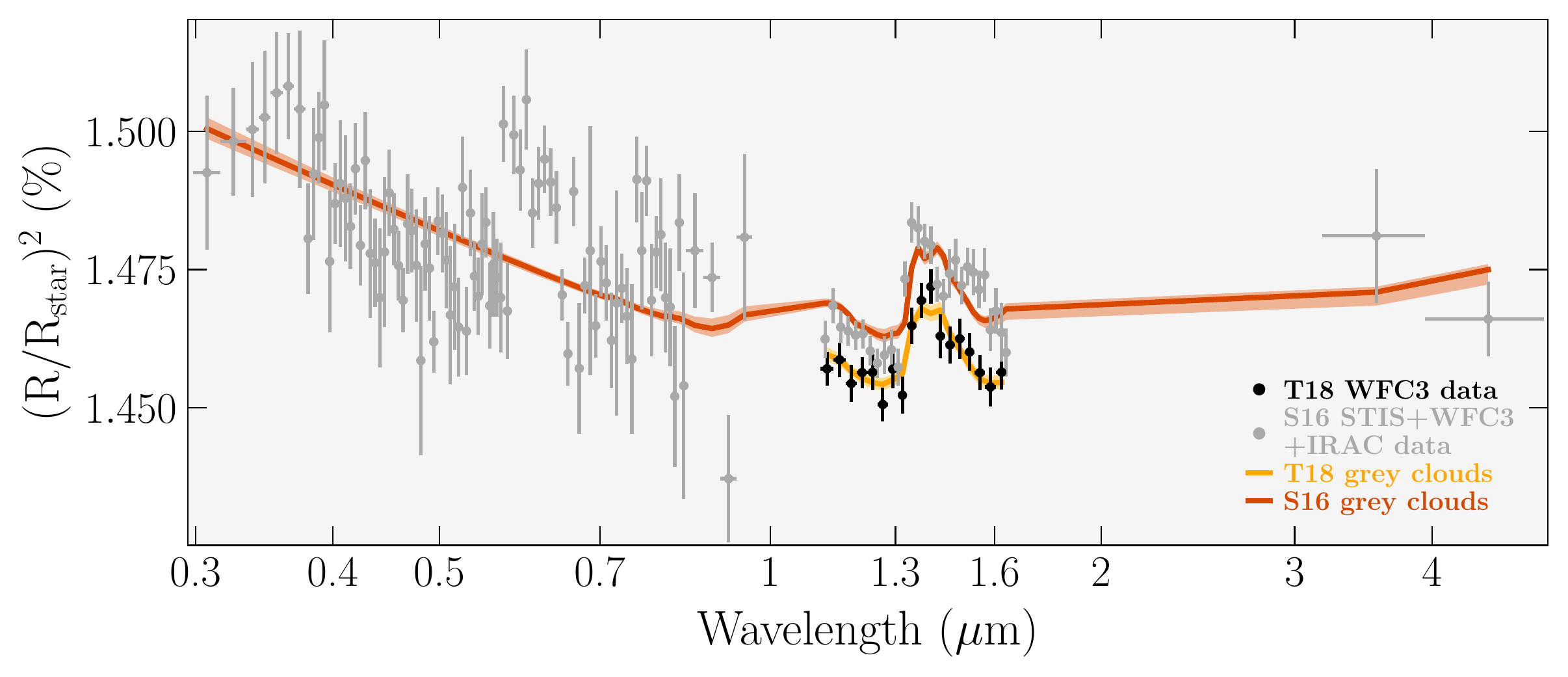}
\end{minipage}
\vspace{-0.1cm}
\caption{HD 209458b full cloud-free (blue, left) and grey-cloud (yellow, right) retrievals using data reduced by \citealt{Tsiaras+18} (T18) vs. cloud-free (orange, left) and grey-cloud (orange, right) retrievals using data reduced by \citealt{Sing+16} (S16). Black points and grey points correspond to data reduced by \citet{Tsiaras+18} and \citet{Sing+16}, respectively. Top corner plots show retrievals using \citet{Tsiaras+18} and \citet{Sing+16} \textit{HST} WFC3 data only, and middle corner plots show retrievals using \citet{Tsiaras+18} \textit{HST} WFC3 and \citet{Sing+16} \textit{HST} STIS, \textit{HST} WFC3, and \textit{Spitzer} IRAC data. Bottom spectra are the best-fit models and their associated 1$\sigma$ uncertainty regions corresponding to the middle corner plots. All spectra are calculated at a resolution of 0.1 cm$^{-1}$ and binned to the data resolution.}
\label{fig:corner_data_reduction_HD209458b}
\end{figure*}

\subsection{Dependence on model properties}

Retrieved \ch{H2O} abundances in the literature may vary in a range between $-6 < \log X_{\ch{H2O}} < -2$ (including uncertainties) for HD 209458b and for WASP-12b. This range encompasses the values found by our current work. Moreover, retrieved temperatures also diverge between different studies, and are generally not consistent with their associated equilibrium temperatures ($1449 \pm 36$ K, $2580 \pm 146$ K, and $1119 \pm 57$ K for HD 209458b, WASP-12b, and WASP-39b, respectively; \citealt{Fu+17}). In order to fairly compare these values, individual assumptions of each work have to be taken into account. For HD 209458b and WASP-12b data, \citet{FisherHeng18} apply an isobaric approximation which has little to no effect on the CIA, resulting in higher \ch{H2O} abundances. \citet{Tsiaras+18} retrieve values from WFC3 spectra assuming an isothermal temperature–pressure profile that includes CIA and Rayleigh scattering, in agreement with our analyses. However, their cloud parameterization is notably different from ours, which may cause discrepancies in retrieved \ch{H2O} abundances \citep{Barstow20}. \citet{Pinhas+19} consider a non-isothermal temperature-pressure profile, and their temperature values in Figure \ref{fig:H2O_T_literature_models} correspond to the temperature at the top of the atmosphere (minimum pressure). Additionally, \citet{Pinhas+19} also include inhomogeneous clouds. Some of these different modelling assumptions could explain the discrepancies in our results for WASP-39b (Figure \ref{fig:H2O_T_literature_models}), though the results for HD 209458b and WASP-12b are consistent within the error bars. \citet{Barstow+17} use a fixed temperature profile for their models, based on assumptions about the planets' Bond albedos. Hence we do not include their temperatures in Figure \ref{fig:H2O_T_literature_models}.

Studies which focused on retrievals for an individual planet rather than a population study (e.g. \citealt{WelbanksMadhusudhan19} and \citealt{MacDonaldMadhusudhan17a} for HD 209458b, and \citealt{Wakeford+18} for WASP-39b) were also represented in Figure \ref{fig:H2O_T_literature_models}. \citet{WelbanksMadhusudhan19} and \citet{MacDonaldMadhusudhan17a} include a non-isothermal $T$-$P$ profile and inhomogeneous cloud coverage, as well as optical data from \citealt{Sing+16}, resulting in narrower uncertainty limits. Yet, our results for log $X_{\ch{H2O}}$ remain consistent with theirs within the error bars. For HD 209458b, \citealt{MacDonaldMadhusudhan17b} find tentative evidence for \ch{NH3} in the spectra, which could also explain some differences in our results.

\subsection{Dependence on data reduction}
\label{sec:data_reduction}

Some of the differences in Figure \ref{fig:H2O_T_literature_models} can be explained by the use of different datasets or different data reductions. We highlight the case of WASP-39b, for which our models present abnormally high \ch{H2O} values due to the limited range of WFC3, which only probes water features. \citet{Wakeford+18} also find an elevated \ch{H2O} abundance for WASP-39b using WFC3 data combined with data from \textit{HST} STIS, VLT FORS2, and \textit{Spitzer} IRAC, which shows these elevated abundance values are not exclusive to our models. On the other hand, \citet{Pinhas+19} find lower \ch{H2O} abundances while also considering spectral data from \textit{HST} STIS, \textit{HST} WFC3 (using different data reductions than \citealt{Wakeford+18}), and \textit{Spitzer} IRAC. The wide scatter in \ch{H2O} abundance across the different studies for WASP-39b demonstrates the necessity of wider wavelength coverage, from individual or multiple instruments, in order to robustly constrain molecular abundances. Further studies using \textit{JWST} spectra with data from 0.5 $\upmu$m up to 5 $\upmu$m \citep[e.g.][]{Ahrer+23, Alderson+23, Feinstein+23, Rustamkulov+23, Lueber+24} are able to infer the presence of clouds, and constrain abundances of \ch{H2O}, \ch{CO2}, \ch{SO2}, and possibly other species with features in the infrared region. However, a recent retrieval study using \textit{JWST} spectra still retrieves high \ch{H2O} abundance values from NIRISS data, where \ch{H2O} is the main opacity source \citep{Lueber+24}.

Finally, we briefly explore different data reductions and how they can affect our retrieval outcomes. We perform retrievals for HD 209458b using the two cited data reductions: by \citet[][used by e.g. \citealt{Barstow+17}, \citealt{Pinhas+19}, \citealt{WelbanksMadhusudhan19}, and \citealt{MacDonaldMadhusudhan17a}]{Sing+16}, and by \citet[][used by e.g. \citealt{FisherHeng18} and this work]{Tsiaras+18}. Since \citet{Sing+16} additionally include \textit{HST} STIS and \textit{Spitzer} IRAC data aside from WFC3 data, we chose to run separate retrievals using only WFC3 and using STIS+WFC3+IRAC combined data. These data sets were tested using the same cloud-free and grey-cloud models presented throughout this work. Retrieval outcomes from the \citet{Sing+16} data are presented together with the \citet{Tsiaras+18} data in Figure \ref{fig:corner_data_reduction_HD209458b} and Table \ref{table:data_reduction_comparison_results}. By comparing the \citet{Tsiaras+18} and \citet{Sing+16} WFC3 data, it becomes clear that the data have a significant vertical offset in this wavelength range, with the \citet{Sing+16} data presenting higher transit depths in comparison to \citet{Tsiaras+18}. To first order, one would expect the retrieval to adjust the reference transit radius $R_{\rm p}$ in order to compensate for the spectral offset. However, the WFC3 spectra also present different shapes at wavelengths above 1.3 $\upmu$m, changing the size and shape of the 1.4 $\upmu$m \ch{H2O} feature. The consequence is a higher scale height for the \citet{Sing+16} spectrum, which leads to higher temperatures in the retrieval of their data. Yet, in the cloud-free case both temperatures retrieved from the \citet{Sing+16} and \citet{Tsiaras+18} data are still low compared to the expected $T_{\rm eq}$ of 1450 K \citep{Fu+17}. Interestingly, for the grey cloud retrieval of the \citet{Sing+16} data, the temperature posterior now encompasses the equilibrium value, suggesting that the data reductions themselves could be an explanation for the low retrieved temperatures in our results. The retrieved abundance of \ch{H2O} is more tightly constrained to the lower values for the \citet{Sing+16} data, especially in the retrieval including grey clouds. The \ch{H2O} abundances retrieved from the \citet{Sing+16} data, presented in Figure \ref{fig:corner_data_reduction_HD209458b} and Table \ref{table:data_reduction_comparison_results}, are fully consistent with results from other studies that use the same data, as exhibited in Figure \ref{fig:H2O_T_literature_models}.

When STIS and IRAC data are included, we verify the wider wavelength coverage is able to provide sharper posteriors for \ch{H2O} abundance and temperature, consequently breaking degeneracies between $X_{\ch{H2O}}$, $T$, and $P_{\rm cloud\mbox{-}top}$, in agreement with \citet{Pinhas+19} and \citet{Fairman+24}. We note that our models include \ch{H2O} as the only species besides background gases \ch{H2} and He, therefore no additional molecular features are seen in the retrieved spectra. The grey cloud retrieval on the combined data displays a different behaviour to the WFC3-only case. Now, the cloud-top pressure is constrained to deep in the atmosphere, indicating a lack of cloud coverage, likely due to the strong Rayleigh slope provided by the STIS data, or to the non-inclusion of optical species such as Na and K. This then results in a similar temperature posterior as the cloud-free case, as expected, tightly constrained around cooler values. Our retrieved \ch{H2O} abundances are consistent with other studies that use \citet{Sing+16} data (see Figure \ref{fig:H2O_T_literature_models}), and with results for Sing WFC3 data only. Overall, the agreement between independent results using spectra reduced by \citet{Sing+16} indicates chemical abundances and temperature posterior constraints are dependent on the data reduction.

It is evident that results are extremely sensitive to each step of the characterisation, from data reduction and calibration to retrieval model assumptions (e.g. if they include Rayleigh scattering, CIA, non-isothermal profile, which chemical species are considered, and how clouds are parameterized). This means even small changes in any step of the procedure may result in inconsistencies. Furthermore, we acknowledge that retrievals using data from WFC3 alone may be unreliable for some planets, highlighting the need to cover a wider wavelength range to better constrain chemical composition and cloud features \citep{BarstowHeng20, Fairman+24}. The \textit{JWST} follow-up of planets previously observed with WFC3 will bring to light the reliability of the constraints from these measurements.

\section{Summary}
\label{sec:summary}

Our main findings include:

\begin{itemize}

\item A non-isobaric treatment is important in establishing chemical abundances in retrievals from \textit{HST} WFC3 transmission data due to the effects of collision-induced absorption (CIA), corroborating the conclusion of \citet{WelbanksMadhusudhan19};

\item The inclusion of the stellar radius and the planetary surface gravity as free parameters in the retrievals does not have a strong effect on the other retrieved parameters, with the exception of planetary radius, as expected;

\item The large number of degeneracies present in WFC3 retrievals can make accurate constraints a challenge. Strong degeneracies exist between \ch{H2O} abundance, temperature, and cloud-top pressure, as well as an expected degeneracy between stellar and planetary radii. The limited wavelength coverage of WFC3 makes it difficult to measure the atmospheric scale height, which in turn affects the retrieved parameters;

\item Retrievals of WFC3 data often obtain temperatures substantially lower than the equilibrium values. This has a number of possible explanations, such as 3D effects \citep{MacDonald+20}, the above-mentioned challenges in measuring the scale height, or the narrow and high-altitude pressure region probed by transmission. These low temperatures could lead to unreliable constraints on chemical abundances. Yet, non-isothermal retrievals from the same data also result in cool temperature profiles.

\item Forcing the temperature priors to higher values can lead to lower \ch{H2O} abundances. This may be compensated by the inclusion of clouds, or by a pressure-dependent planetary surface gravity when clouds are not included. However, in this case the temperature posterior hits the lower limit of the prior, and the fit to the data is generally worse. It is possible that the cause of some of these low temperatures could be the data quality, since the low-resolution of WFC3 makes the retrievals highly sensitive to shifts in the data points.

\item Differences in data reductions and the inclusion of wider-wavelength data can cause major differences in the retrieval results. This indicates that singular retrievals on WFC3 data alone are unlikely to be reliable.

\end{itemize}

In conclusion, our study advocates for the importance of wide wavelength coverage for accurately characterising the atmospheres of even the most observable exoplanetary atmospheres. However, we cannot rely on \textit{JWST} alone. \citet{Fairman+24} already demonstrated the importance of optical data for constraining atmospheric cloud properties, which in turn will affect spectra in the wavelengths covered by \textit{JWST}. We also stress the importance of understanding the many degeneracies that exist in retrievals. Additionally, \citet{Nixon+24} discuss the need for accounting for model uncertainty when reporting constraints, which has a further effect of increasing abundance uncertainties. By combining data from multiple instruments, and fully understanding the limitations of our atmospheric retrievals, we can hope to build the most accurate picture of an exoplanet's atmosphere.

\section*{Acknowledgements}

The authors thank the anonymous referee for their helpful insights. We thank Jens Hoeijmakers and Bibiana Prinoth for setting up the GPU at Lund University, allowing us to run \texttt{BeAR}. We thank Johannes Buchner for the Nested Sampling Monte Carlo library. We thank Anna Lueber for useful feedback. A.N. acknowledges financial support from The Fund of the Walter Gyllenberg Foundation. C.F. acknowledges financial support from the Swiss National Science Foundation (SNSF) Mobility Fellowship under grant no. P500PT\_203110 and the European Research Council (ERC) under the European Union’s Horizon 2020 research and innovation program under grant agreement no 805445. L.G. acknowledges financial support from Fundação Carlos Chagas Filho de Amparo à Pesquisa do Estado do Rio de Janeiro (FAPERJ), through the ARC research grant E-26/211.386/2019. B.T. acknowledges the financial support from the Wenner-Gren Foundation (WGF2022-0041). K.H. acknowledges partial funding support from an ERC Consolidator grant (EXOKLEIN) and an ERC Synergy grant (Geoastronomy).

\section*{Data Availability}

The open-source atmospheric code \texttt{BeAR} \citep{Kitzmann+20} is publicly available at \hyperlink{https://github.com/newstrangeworlds/bear}{https://github.com/newstrangeworlds/bear}. The open-source nested-sampling package \textsc{pymultinest} \citep{Buchner+14} is available at \hyperlink{https://johannesbuchner.github.io/PyMultiNest/}{https://johannesbuchner.github.io/PyMultiNest/}. Retrieval outcomes for the 42 \textit{HST} WFC3 spectra using \texttt{BeAR} are available online at \hyperlink{https://doi.org/10.5281/zenodo.14982772}{https://doi.org/10.5281/zenodo.14982772}.

\bibliographystyle{mnras}
\bibliography{main}

\clearpage

\appendix

\section{Supplementary Results}
\label{appendix:supplementary_results}

The following figures and tables display the retrieval results using \texttt{BeAR} cloud free (blue), grey cloud (yellow), non-grey cloud (red), and flat line (grey) models for each of the 38 \textit{HST} WFC3 transmission spectra examined in this work. All retrievals were calculated using a spectral resolution of 0.1 cm$^{-1}$.

Table \ref{table:input_parameters} lists the input values used in our retrievals. Figures \ref{fig:bayesian_comparison_all1} and \ref{fig:bayesian_comparison_all2} show the Bayesian comparison between the four (cloud free, grey clouds, non-grey clouds, and flat line) models for all objects. Tables \ref{table:retrieval_results_all1} and \ref{table:retrieval_results_all2} present the values for the free parameters retrieved by the cloud free, grey cloud, and non-grey cloud models for all objects. Tables \ref{table:nonisothermal_results} and \ref{table:tight_T_prior_results} display the values for the free parameters retrieved by the cloud free, grey cloud, and non-grey cloud models for our case studies: HD 209458b, WASP-12b, and WASP-39b using non-isothermal models and tight temperature priors, respectively. Table \ref{table:data_reduction_comparison_results} shows values for the free parameters retrieved by the cloud free and grey cloud models using data reduced by \citet{Sing+16}.

\begin{table*}
\caption{Input parameters used in our retrievals. References for $R_{\rm p}$, $R_{\rm star}$, and log $g_{\rm p}$ are listed in the ``Reference'' column. $T_{\rm eq}$ for all objects are from \citet{Fu+17}.}
\label{table:input_parameters}
\resizebox{\textwidth}{!}{
\begin{tabular}{lccccc}
\hline
Planet & $T_{\rm eq}$ (K) & $R_{\rm p}$ (R$_{\rm Jup}$) & $R_{\rm star}$ (R$_\odot$) & log $g_{\rm p}$ (cm s$^{-2}$) & Reference \\ \hline
GJ 436b & $633 \pm 58$ & $0.369 \pm 0.015$ & $0.455 \pm 0.018$ & $3.120 \pm 0.030$ & \citet{vonBraun+12} \\
GJ 1214b & $573 \pm 35$ & $0.243 \pm 0.021$ & $0.211 \pm 0.011$ & $2.885 \pm 0.067$ & \citet{AngladaEscude+13} \\
GJ 3470b & $692 \pm 101$ & $0.346 \pm 0.029$ & $0.480 \pm 0.040$ & $2.830 \pm 0.110$ & \citet{Biddle+14} \\
HAT-P-1b & $1320 \pm 103$ & $1.225 \pm 0.059$ & $1.1150 \pm 0.050$ & $2.873 \pm 0.010$ & \citet{Johnson+08, Nikolov+14} \\
HAT-P-3b & $1127 \pm 68$ & $0.827 \pm 0.055$ & $0.799 \pm 0.039$ & $3.330 \pm 0.058$ & \citet{Chan+11} \\
HAT-P-11b & $856 \pm 37$ & $0.422 \pm 0.014$ & $0.750 \pm 0.020$ & $3.050 \pm 0.060$ & \citet{Bakos+10} \\
HAT-P-12b & $958 \pm 28$ & $0.959^{+0.029}_{-0.021}$ & $0.701^{+0.017}_{-0.012}$ & $2.750 \pm 0.030$ & \citet{hartman+09} \\
HAT-P-17b & $780 \pm 34$ & $1.010 \pm 0.029$ & $0.838 \pm 0.021$ & $3.110 \pm 0.020$ & \citet{Howard+12} \\
HAT-P-18b & $843 \pm 35$ & $0.947 \pm 0.044$ & $0.717 \pm 0.026$ & $2.734 \pm 0.044$ & \citet{Esposito+14} \\
HAT-P-26b & $980 \pm 56$ & $0.565^{+0.072}_{-0.032}$ & $0.788^{+0.098}_{-0.043}$ & $2.650 \pm 0.090$ & \citet{Hartman+11a} \\
HAT-P-32b & $1784 \pm 58$ & $1.789 \pm 0.025$ & $1.219 \pm 0.016$ & $2.820 \pm 0.080$ & \citet{Hartman+11b} \\
HAT-P-38b & $1080 \pm 78$ & $0.825^{+0.092}_{-0.063}$ & $0.923^{+0.096}_{-0.067}$ & $2.990 \pm 0.080$ & \citet{Sato+12} \\
HAT-P-41b & $1937 \pm 74$ & $1.685^{+0.076}_{-0.051}$ & $1.683^{+0.058}_{-0.036}$ & $2.840 \pm 0.060$ & \citet{Hartman+12} \\
HD 97658b & $753 \pm 33$ & $0.200^{+0.009}_{-0.008}$ & $0.741^{+0.024}_{-0.023}$ & $3.166 \pm 0.060$ & \citet{VanGrootel+14} \\
HD 149026b & $1627 \pm 83$ & $0.654^{+0.060}_{-0.045}$ & $1.368^{+0.12}_{-0.083}$ & $3.360 \pm 0.066$ & \citet{Torres+08} \\
HD 189733b & $1201 \pm 51$ & $1.216 \pm 0.024$ & $0.805 \pm 0.016$ & $3.290 \pm 0.020$ & \citet{Boyajian+15} \\
HD 209458b & $1449 \pm 36$ & $1.451 \pm 0.074$ & $1.203 \pm 0.061$ & $2.880 \pm 0.070$ & \citet{Boyajian+15} \\
WASP-12b & $2580 \pm 146$ & $1.790 \pm 0.090$ & $1.570 \pm 0.070$ & $2.990 \pm 0.030$ & \citet{Hebb+09} \\
WASP-17b & $1632 \pm 126$ & $1.932 \pm 0.053$ & $1.583 \pm 0.041$ & $2.500 \pm 0.027$ & \citet{Southworth+12} \\
WASP-19b & $2037 \pm 156$ & $1.395 \pm 0.023$ & $1.004 \pm 0.016$ & $3.152 \pm 0.008$ & \citet{Tregloan-Reed+13} \\
WASP-29b & $963 \pm 69$ & $0.792^{+0.056}_{-0.035}$ & $0.808 \pm 0.044$ & $2.950 \pm 0.050$ & \citet{Hellier+10} \\
WASP-31b & $1576 \pm 58$ & $1.549 \pm 0.050$ & $1.252 \pm 0.033$ & $2.659 \pm 0.036$ & \citet{Anderson+11} \\
WASP-39b & $1119 \pm 57$ & $1.270 \pm 0.040$ & $0.918^{+0.022}_{-0.019}$ & $2.617 \pm 0.065$ & \citet{Maciejewski+16} \\
WASP-43b & $1374 \pm 147$ & $1.060 \pm 0.050$ & $0.670 \pm 0.040$ & $ 3.672 \pm 0.070$ & \citet{Hellier+11} \\
WASP-52b & $1300 \pm 115$ & $1.270 \pm 0.030$ & $0.790 \pm 0.020$ & $2.810 \pm 0.030$ & \citet{Hebrard+13} \\
WASP-63b & $1508 \pm 69$ & $1.430^{+0.100}_{-0.060}$ & $1.880^{+0.100}_{-0.060}$ & $2.620 \pm 0.050$ & \citet{Hellier+12} \\
WASP-67b & $1026 \pm 59$ & $1.400^{+0.300}_{-0.200}$ & $0.870 \pm 0.040$ & $2.700 \pm 0.015$ & \citet{Hellier+12} \\
WASP-69b & $964 \pm 38$ & $1.057 \pm 0.047$ & $0.813 \pm 0.028$ & $2.726 \pm 0.046$ & \citet{Anderson+14} \\
WASP-74b & $1915 \pm 116$ & $1.560 \pm 0.060$ & $1.640 \pm 0.050$ & $2.950 \pm 0.020$ & \citet{Hellier+15} \\
WASP-76b & $2206 \pm 95$ & $1.830^{+0.040}_{-0.060}$ & $1.730 \pm 0.040$ & $2.800 \pm 0.020$ & \citet{West+16} \\
WASP-80b & $824 \pm 58$ & $0.999^{+0.030}_{-0.031}$ & $0.586^{+0.017}_{-0.018}$ & $3.145 \pm 0.015$ & \citet{Triaud+15} \\
WASP-101b & $1552 \pm 81$ & $1.410 \pm 0.050$ & $1.290 \pm 0.040$ & $2.760 \pm 0.040$ & \citet{Hellier+14} \\
WASP-121b & $2358 \pm 122$ & $1.865 \pm 0.044$ & $1.458 \pm 0.030$ & $2.973 \pm 0.017$ & \citet{Delrez+16} \\
XO-1b & $1196 \pm 60$ & $1.206^{+0.047}_{-0.042}$ & $0.934^{+0.037}_{-0.032}$ & $3.211 \pm 0.040$ & \citet{Torres+08} \\
TRAPPIST-1d (\ch{N2}) & $288$ & $0.070 \pm 0.002$ & $0.121 \pm 0.003$ & $2.676 \pm 0.045$ & \citet{VanGrootel+18, Grimm+18} \\
TRAPPIST-1d (\ch{H2}) & $288$ & $0.070 \pm 0.002$ & $0.121 \pm 0.003$ & $2.676 \pm 0.045$ & \citet{VanGrootel+18, Grimm+18} \\
TRAPPIST-1e (\ch{N2}) & $251$ & $0.081 \pm 0.002$ & $0.121 \pm 0.003$ & $2.960 \pm 0.031$ & \citet{VanGrootel+18, Grimm+18} \\
TRAPPIST-1e (\ch{H2}) & $251$ & $0.081 \pm 0.002$ & $0.121 \pm 0.003$ & $2.960 \pm 0.031$ & \citet{VanGrootel+18, Grimm+18} \\
TRAPPIST-1f (\ch{N2}) & $219$ & $0.093 \pm 0.003$ & $0.121 \pm 0.003$ & $2.923 \pm 0.020$ & \citet{VanGrootel+18, Grimm+18} \\
TRAPPIST-1f (\ch{H2}) & $219$ & $0.102 \pm 0.003$ & $0.121 \pm 0.003$ & $2.923 \pm 0.020$ & \citet{VanGrootel+18, Grimm+18} \\
TRAPPIST-1g (\ch{N2}) & $199$ & $0.102 \pm 0.003$ & $0.121 \pm 0.003$ & $2.931 \pm 0.020$ & \citet{VanGrootel+18, Grimm+18} \\
TRAPPIST-1g (\ch{H2}) & $199$ & $0.102 \pm 0.003$ & $0.121 \pm 0.003$ & $2.931 \pm 0.020$ & \citet{VanGrootel+18, Grimm+18} \\ \hline
\end{tabular}
}
\end{table*}

\begin{figure*}
\centering
\includegraphics[width=\textwidth]{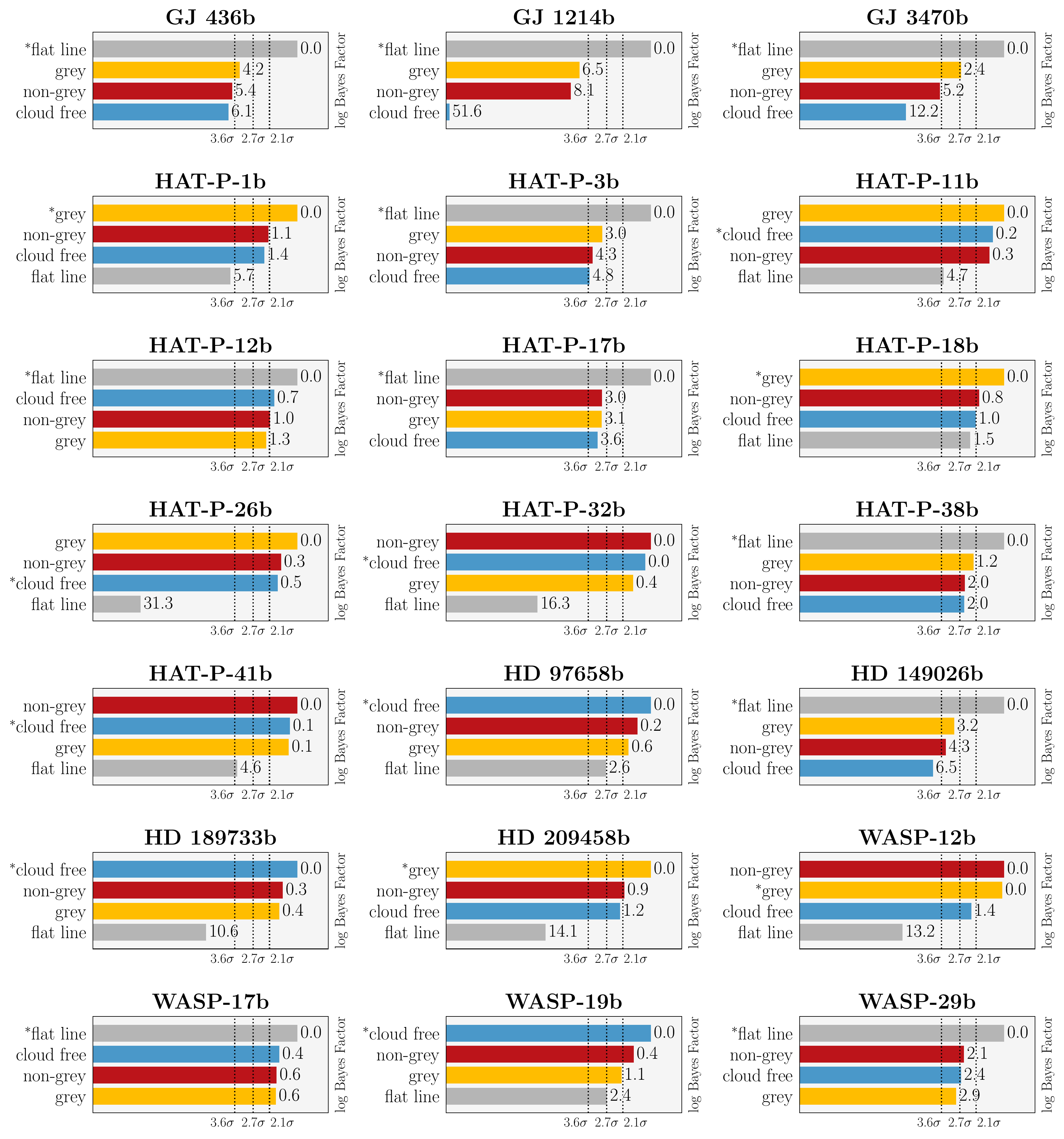}
\vspace{-0.5cm}
\caption{Bayesian comparison between cloud-free (blue), grey-cloud (yellow), non-grey cloud (red), and flat-line (grey) models. For each object, models are ordered from highest (top) to lowest (bottom) Bayesian evidence, and consequently from most favoured (highest evidence) to least favoured in comparison to the highest evidence model. Values on the right side of each bar show the corresponding natural logarithm of the Bayes factor relative to the highest-evidence model. Vertical dotted lines show significances of 3.6$\sigma$, 2.7$\sigma$, and 2.1$\sigma$, corresponding to $\log$ Bayes factors of 5.0, 2.5, and 1.0, considered ``strong'', ``moderate'', and ``weak'' evidences compared to the highest-evidence model, according to \citet{Trotta08}. Therefore, the simplest models with $\log$ Bayes factor < 1 are considered favoured, and are denoted by *. All runs are calculated at a spectral resolution of 0.1 cm$^{-1}$.}
\label{fig:bayesian_comparison_all1}
\end{figure*}

\begin{figure*}
\centering
\includegraphics[width=\textwidth]{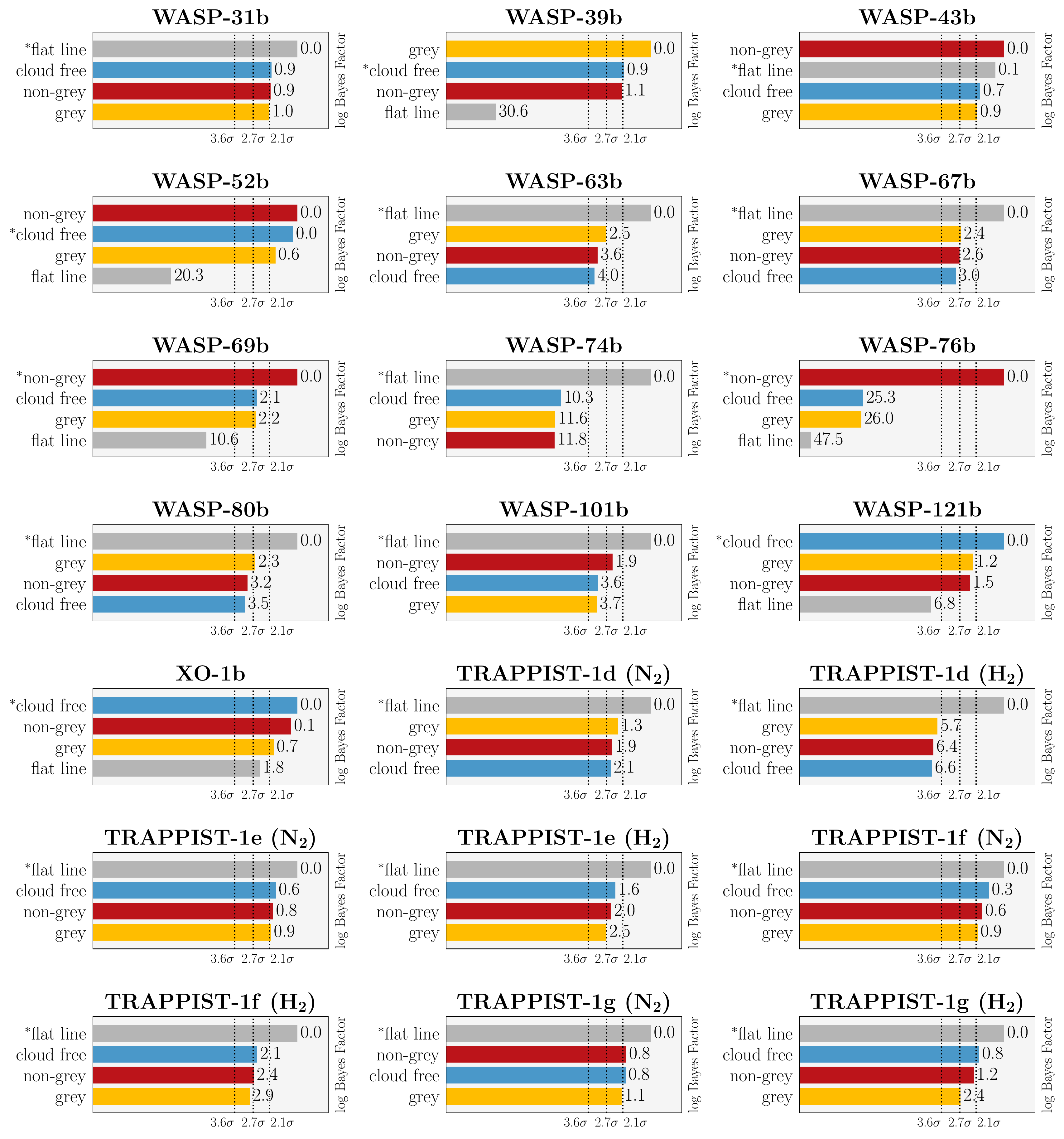}
\vspace{-0.5cm}
\caption{Continuation of Figure \ref{fig:bayesian_comparison_all1}.}
\label{fig:bayesian_comparison_all2}
\end{figure*}

\begin{table*}
\centering
\caption{Free parameters retrieved by our \texttt{BeAR} cloud free, grey cloud, and non-grey cloud models, for all objects. If the best-fit model for a planet is not a flat line, the model is denoted by * (see Figures \ref{fig:bayesian_comparison_all1} and \ref{fig:bayesian_comparison_all2}). Values correspond to the median of the posterior distribution, and uncertainties correspond to the 16th and 84th quantiles of the distribution (i.e. $\pm$1$\sigma$).}
\label{table:retrieval_results_all1}
\resizebox{0.7\textwidth}{!}{
\begin{tabular}{lccccccc}
\hline
Planet & Model & log $g_{\rm p}$ (cm s$^{-2}$) & $R_{\rm star}$ (R$_\odot$) & $R_{\rm p}$ (R$_{\rm Jup}$) & log $X_{\ch{H2O}}$ & $T$ (K) & log $P_{\rm cloud\mbox{-}top}$ (bar) \\
\hline
\multirow{3}{*}{GJ 436b} & cloud free & $3.13^{+0.03}_{-0.03}$ & $0.459^{+0.011}_{-0.010}$ & $0.369^{+0.009}_{-0.008}$ & $-6.84^{+2.59}_{-0.70}$ & $285^{+62}_{-167}$ & $-$ \\
& grey clouds & $3.12^{+0.03}_{-0.02}$ & $0.458^{+0.008}_{-0.011}$ & $0.363^{+0.005}_{-0.007}$ & $-8.42^{+2.16}_{-2.24}$ & $397^{+124}_{-211}$ & $-0.82^{+1.62}_{-0.58}$ \\
& non-grey clouds & $3.13^{+0.02}_{-0.02}$ & $0.460^{+0.009}_{-0.009}$ & $0.368^{+0.007}_{-0.007}$ & $-7.45^{+2.48}_{-1.19}$ & $343^{+98}_{-183}$ & $-0.82^{+1.31}_{-0.65}$ \\ \hline
\multirow{3}{*}{GJ 1214b} & cloud free & $3.29^{+0.03}_{-0.03}$ & $0.226^{+0.009}_{-0.004}$ & $0.255^{+0.010}_{-0.005}$ & $-5.86^{+0.08}_{-0.28}$ & $209^{+7}_{-5}$ & $-$ \\
& grey clouds & $2.91^{+0.06}_{-0.05}$ & $0.215^{+0.005}_{-0.006}$ & $0.229^{+0.004}_{-0.006}$ & $-7.14^{+2.87}_{-2.00}$ & $286^{+55}_{-92}$ & $-1.87^{+1.09}_{-0.58}$ \\
& non-grey clouds & $2.91^{+0.04}_{-0.05}$ & $0.216^{+0.006}_{-0.005}$ & $0.230^{+0.005}_{-0.007}$ & $-7.43^{+2.35}_{-1.68}$ & $301^{+62}_{-101}$ & $-1.62^{+0.86}_{-0.44}$ \\ \hline
\multirow{3}{*}{GJ 3470b} & cloud free & $3.10^{+0.06}_{-0.05}$ & $0.489^{+0.020}_{-0.014}$ & $0.361^{+0.015}_{-0.010}$ & $-5.92^{+0.25}_{-1.40}$ & $811^{+554}_{-107}$ & $-$ \\
& grey clouds & $2.88^{+0.09}_{-0.07}$ & $0.492^{+0.021}_{-0.020}$ & $0.342^{+0.014}_{-0.018}$ & $-4.49^{+1.47}_{-1.42}$ & $496^{+165}_{-247}$ & $-2.12^{+0.89}_{-0.82}$ \\
& non-grey clouds & $2.88^{+0.07}_{-0.07}$ & $0.490^{+0.021}_{-0.018}$ & $0.351^{+0.017}_{-0.012}$ & $-4.56^{+1.37}_{-1.33}$ & $412^{+110}_{-177}$ & $-1.66^{+0.76}_{-0.54}$ \\ \hline
\multirow{3}{*}{HAT-P-1b} & cloud free & $2.88^{+0.01}_{-0.01}$ & $1.105^{+0.023}_{-0.021}$ & $1.240^{+0.025}_{-0.024}$ & $-1.91^{+2.27}_{-0.79}$ & $532^{+132}_{-205}$ & $-$ \\
& $^*$grey clouds & $2.87^{+0.01}_{-0.01}$ & $1.125^{+0.028}_{-0.023}$ & $1.232^{+0.035}_{-0.030}$ & $-3.08^{+1.62}_{-1.75}$ & $905^{+320}_{-579}$ & $-1.69^{+1.11}_{-1.31}$ \\
& non-grey clouds & $2.87^{+0.01}_{-0.01}$ & $1.115^{+0.027}_{-0.019}$ & $1.249^{+0.030}_{-0.022}$ & $-2.53^{+2.04}_{-1.23}$ & $540^{+135}_{-305}$ & $-1.05^{+1.26}_{-1.20}$ \\ \hline
\multirow{3}{*}{HAT-P-3b} & cloud free & $3.35^{+0.06}_{-0.05}$ & $0.802^{+0.022}_{-0.017}$ & $0.855^{+0.024}_{-0.018}$ & $-7.34^{+3.11}_{-3.32}$ & $624^{+332}_{-591}$ & $-$ \\
& grey clouds & $3.34^{+0.05}_{-0.05}$ & $0.803^{+0.025}_{-0.019}$ & $0.837^{+0.029}_{-0.025}$ & $-7.67^{+2.55}_{-2.54}$ & $1191^{+609}_{-1095}$ & $-2.06^{+1.11}_{-1.12}$ \\
& non-grey clouds & $3.34^{+0.05}_{-0.04}$ & $0.803^{+0.020}_{-0.016}$ & $0.851^{+0.023}_{-0.017}$ & $-7.66^{+2.55}_{-2.22}$ & $865^{+447}_{-594}$ & $-1.35^{+1.30}_{-1.22}$ \\ \hline
\multirow{3}{*}{HAT-P-11b} & $^*$cloud free & $3.08^{+0.05}_{-0.05}$ & $0.752^{+0.014}_{-0.012}$ & $0.424^{+0.008}_{-0.007}$ & $-4.77^{+0.68}_{-1.91}$ & $669^{+302}_{-242}$ & $-$ \\
& grey clouds & $3.08^{+0.05}_{-0.04}$ & $0.755^{+0.013}_{-0.012}$ & $0.419^{+0.007}_{-0.009}$ & $-4.31^{+1.00}_{-1.82}$ & $722^{+274}_{-295}$ & $-0.65^{+0.85}_{-0.79}$ \\
& non-grey clouds & $3.08^{+0.04}_{-0.05}$ & $0.752^{+0.012}_{-0.011}$ & $0.423^{+0.008}_{-0.007}$ & $-4.86^{+0.56}_{-1.28}$ & $705^{+276}_{-200}$ & $-0.90^{+1.80}_{-1.16}$ \\ \hline
\multirow{3}{*}{HAT-P-12b} & cloud free & $2.76^{+0.03}_{-0.03}$ & $0.718^{+0.004}_{-0.006}$ & $0.946^{+0.005}_{-0.008}$ & $-4.84^{+0.59}_{-0.69}$ & $308^{+75}_{-137}$ & $-$ \\
& grey clouds & $2.76^{+0.03}_{-0.02}$ & $0.718^{+0.004}_{-0.005}$ & $0.944^{+0.004}_{-0.006}$ & $-4.68^{+0.64}_{-0.98}$ & $310^{+73}_{-118}$ & $-0.08^{+0.77}_{-0.64}$ \\
& non-grey clouds & $2.76^{+0.02}_{-0.02}$ & $0.717^{+0.004}_{-0.005}$ & $0.945^{+0.004}_{-0.007}$ & $-4.78^{+0.55}_{-0.58}$ & $312^{+71}_{-107}$ & $-0.93^{+1.66}_{-1.10}$ \\ \hline
\multirow{3}{*}{HAT-P-17b} & cloud free & $3.11^{+0.02}_{-0.02}$ & $0.850^{+0.009}_{-0.011}$ & $0.996^{+0.010}_{-0.014}$ & $-4.84^{+1.05}_{-1.51}$ & $370^{+120}_{-222}$ & $-$ \\
& grey clouds & $3.11^{+0.02}_{-0.02}$ & $0.851^{+0.008}_{-0.011}$ & $0.994^{+0.008}_{-0.011}$ & $-4.97^{+3.45}_{-1.87}$ & $441^{+153}_{-286}$ & $-1.02^{+1.23}_{-0.96}$ \\
& non-grey clouds & $3.11^{+0.02}_{-0.02}$ & $0.849^{+0.007}_{-0.010}$ & $0.993^{+0.007}_{-0.010}$ & $-4.90^{+1.47}_{-1.37}$ & $405^{+132}_{-193}$ & $-1.38^{+1.22}_{-1.12}$ \\ \hline
\multirow{3}{*}{HAT-P-18b} & cloud free & $2.77^{+0.04}_{-0.04}$ & $0.726^{+0.019}_{-0.016}$ & $0.957^{+0.025}_{-0.021}$ & $-1.93^{+2.06}_{-0.81}$ & $232^{+22}_{-49}$ & $-$ \\
& $^*$grey clouds & $2.75^{+0.04}_{-0.04}$ & $0.719^{+0.012}_{-0.015}$ & $0.933^{+0.017}_{-0.020}$ & $-2.82^{+1.55}_{-1.15}$ & $396^{+141}_{-163}$ & $-1.56^{+0.94}_{-0.72}$ \\
& non-grey cloud & $2.76^{+0.03}_{-0.03}$ & $0.720^{+0.013}_{-0.014}$ & $0.947^{+0.019}_{-0.019}$ & $-3.15^{+1.25}_{-1.77}$ & $249^{+35}_{-140}$ & $-1.21^{+1.20}_{-1.10}$ \\ \hline
\multirow{3}{*}{HAT-P-26b} & $^*$cloud free & $2.73^{+0.07}_{-0.06}$ & $0.846^{+0.029}_{-0.041}$ & $0.561^{+0.019}_{-0.027}$ & $-3.48^{+0.77}_{-1.21}$ & $520^{+108}_{-142}$ & $-$ \\
& grey clouds & $2.70^{+0.06}_{-0.06}$ & $0.856^{+0.029}_{-0.033}$ & $0.554^{+0.014}_{-0.023}$ & $-3.27^{+0.88}_{-1.09}$ & $587^{+128}_{-140}$ & $-0.82^{+0.77}_{-0.90}$ \\
& non-grey clouds & $2.72^{+0.06}_{-0.05}$ & $0.841^{+0.023}_{-0.030}$ & $0.557^{+0.015}_{-0.020}$ & $-3.42^{+0.66}_{-0.90}$ & $525^{+107}_{-109}$ & $-0.91^{+1.56}_{-1.12}$ \\ \hline
\multirow{3}{*}{HAT-P-32b} & $^*$cloud free & $2.87^{+0.08}_{-0.06}$ & $1.222^{+0.009}_{-0.010}$ & $1.784^{+0.013}_{-0.016}$ & $-2.86^{+1.28}_{-1.28}$ & $610^{+132}_{-143}$ & $-$ \\
& grey clouds & $2.86^{+0.07}_{-0.06}$ & $1.226^{+0.010}_{-0.009}$ & $1.782^{+0.012}_{-0.014}$ & $-2.88^{+1.16}_{-1.09}$ & $642^{+144}_{-176}$ & $-0.55^{+1.12}_{-0.95}$ \\
& non-grey clouds & $2.86^{+0.06}_{-0.06}$ & $1.221^{+0.007}_{-0.009}$ & $1.782^{+0.010}_{-0.015}$ & $-3.10^{+1.09}_{-1.26}$ & $615^{+129}_{-151}$ & $-0.95^{+1.50}_{-1.15}$ \\ \hline
\multirow{3}{*}{HAT-P-38b} & cloud free & $3.03^{+0.07}_{-0.06}$ & $0.942^{+0.051}_{-0.046}$ & $0.836^{+0.046}_{-0.042}$ & $-5.83^{+0.52}_{-0.76}$ & $1199^{+376}_{-349}$ & $-$ \\
& grey clouds & $3.02^{+0.06}_{-0.05}$ & $0.952^{+0.039}_{-0.041}$ & $0.808^{+0.028}_{-0.042}$ & $-5.20^{+0.96}_{-1.51}$ & $1539^{+713}_{-875}$ & $-1.42^{+1.29}_{-1.01}$ \\
& non-grey clouds & $3.03^{+0.06}_{-0.06}$ & $0.939^{+0.043}_{-0.041}$ & $0.831^{+0.038}_{-0.037}$ & $-5.79^{+0.52}_{-0.72}$ & $1204^{+367}_{-297}$ & $-1.00^{+1.73}_{-1.21}$ \\ \hline
\multirow{3}{*}{HAT-P-41b} & $^*$cloud free & $2.87^{+0.06}_{-0.05}$ & $1.720^{+0.019}_{-0.025}$ & $1.658^{+0.016}_{-0.026}$ & $-4.61^{+0.80}_{-2.03}$ & $1064^{+293}_{-358}$ & $-$ \\
& grey clouds & $2.87^{+0.05}_{-0.04}$ & $1.725^{+0.021}_{-0.022}$ & $1.654^{+0.013}_{-0.018}$ & $-4.20^{+0.95}_{-1.72}$ & $1133^{+299}_{-329}$ & $-0.67^{+1.03}_{-1.00}$ \\
& non-grey clouds & $2.87^{+0.05}_{-0.04}$ & $1.717^{+0.015}_{-0.020}$ & $1.656^{+0.013}_{-0.019}$ & $-4.73^{+0.65}_{-1.27}$ & $1139^{+323}_{-290}$ & $-0.74^{+1.65}_{-1.00}$ \\ \hline
\multirow{3}{*}{HD 97658b} & $^*$cloud free & $3.21^{+0.05}_{-0.05}$ & $0.736^{+0.015}_{-0.012}$ & $0.205^{+0.004}_{-0.003}$ & $-9.26^{+1.83}_{-1.90}$ & $1625^{+217}_{-231}$ & $-$ \\
& grey clouds & $3.20^{+0.05}_{-0.04}$ & $0.744^{+0.015}_{-0.016}$ & $0.202^{+0.005}_{-0.004}$ & $-9.08^{+1.75}_{-1.75}$ & $1849^{+338}_{-447}$ & $-0.01^{+0.22}_{-0.58}$ \\
& non-grey clouds & $3.20^{+0.05}_{-0.04}$ & $0.737^{+0.013}_{-0.011}$ & $0.205^{+0.004}_{-0.003}$ & $-9.09^{+1.68}_{-1.61}$ & $1635^{+213}_{-205}$ & $-0.89^{+1.87}_{-1.29}$ \\ \hline
\multirow{3}{*}{HD 149026b} & cloud free & $3.38^{+0.06}_{-0.06}$ & $1.447^{+0.010}_{-0.010}$ & $0.710^{+0.005}_{-0.003}$ & $-4.09^{+2.23}_{-2.44}$ & $815^{+327}_{-546}$ & $-$ \\
& grey clouds & $3.35^{+0.05}_{-0.05}$ & $1.518^{+0.028}_{-0.029}$ & $0.701^{+0.014}_{-0.008}$ & $-4.59^{+4.62}_{-1.42}$ & $2424^{+531}_{-392}$ & $-3.09^{+0.56}_{-0.76}$ \\
& non-grey clouds & $3.35^{+0.05}_{-0.05}$ & $1.515^{+0.026}_{-0.026}$ & $0.699^{+0.013}_{-0.009}$ & $-4.00^{+3.17}_{-0.99}$ & $2412^{+482}_{-395}$ & $-3.05^{+0.58}_{-0.80}$ \\ \hline
\multirow{3}{*}{HD 189733b} & $^*$cloud free & $3.29^{+0.02}_{-0.02}$ & $0.806^{+0.010}_{-0.009}$ & $1.215^{+0.015}_{-0.015}$ & $-2.58^{+1.07}_{-0.95}$ & $504^{+90}_{-113}$ & $-$ \\
& grey clouds & $3.29^{+0.02}_{-0.02}$ & $0.806^{+0.009}_{-0.009}$ & $1.215^{+0.013}_{-0.013}$ & $-2.43^{+1.04}_{-0.78}$ & $518^{+89}_{-127}$ & $-0.01^{+0.83}_{-0.66}$ \\
& non-grey clouds & $3.29^{+0.02}_{-0.02}$ & $0.806^{+0.008}_{-0.008}$ & $1.215^{+0.012}_{-0.013}$ & $-2.61^{+1.03}_{-0.79}$ & $507^{+81}_{-127}$ & $-0.90^{+1.66}_{-1.20}$ \\ \hline
\multirow{3}{*}{HD 209458b} & cloud free & $2.93^{+0.05}_{-0.05}$ & $1.226^{+0.027}_{-0.034}$ & $1.424^{+0.030}_{-0.040}$ & $-5.26^{+0.31}_{-0.66}$ & $752^{+219}_{-126}$ & $-$ \\
& $^*$grey clouds & $2.90^{+0.05}_{-0.05}$ & $1.224^{+0.021}_{-0.024}$ & $1.410^{+0.020}_{-0.025}$ & $-4.25^{+1.02}_{-1.36}$ & $717^{+207}_{-215}$ & $-1.19^{+1.06}_{-0.87}$ \\
& non-grey clouds & $2.93^{+0.05}_{-0.05}$ & $1.216^{+0.018}_{-0.024}$ & $1.413^{+0.022}_{-0.026}$ & $-5.20^{+0.34}_{-0.70}$ & $743^{+189}_{-129}$ & $-0.74^{+1.71}_{-1.10}$ \\ \hline
\multirow{3}{*}{WASP-12b} & cloud free & $2.99^{+0.03}_{-0.03}$ & $1.570^{+0.045}_{-0.036}$ & $1.814^{+0.053}_{-0.042}$ & $-4.54^{+0.53}_{-0.79}$ & $964^{+192}_{-206}$ & $-$ \\
& $^*$grey clouds & $2.99^{+0.02}_{-0.02}$ & $1.572^{+0.037}_{-0.038}$ & $1.793^{+0.046}_{-0.047}$ & $-3.85^{+0.97}_{-1.28}$ & $1251^{+296}_{-313}$ & $-1.40^{+1.43}_{-0.87}$ \\
& non-grey clouds & $2.99^{+0.02}_{-0.03}$ & $1.577^{+0.037}_{-0.031}$ & $1.813^{+0.045}_{-0.034}$ & $-4.39^{+0.57}_{-1.07}$ & $1149^{+239}_{-245}$ & $-1.36^{+1.18}_{-0.76}$ \\ \hline
\multirow{3}{*}{WASP-17b} & cloud free & $2.51^{+0.02}_{-0.02}$ & $1.629^{+0.019}_{-0.022}$ & $1.904^{+0.016}_{-0.027}$ & $-3.51^{+1.44}_{-1.75}$ & $508^{+158}_{-299}$ & $-$ \\
& grey clouds & $2.50^{+0.02}_{-0.02}$ & $1.628^{+0.018}_{-0.019}$ & $1.899^{+0.013}_{-0.021}$ & $-3.41^{+1.30}_{-1.34}$ & $509^{+149}_{-211}$ & $-0.60^{+0.86}_{-0.90}$ \\
& non-grey clouds & $2.50^{+0.02}_{-0.02}$ & $1.626^{+0.018}_{-0.017}$ & $1.901^{+0.014}_{-0.019}$ & $-3.57^{+1.21}_{-1.39}$ & $506^{+165}_{-255}$ & $-0.92^{+1.53}_{-1.12}$ \\ \hline
\multirow{3}{*}{WASP-19b} & $^*$cloud free & $3.15^{+0.01}_{-0.01}$ & $1.028^{+0.006}_{-0.007}$ & $1.379^{+0.004}_{-0.008}$ & $-5.04^{+0.60}_{-0.73}$ & $1828^{+479}_{-441}$ & $-$ \\
& grey clouds & $3.15^{+0.01}_{-0.01}$ & $1.027^{+0.006}_{-0.006}$ & $1.379^{+0.005}_{-0.007}$ & $-4.89^{+0.63}_{-0.96}$ & $1712^{+501}_{-422}$ & $0.21^{+0.58}_{-0.47}$ \\
& non-grey clouds & $3.15^{+0.01}_{-0.01}$ & $1.027^{+0.006}_{-0.006}$ & $1.379^{+0.005}_{-0.007}$ & $-4.92^{+0.59}_{-0.81}$ & $1724^{+446}_{-419}$ & $-0.86^{+1.84}_{-1.20}$ \\ \hline
\multirow{3}{*}{WASP-29b} & cloud free & $2.95^{+0.05}_{-0.05}$ & $0.827^{+0.018}_{-0.025}$ & $0.781^{+0.017}_{-0.023}$ & $-8.54^{+2.25}_{-1.84}$ & $364^{+117}_{-181}$ & $-$ \\
& grey clouds & $2.95^{+0.04}_{-0.04}$ & $0.826^{+0.016}_{-0.025}$ & $0.778^{+0.013}_{-0.022}$ & $-8.69^{+2.06}_{-1.89}$ & $404^{+141}_{-210}$ & $0.17^{+0.65}_{-0.53}$ \\
& non-grey clouds & $2.95^{+0.04}_{-0.04}$ & $0.823^{+0.013}_{-0.017}$ & $0.776^{+0.012}_{-0.016}$ & $-8.13^{+2.20}_{-1.51}$ & $389^{+126}_{-194}$ & $-1.37^{+1.46}_{-1.35}$ \\
\hline
\end{tabular}
}
\end{table*}

\begin{table*}
\centering
\caption{Continuation of Table \ref{table:retrieval_results_all1}.}
\label{table:retrieval_results_all2}
\resizebox{0.75\textwidth}{!}{
\begin{tabular}{lccccccc}
\hline
Planet & Model & log $g_{\rm p}$ (cm s$^{-2}$) & $R_{\rm star}$ (R$_\odot$) & $R_{\rm p}$ (R$_{\rm Jup}$) & log $X_{\ch{H2O}}$ & $T$ (K) & log $P_{\rm cloud\mbox{-}top}$ (bar) \\
\hline
\multirow{3}{*}{WASP-31b} & cloud free & $2.66^{+0.03}_{-0.03}$ & $1.266^{+0.017}_{-0.021}$ & $1.527^{+0.019}_{-0.026}$ & $-4.42^{+0.98}_{-1.51}$ & $454^{+175}_{-306}$ & $-$ \\
& grey clouds & $2.66^{+0.03}_{-0.03}$ & $1.270^{+0.018}_{-0.019}$ & $1.523^{+0.016}_{-0.025}$ & $-3.87^{+1.39}_{-1.47}$ & $457^{+154}_{-261}$ & $-0.58^{+1.18}_{-0.94}$ \\
& non-grey clouds & $2.66^{+0.03}_{-0.03}$ & $1.265^{+0.014}_{-0.019}$ & $1.525^{+0.016}_{-0.024}$ & $-4.49^{+0.89}_{-1.28}$ & $484^{+183}_{-263}$ & $-1.00^{+1.65}_{-1.25}$ \\ \hline
\multirow{3}{*}{WASP-39b} & $^*$cloud free & $2.70^{+0.05}_{-0.04}$ & $0.917^{+0.020}_{-0.019}$ & $1.274^{+0.028}_{-0.025}$ & $-1.16^{+0.58}_{-0.12}$ & $408^{+89}_{-85}$ & $-$ \\
& grey clouds & $2.67^{+0.05}_{-0.04}$ & $0.929^{+0.023}_{-0.022}$ & $1.275^{+0.027}_{-0.023}$ & $-1.42^{+1.56}_{-0.36}$ & $493^{+132}_{-194}$ & $-1.21^{+1.32}_{-1.43}$ \\
& non-grey clouds & $2.68^{+0.04}_{-0.04}$ & $0.917^{+0.019}_{-0.015}$ & $1.275^{+0.026}_{-0.021}$ & $-1.38^{+1.19}_{-0.29}$ & $361^{+60}_{-70}$ & $-0.97^{+1.63}_{-1.18}$ \\ \hline
\multirow{3}{*}{WASP-43b} & cloud free & $3.68^{+0.06}_{-0.06}$ & $0.675^{+0.015}_{-0.022}$ & $1.045^{+0.024}_{-0.034}$ & $-3.09^{+1.27}_{-1.33}$ & $471^{+125}_{-153}$ & $-$ \\
& grey clouds & $3.67^{+0.06}_{-0.06}$ & $0.673^{+0.012}_{-0.017}$ & $1.041^{+0.018}_{-0.026}$ & $-3.00^{+1.35}_{-1.19}$ & $518^{+137}_{-231}$ & $-0.28^{+1.08}_{-0.80}$ \\
& non-grey clouds & $3.68^{+0.06}_{-0.05}$ & $0.673^{+0.012}_{-0.017}$ & $1.041^{+0.018}_{-0.026}$ & $-3.28^{+1.22}_{-1.22}$ & $510^{+131}_{-193}$ & $-1.62^{+1.34}_{-1.57}$ \\ \hline
\multirow{3}{*}{WASP-52b} & $^*$cloud free & $2.80^{+0.03}_{-0.03}$ & $0.796^{+0.008}_{-0.012}$ & $1.259^{+0.013}_{-0.020}$ & $-3.16^{+0.92}_{-1.03}$ & $397^{+73}_{-115}$ & $-$ \\
& grey clouds & $2.80^{+0.03}_{-0.03}$ & $0.796^{+0.008}_{-0.012}$ & $1.258^{+0.012}_{-0.020}$ & $-3.03^{+0.89}_{-0.92}$ & $396^{+67}_{-104}$ & $0.12^{+0.65}_{-0.55}$ \\
& non-grey clouds & $2.81^{+0.03}_{-0.03}$ & $0.794^{+0.006}_{-0.009}$ & $1.256^{+0.010}_{-0.015}$ & $-3.19^{+0.88}_{-0.87}$ & $402^{+71}_{-109}$ & $-0.89^{+1.76}_{-1.19}$ \\ \hline
\multirow{3}{*}{WASP-63b} & cloud free & $2.64^{+0.05}_{-0.04}$ & $1.896^{+0.045}_{-0.052}$ & $1.433^{+0.036}_{-0.040}$ & $-4.38^{+1.22}_{-1.86}$ & $301^{+72}_{-238}$ & $-$ \\
& grey clouds & $2.63^{+0.04}_{-0.04}$ & $1.911^{+0.041}_{-0.046}$ & $1.415^{+0.026}_{-0.034}$ & $-4.40^{+2.63}_{-1.62}$ & $485^{+164}_{-236}$ & $-1.77^{+1.08}_{-0.85}$ \\
& non-grey clouds & $2.64^{+0.04}_{-0.04}$ & $1.891^{+0.039}_{-0.045}$ & $1.418^{+0.027}_{-0.038}$ & $-4.49^{+1.22}_{-1.38}$ & $378^{+131}_{-216}$ & $-1.38^{+1.19}_{-1.16}$ \\ \hline
\multirow{3}{*}{WASP-67b} & cloud free & $2.70^{+0.02}_{-0.01}$ & $0.873^{+0.028}_{-0.029}$ & $1.367^{+0.044}_{-0.047}$ & $-5.33^{+0.67}_{-1.02}$ & $536^{+207}_{-228}$ & $-$ \\
& grey clouds & $2.70^{+0.01}_{-0.01}$ & $0.871^{+0.026}_{-0.027}$ & $1.344^{+0.054}_{-0.048}$ & $-5.11^{+1.56}_{-1.49}$ & $612^{+249}_{-381}$ & $-0.93^{+1.69}_{-1.25}$ \\
& non-grey clouds & $2.70^{+0.01}_{-0.01}$ & $0.872^{+0.026}_{-0.024}$ & $1.363^{+0.041}_{-0.039}$ & $-5.29^{+0.74}_{-0.83}$ & $548^{+197}_{-217}$ & $-1.07^{+1.63}_{-1.24}$ \\ \hline
\multirow{3}{*}{WASP-69b} & cloud free & $2.76^{+0.04}_{-0.04}$ & $0.831^{+0.012}_{-0.016}$ & $1.032^{+0.015}_{-0.020}$ & $-4.41^{+0.37}_{-0.39}$ & $228^{+21}_{-40}$ & $-$ \\
& grey clouds & $2.75^{+0.04}_{-0.04}$ & $0.829^{+0.009}_{-0.013}$ & $1.026^{+0.010}_{-0.015}$ & $-4.49^{+0.47}_{-0.48}$ & $250^{+35}_{-88}$ & $-0.26^{+0.41}_{-0.66}$ \\
& $^*$non-grey clouds & $2.74^{+0.04}_{-0.04}$ & $0.830^{+0.007}_{-0.008}$ & $1.022^{+0.007}_{-0.008}$ & $-3.70^{+0.84}_{-0.94}$ & $282^{+51}_{-89}$ & $-1.64^{+0.49}_{-0.52}$ \\ \hline
\multirow{3}{*}{WASP-74b} & cloud free & $2.95^{+0.01}_{-0.02}$ & $1.626^{+0.005}_{-0.006}$ & $1.505^{+0.004}_{-0.006}$ & $-4.59^{+0.92}_{-1.17}$ & $386^{+121}_{-182}$ & $-$ \\
& grey clouds & $2.96^{+0.02}_{-0.01}$ & $1.625^{+0.004}_{-0.005}$ & $1.505^{+0.003}_{-0.005}$ & $-4.80^{+1.47}_{-1.21}$ & $391^{+114}_{-170}$ & $-0.13^{+0.76}_{-0.65}$ \\
& non-grey clouds & $2.95^{+0.01}_{-0.01}$ & $1.624^{+0.004}_{-0.005}$ & $1.505^{+0.003}_{-0.004}$ & $-4.76^{+1.00}_{-0.94}$ & $394^{+112}_{-156}$ & $-1.17^{+1.45}_{-1.19}$ \\ \hline
\multirow{3}{*}{WASP-76b} & cloud free & $2.80^{+0.02}_{-0.02}$ & $1.800^{+0.023}_{-0.035}$ & $1.804^{+0.023}_{-0.035}$ & $-2.03^{+1.16}_{-0.61}$ & $416^{+57}_{-62}$ & $-$ \\
& grey clouds & $2.80^{+0.02}_{-0.02}$ & $1.794^{+0.018}_{-0.030}$ & $1.797^{+0.018}_{-0.031}$ & $-1.94^{+1.04}_{-0.54}$ & $422^{+59}_{-62}$ & $0.22^{+0.55}_{-0.49}$ \\
& $^*$non-grey clouds & $2.81^{+0.02}_{-0.02}$ & $1.841^{+0.016}_{-0.014}$ & $1.786^{+0.010}_{-0.015}$ & $-4.45^{+0.39}_{-0.65}$ & $1313^{+291}_{-223}$ & $-2.53^{+0.55}_{-0.34}$ \\ \hline
\multirow{3}{*}{WASP-80b} & cloud free & $3.15^{+0.01}_{-0.01}$ & $0.594^{+0.007}_{-0.009}$ & $0.985^{+0.011}_{-0.015}$ & $-5.28^{+0.51}_{-0.66}$ & $425^{+163}_{-226}$ & $-$ \\
& grey clouds & $3.14^{+0.01}_{-0.01}$ & $0.594^{+0.005}_{-0.008}$ & $0.979^{+0.007}_{-0.011}$ & $-4.46^{+1.13}_{-1.54}$ & $468^{+162}_{-196}$ & $-1.22^{+1.07}_{-0.74}$ \\
& non-grey clouds & $3.15^{+0.01}_{-0.01}$ & $0.591^{+0.004}_{-0.006}$ & $0.980^{+0.007}_{-0.010}$ & $-5.23^{+0.53}_{-0.64}$ & $429^{+146}_{-197}$ & $-0.72^{+1.54}_{-0.92}$ \\ \hline
\multirow{3}{*}{WASP-101b} & cloud free & $2.76^{+0.04}_{-0.04}$ & $1.315^{+0.014}_{-0.021}$ & $1.382^{+0.015}_{-0.022}$ & $-5.53^{+1.70}_{-1.17}$ & $275^{+52}_{-151}$ & $-$ \\
& grey clouds & $2.76^{+0.04}_{-0.03}$ & $1.318^{+0.015}_{-0.021}$ & $1.379^{+0.012}_{-0.019}$ & $-5.88^{+2.71}_{-1.57}$ & $324^{+89}_{-206}$ & $-0.32^{+1.57}_{-0.80}$ \\
& non-grey clouds & $2.77^{+0.03}_{-0.03}$ & $1.323^{+0.014}_{-0.016}$ & $1.378^{+0.012}_{-0.016}$ & $-5.45^{+3.05}_{-1.73}$ & $344^{+89}_{-165}$ & $-2.35^{+0.81}_{-0.88}$ \\ \hline
\multirow{3}{*}{WASP-121b} & $^*$cloud free & $2.98^{+0.02}_{-0.01}$ & $1.569^{+0.005}_{-0.007}$ & $1.826^{+0.004}_{-0.008}$ & $-3.05^{+0.78}_{-0.89}$ & $717^{+107}_{-108}$ & $-$ \\
& grey clouds & $2.97^{+0.01}_{-0.01}$ & $1.569^{+0.004}_{-0.005}$ & $1.826^{+0.003}_{-0.005}$ & $-3.03^{+0.77}_{-0.80}$ & $700^{+106}_{-121}$ & $-0.00^{+0.58}_{-0.61}$ \\
& non-grey clouds & $2.97^{+0.01}_{-0.01}$ & $1.568^{+0.004}_{-0.004}$ & $1.825^{+0.003}_{-0.004}$ & $-3.04^{+0.69}_{-0.70}$ & $692^{+111}_{-106}$ & $-0.86^{+1.43}_{-1.00}$ \\ \hline
\multirow{3}{*}{XO-1b} & $^*$cloud free & $3.22^{+0.04}_{-0.03}$ & $0.932^{+0.009}_{-0.011}$ & $1.183^{+0.012}_{-0.014}$ & $-2.34^{+1.51}_{-0.95}$ & $671^{+184}_{-171}$ & $-$ \\
& grey clouds & $3.22^{+0.03}_{-0.03}$ & $0.932^{+0.008}_{-0.009}$ & $1.181^{+0.009}_{-0.012}$ & $-2.35^{+1.47}_{-0.90}$ & $686^{+172}_{-195}$ & $-0.32^{+1.03}_{-0.79}$ \\
& non-grey clouds & $3.22^{+0.03}_{-0.03}$ & $0.931^{+0.008}_{-0.009}$ & $1.181^{+0.010}_{-0.011}$ & $-2.93^{+1.29}_{-1.18}$ & $709^{+178}_{-234}$ & $-1.54^{+1.45}_{-1.54}$ \\ \hline
\multirow{3}{*}{TRAPPIST-1d (\ch{N2})} & cloud free & $2.68^{+0.04}_{-0.04}$ & $0.117^{+0.002}_{-0.002}$ & $0.071^{+0.001}_{-0.001}$ & $-6.87^{+3.08}_{-3.60}$ & $1005^{+521}_{-1250}$ - \\
& grey clouds & $2.68^{+0.04}_{-0.04}$ & $0.121^{+0.002}_{-0.003}$ & $0.070^{+0.001}_{-0.001}$ & $-7.37^{+2.70}_{-3.31}$ & $895^{+400}_{-912}$ & $-0.97^{+1.68}_{-1.09}$ \\
& non-grey clouds & $2.68^{+0.04}_{-0.04}$ & $0.118^{+0.002}_{-0.003}$ & $0.071^{+0.001}_{-0.001}$ & $-6.66^{+3.08}_{-3.13}$ & $865^{+391}_{-855}$ & $-1.34^{+1.42}_{-1.32}$ \\ \hline
\multirow{3}{*}{TRAPPIST-1d (\ch{H2})} & cloud free & $2.73^{+0.04}_{-0.04}$ & $0.123^{+0.002}_{-0.002}$ & $0.069^{+0.001}_{-0.001}$ & $-6.20^{+0.81}_{-0.61}$ & $219^{+14}_{-31}$ & $-$ \\
& grey clouds & $2.72^{+0.04}_{-0.04}$ & $0.125^{+0.002}_{-0.002}$ & $0.069^{+0.001}_{-0.001}$ & $-7.89^{+2.77}_{-2.00}$ & $217^{+12}_{-26}$ & $-0.30^{+0.29}_{-0.30}$ \\
& non-grey clouds & $2.72^{+0.04}_{-0.04}$ & $0.124^{+0.002}_{-0.002}$ & $0.069^{+0.001}_{-0.001}$ & $-6.52^{+3.03}_{-0.82}$ & $218^{+13}_{-25}$ & $-0.45^{+1.51}_{-0.53}$ \\ \hline
\multirow{3}{*}{TRAPPIST-1e (\ch{N2})} & cloud free & $2.96^{+0.03}_{-0.03}$ & $0.119^{+0.002}_{-0.001}$ & $0.082^{+0.001}_{-0.001}$ & $-6.49^{+3.74}_{-3.58}$ & $1640^{+943}_{-977}$ & $-$ \\
& grey clouds & $2.96^{+0.03}_{-0.03}$ & $0.121^{+0.002}_{-0.002}$ & $0.081^{+0.001}_{-0.001}$ & $-6.63^{+3.30}_{-3.50}$ & $1105^{+577}_{-1033}$ & $-0.56^{+1.72}_{-1.00}$ \\
& non-grey clouds & $2.96^{+0.03}_{-0.03}$ & $0.120^{+0.002}_{-0.001}$ & $0.082^{+0.001}_{-0.001}$ & $-6.68^{+3.27}_{-3.61}$ & $1553^{+863}_{-933}$ & $-1.20^{+1.67}_{-1.40}$ \\ \hline
\multirow{3}{*}{TRAPPIST-1e (\ch{H2})} & cloud free & $2.97^{+0.03}_{-0.03}$ & $0.121^{+0.002}_{-0.002}$ & $0.081^{+0.001}_{-0.001}$ & $-6.00^{+4.10}_{-3.96}$ & $343^{+106}_{-167}$ & $-$ \\
& grey clouds & $2.96^{+0.03}_{-0.03}$ & $0.123^{+0.002}_{-0.002}$ & $0.080^{+0.001}_{-0.001}$ & $-6.15^{+3.85}_{-3.61}$ & $316^{+86}_{-149}$ & $0.43^{+0.54}_{-0.36}$ \\
& non-grey clouds & $2.97^{+0.03}_{-0.03}$ & $0.121^{+0.002}_{-0.002}$ & $0.081^{+0.001}_{-0.001}$ & $-6.27^{+3.66}_{-3.68}$ & $328^{+91}_{-157}$ & $-1.25^{+1.75}_{-1.67}$ \\ \hline
\multirow{3}{*}{TRAPPIST-1f (\ch{N2})} & cloud free & $2.92^{+0.02}_{-0.02}$ & $0.121^{+0.002}_{-0.002}$ & $0.093^{+0.002}_{-0.002}$ & $-6.54^{+3.54}_{-3.84}$ & $1625^{+942}_{-963}$ & $-$ \\
& grey clouds & $2.92^{+0.02}_{-0.02}$ & $0.122^{+0.002}_{-0.002}$ & $0.092^{+0.001}_{-0.002}$ & $-6.71^{+3.33}_{-3.55}$ & $981^{+518}_{-1093}$ & $-0.49^{+1.62}_{-0.98}$ \\
& non-grey clouds & $2.92^{+0.02}_{-0.02}$ & $0.121^{+0.002}_{-0.002}$ & $0.093^{+0.002}_{-0.002}$ & $-6.48^{+3.36}_{-3.43}$ & $1570^{+902}_{-967}$ & $-1.12^{+1.79}_{-1.38}$ \\ \hline
\multirow{3}{*}{TRAPPIST-1f (\ch{H2})} & cloud free & $2.93^{+0.02}_{-0.02}$ & $0.122^{+0.002}_{-0.002}$ & $0.092^{+0.002}_{-0.002}$ & $-5.86^{+4.22}_{-3.97}$ & $329^{+93}_{-169}$ & $-$ \\
& grey clouds & $2.93^{+0.02}_{-0.02}$ & $0.123^{+0.002}_{-0.002}$ & $0.092^{+0.001}_{-0.002}$ & $-6.12^{+3.92}_{-3.87}$ & $304^{+75}_{-138}$ & $0.39^{+0.51}_{-0.38}$ \\
& non-grey clouds & $2.93^{+0.02}_{-0.02}$ & $0.122^{+0.002}_{-0.002}$ & $0.092^{+0.001}_{-0.002}$ & $-6.00^{+3.92}_{-3.70}$ & $316^{+84}_{-155}$ & $-1.16^{+1.82}_{-1.56}$ \\ \hline
\multirow{3}{*}{TRAPPIST-1g (\ch{N2})} & cloud free & $2.93^{+0.02}_{-0.02}$ & $0.120^{+0.002}_{-0.002}$ & $0.103^{+0.002}_{-0.001}$ & $-6.76^{+3.42}_{-3.58}$ & $1689^{+1010}_{-953}$ & $-$ \\
& grey clouds & $2.93^{+0.02}_{-0.02}$ & $0.122^{+0.002}_{-0.002}$ & $0.102^{+0.002}_{-0.002}$ & $-6.47^{+3.27}_{-3.45}$ & $1125^{+580}_{-997}$ & $-0.55^{+1.84}_{-0.99}$ \\
& non-grey clouds & $2.93^{+0.02}_{-0.02}$ & $0.120^{+0.002}_{-0.002}$ & $0.103^{+0.002}_{-0.001}$ & $-6.27^{+3.49}_{-3.40}$ & $1550^{+866}_{-932}$ & $-1.13^{+1.72}_{-1.34}$ \\ \hline
\multirow{3}{*}{TRAPPIST-1g (\ch{H2})} & cloud free & $2.93^{+0.02}_{-0.02}$ & $0.122^{+0.002}_{-0.002}$ & $0.102^{+0.002}_{-0.002}$ & $-6.85^{+3.45}_{-4.35}$ & $422^{+152}_{-210}$ & $-$ \\
& grey clouds & $2.93^{+0.02}_{-0.02}$ & $0.123^{+0.002}_{-0.002}$ & $0.101^{+0.001}_{-0.002}$ & $-6.41^{+3.71}_{-3.67}$ & $366^{+122}_{-183}$ & $0.49^{+0.55}_{-0.36}$ \\
& non-grey clouds & $2.93^{+0.02}_{-0.02}$ & $0.122^{+0.002}_{-0.002}$ & $0.102^{+0.002}_{-0.002}$ & $-6.49^{+3.58}_{-3.79}$ & $411^{+141}_{-193}$ & $-1.20^{+1.80}_{-1.51}$ \\ \hline
\end{tabular}
}
\end{table*}

\begin{table*}
\centering
\caption{Free parameters retrieved using cloud-free and grey-cloud non-isothermal models (see Section \ref{sec:nonisothermal_tests}).}
\label{table:nonisothermal_results}
\begin{tabular}{lcccccc}
\hline
Planet & Model & log $g_{\rm p}$ (cm s$^{-2}$) & $R_{\rm star}$ (R$_\odot$) & $R_{\rm p}$ (R$_{\rm Jup}$) & log $X_{\ch{H2O}}$ & log $P_{\rm cloud\mbox{-}top}$ (bar) \\ \hline
\multirow{2}{*}{HD 209458b} & cloud free & $2.92^{+0.05}_{-0.05}$ & $1.22^{+0.02}_{-0.02}$ & $1.41^{+0.03}_{-0.02}$ & $-5.07^{+0.50}_{-0.36}$ & $-$ \\
& grey clouds & $2.9^{+0.05}_{-0.05}$ & $1.23^{+0.02}_{-0.02}$ & $1.41^{+0.03}_{-0.02}$ & $-4.44^{+1.15}_{-0.77}$ & $-0.93^{+0.62}_{-0.65}$ \\ \hline
\multirow{2}{*}{WASP-12b} & cloud free & $2.99^{+0.03}_{-0.03}$ & $1.57^{+0.03}_{-0.04}$ & $1.81^{+0.04}_{-0.04}$ & $-4.5^{+0.77}_{-0.49}$ & $-$ \\
& grey clouds & $2.99^{+0.02}_{-0.02}$ & $1.58^{+0.03}_{-0.04}$ & $1.79^{+0.04}_{-0.05}$ & $-4.25^{+0.95}_{-0.65}$ & $-1.07^{+0.57}_{-0.70}$ \\ \hline
\multirow{2}{*}{WASP-39b} & cloud free & $2.67^{+0.04}_{-0.04}$ & $0.94^{+0.02}_{-0.02}$ & $1.27^{+0.02}_{-0.03}$ & $-1.46^{+0.37}_{-1.14}$ & $-$ \\
& grey clouds & $2.67^{+0.04}_{-0.04}$ & $0.94^{+0.02}_{-0.02}$ & $1.28^{+0.02}_{-0.03}$ & $-1.51^{+0.40}_{-1.27}$ & $-0.79^{+0.98}_{-1.01}$ \\ \hline 
\end{tabular}
\end{table*}

\begin{table*}
\centering
\caption{Free parameters retrieved using cloud-free and grey-cloud models with tight temperature priors (see Section \ref{sec:tight_temp}).}
\label{table:tight_T_prior_results}
\begin{tabular}{lccccccc}
\hline
Planet & Model & log $g_{\rm p}$ (cm s$^{-2}$) & $R_{\rm star}$ (R$_\odot$) & $R_{\rm p}$ (R$_{\rm Jup}$) & log $X_{\ch{H2O}}$ & $T$ (K) & log $P_{\rm cloud\mbox{-}top}$ (bar) \\ \hline
\multirow{2}{*}{HD 209458b} & cloud free & $3.04^{+0.03}_{-0.04}$ & $1.25^{+0.03}_{-0.04}$ & $1.45^{+0.03}_{-0.04}$ & $-5.94^{+0.14}_{-0.14}$ & $1360^{+25}_{-13}$ & $-$ \\
& grey clouds & $2.9^{+0.05}_{-0.05}$ & $1.25^{+0.02}_{-0.02}$ & $1.4^{+0.02}_{-0.01}$ & $-5.11^{+1.09}_{-0.62}$ & $1417^{+75}_{-51}$ & $-1.59^{+0.63}_{-1.10}$ \\ \hline
\multirow{2}{*}{WASP-12b} & cloud free & $3.03^{+0.02}_{-0.02}$ & $1.6^{+0.03}_{-0.04}$ & $1.83^{+0.03}_{-0.05}$ & $-6.16^{+0.17}_{-0.19}$ & $2176^{+51}_{-24}$ & $-$ \\
& grey clouds & $2.99^{+0.02}_{-0.02}$ & $1.59^{+0.03}_{-0.03}$ & $1.75^{+0.04}_{-0.03}$ & $-4.41^{+1.02}_{-0.84}$ & $2276^{+168}_{-91}$ & $-2.13^{+0.85}_{-1.03}$ \\ \hline
\multirow{2}{*}{WASP-39b} & cloud free & $2.73^{+0.03}_{-0.04}$ & $0.94^{+0.01}_{-0.02}$ & $1.28^{+0.02}_{-0.03}$ & $-5.43^{+0.17}_{-0.17}$ & $966^{+27}_{-13}$ & $-$ \\
& grey clouds & $2.65^{+0.04}_{-0.05}$ & $0.98^{+0.02}_{-0.03}$ & $1.27^{+0.02}_{-0.03}$ & $-3.94^{+0.92}_{-0.86}$ & $1045^{+107}_{-66}$ & $-2.35^{+0.85}_{-0.91}$ \\ \hline 
\end{tabular}
\end{table*}

\begin{table*}
\centering
\caption{Free parameters retrieved using cloud-free and grey-cloud models from \citet{Sing+16} data (see Section \ref{sec:data_reduction}).}
\label{table:data_reduction_comparison_results}
\begin{tabular}{lcccccccc}
\hline
Planet & Data & Model & log $g_{\rm p}$ (cm s$^{-2}$) & $R_{\rm star}$ (R$_\odot$) & $R_{\rm p}$ (R$_{\rm Jup}$) & log $X_{\ch{H2O}}$ & $T$ (K) & log $P_{\rm cloud\mbox{-}top}$ (bar) \\ \hline
\multirow{2}{*}{HD 209458b} & \multirow{2}{*}{WFC3 only} & cloud free & $3.0^{+0.04}_{-0.05}$ & $1.24^{+0.04}_{-0.03}$ & $1.44^{+0.04}_{-0.04}$ & $-5.49^{+0.21}_{-0.19}$ & $1081^{+97}_{-94}$ & $-$ \\
& & grey clouds & $2.92^{+0.04}_{-0.05}$ & $1.25^{+0.02}_{-0.02}$ & $1.4^{+0.02}_{-0.01}$ & $-5.38^{+0.66}_{-0.41}$ & $1783^{+236}_{-304}$ & $-1.35^{+0.47}_{-0.64}$ \\ \hline
\multirow{2}{*}{HD 209458b} & \multirow{2}{*}{STIS+WFC3+IRAC} & cloud free & $3.04^{+0.02}_{-0.03}$ & $1.23^{+0.02}_{-0.03}$ & $1.44^{+0.03}_{-0.04}$ & $-5.25^{+0.13}_{-0.11}$ & $907^{+49}_{-36}$ & $-$ \\
& & grey clouds & $3.03^{+0.03}_{-0.06}$ & $1.23^{+0.02}_{-0.02}$ & $1.43^{+0.03}_{-0.03}$ & $-5.23^{+0.15}_{-0.12}$ & $899^{+57}_{-95}$ & $-0.11^{+0.64}_{-0.37}$ \\ \hline
\end{tabular}
\end{table*}

\bsp
\label{lastpage}

\end{document}